\documentclass[suppldata]{interact}

\usepackage{epstopdf}% To incorporate .eps illustrations using PDFLaTeX, etc.
\usepackage[caption=false]{subfig}% Support for small, `sub' figures and tables

\usepackage[numbers,sort&compress]{natbib}% Citation support using natbib.sty
\bibpunct[, ]{[}{]}{,}{n}{,}{,}% Citation support using natbib.sty
\theoremstyle{plain}% Theorem-like structures provided by amsthm.sty

\theoremstyle{definition}

\theoremstyle{remark}

\usepackage[hyphens]{url}
\usepackage{float,epsfig,epstopdf}
\usepackage{graphicx}
\usepackage{amssymb}
\usepackage{amsmath}
\usepackage{bm}
\usepackage[colorlinks,citecolor=blue,linkcolor=blue,urlcolor=blue]{hyperref}
\usepackage{arydshln}
\usepackage{appendix}
\usepackage{pdflscape}
\usepackage{cancel}
\usepackage{multirow}

\begin{document}

	\title{A time-varying bivariate copula joint model for longitudinal and time-to-event data}
	\author	{
	\name{Zili Zhang, Christiana Charalambous and Peter Foster}
	\affil{Department of Mathematics, University of Manchester, Manchester M13 9PL, UK}
}
	\maketitle
	
	\begin{abstract}
A time-varying bivariate copula joint model, which models the repeatedly measured longitudinal outcome at each time point and the survival data jointly by both the random effects and time-varying bivariate copulas, is proposed in this paper. A regular joint model normally supposes there exist subject-specific latent random effects or classes shared by the longitudinal and time-to-event processes and the two processes are conditionally independent given these latent variables. Under this assumption, the joint likelihood of the two processes is straightforward to derive and their association, as well as heterogeneity among the population, are naturally introduced by the unobservable latent variables. However, because of the unobservable nature of these latent variables, the conditional independence assumption is difficult to verify. Therefore, besides the random effects, a time-varying bivariate copula is introduced to account for the extra time-dependent association between the two processes. The proposed model includes a regular joint model as a special case under some copulas. Simulation studies indicates the parameter estimators in the proposed model are robust against copula misspecification and it has superior performance in predicting survival probabilities compared to the regular joint model. A real data application on the Primary biliary cirrhosis (PBC) data is performed.
	\end{abstract}

	
	\section{Introduction}
	Joint modelling of longitudinal measurements and time-to-event data has become increasingly popular in recent decades, as previous studies indicate separately modelling (Tsiatis, DeGruttola and Wulfsohn, 1995\cite{tsi95} and Guo and Carlin, 2004\cite{guo04}) or a two-stage modelling (Wulfsohn and Tsiatis, 1997\cite{wul97}) of the two processes can result in biased estimations due to not fully considering the association between the two sub-models. The classical joint model in Faucett and Thomas, (1996)\cite{fau96} and Wulfsohn and Tsiatis (1997)\cite{wul97} assumes a linear mixed effects model and a proportional hazard model for the  longitudinal and survival sub-models, respectively. Thereafter, joint models with non-linear mixed longitudinal process modelled by cubic B-splines or functional principal components have been 
	 proposed by Brown \textit{et al.} (2005)\cite{bro05}, Yao (2007)\cite{yao07} and Li \textit{et al.} (2021)\cite{li21}. Li and Luo (2017)\cite{li17} and (2019)\cite{li19} incorporate functional covariates, which is a continuous curve over some domain, into the joint model. A latent class joint model is discussed by Lin \textit{et al.} (2002)\cite{lin02}, (2004)\cite{lin04} and  Proust-Lima \textit{et al.} (2009)\cite{pro09}. Excellent overviews on joint modelling can be found in Tsiatis and Davidian (2004)\cite{tsi04},  Ibrahim \textit{et al.} (2010)\cite{ibr10}, Papageorgiou \textit{et al.} (2019)\cite{pap19} and Alsefri \textit{et al.} (2020)\cite{als20}. Despite the diverse extension of the joint modelling framework in recent years, conditional independence of the two sub-models given the latent variables is an assumption which has always remained stable. This assumption is tricky to verify as the latent variables are not observable. Roy (2003)\cite{roy03}, Guo \textit{et al.} (2006)\cite{guo06} and Jacqmin-Gadda \textit{et al.} (2010)\cite{jac10} proposed some tests to assess conditional independence given latent classes via a score test or testing the dependency within each class after randomly allocating the subjects into classes using the estimated posterior class-membership probabilities, but these tests apply only to the latent class joint model. 
	 
	 Emura \textit{et al.} (2017)\cite{emu17} pointed out  joint analysis of two survival outcomes, such as death and relapse of cancer, by a joint frailty model (Rondeau \textit{et al.,} 2015\cite{ron15}), which assumes conditional independence between the two survival outcomes given a shared study-specific frailty term, may not be sufficient to account for all the dependency between outcomes in a subject level, especially when the covariates information are insufficiently collected. For this reason, they introduced a joint frailty-copula model, which associates the two survival outcomes by both shared frailty terms and copula. It is reasonable to have the same doubts about conditional independence in the joint modelling of longitudinal and time-to-event data when there are only a few covariates or the latent variable structure is too simple. To alleviate this issue, Henderson \textit{et al.} (2000)\cite{hen00} replaced the simple shared random effects in the two sub-models by a combination of correlated random effects and a mean-zero bivariate Gaussian stochastic process, representing a long-term trend and local variation at subject-level. However, the infinite dimension of the stochastic process comes with intensive computation and the baseline hazard function is only allowed to be modelled nonparametrically by taking mass at each failure time due to integrability issues. In addition, Hsieh \textit{et al.} (2006)\cite{hsi06} pointed out the unspecified baseline hazard leads to underestimation of the standard errors for parameters in the EM algorithm. Similar to Henderson \textit{et al.} (2000)\cite{hen00}, Wang \textit{et al.} (2001)\cite{wan01} applied a random intercept and an integrated Ornstein-Uhlenbeck stochastic process for the latent variables. As a compromise, the stochastic process was treated as a piecewise constant function for estimation. In more recent work, Dutta \textit{et al.} (2021)\cite{dut21} modelled the joint distribution of log-transformed survival time and the longitudinal measurement, conditional on the shared random effects, by a multivariate normal distribution. This model introduced dependency between the two processes by both the random effects and covariance matrix of the multivariate normal distribution, but the assumption of log normal distribution on the event time may be too restrictive.

   Associating the longitudinal and survival sub-models by copula (Hofert \textit{et al.,} 2018)\cite{hof18} has also been considered. Rizopoulos \textit{et al.} (2008a\cite{riz08a}, 2008b\cite{riz08b}) and Malehi \textit{et al.} (2015)\cite{mal15} applied copulas to model the joint distribution of the latent random effects, which offered greater flexibility of association compared with the traditional shared random effects joint model, by allowing different copulas and marginal distributions for random effect components. However, conditional independence is still assumed in these models. Applying multivariate copulas directly on marginals of all the longitudinal measurements and event time data for each subject was considered by Ganjali and Baghfalaki (2015)\cite{gan15}, Zhang \textit{et al.}(2023a)\cite{zha23a} and Cho \textit{et al.}(2024)\cite{cho24}. The non-linear correlation introduced by the copula between the two sub-models is different to the linear one arising from latent random effects, but the lack of random effects in the joint models compromises their capability to capture the longitudinal trajectories at the subject level. For this reason, Zhang \textit{et al.}(2023b)\cite{zha23b} further proposed modelling the joint distribution of all the longitudinal and survival outcomes in each subject, given the random effects, by a multivariate Gaussian copula. This model has two layers of correlation between the two sub-models and a regular joint model of conditional independence becomes a specially case when the correlation matrix in the multivariate Gaussian copula is an identity matrix. However, in these multivariate copula approaches of joint modelling, the difficulty of modelling the correlation matrix in the copula increases with the number of repeated longitudinal measurements. The approach of Suresh \textit{et al.}(2021)\cite{sur21a} can solve this issue by applying a bivariate Gaussian copula on the conditional joint distribution of survival time and a longitudinal measurement at a single time point given the subject being alive by that time point, since the correlation matrix of a two-dimensional copula is easier to model. But their model still falls under the marginal models as Zhang \textit{et al.}(2023a)\cite{zha23a}  for not including the latent effects.
   
   In this paper, we propose a  time-varying bivariate copula joint model by modelling the joint distribution of survival time and a single longitudinal measurement, conditional on the random effects and that the subject survives beyond the time point of this longitudinal observation, through a  time-varying bivariate copula. The time-invariant random effects introduce the dominant or long-term association between the two sub-models, while the time-varying bivariate copula captures the residual correlation from the local biological variation. A cubic B-spline function is applied to characteristic the possible dynamic nature of the correlation structure in the bivariate copula. 
   
   The remainder of the paper is organised as followed. Section 2 describes the notations and specification of the proposed model. In Section 3, simulations are conducted to assess the performance of the proposed model in terms of parameter estimation and dynamic prediction of survival probabilities and its performance under copula misspecification is investigated. A real data application is carried out in Section 4. The limitations of the proposed model are discussed and some future work is suggested in Section 5.

	\section{Time-varying bivariate copula joint model framework}
Suppose there are $n$ subjects being followed over a period of time. For the $i$th, $i=1,...,n,$ subject, its longitudinal observations $\bm Y_{i}=\left\{Y_{i1}=Y_{i}(s_{i1}),...,Y_{in_{i}}=Y_{i}(s_{in_{i}})\right\}^{'}$  are measured  intermittently at time points $\bm s_{i}=(s_{i1},...,s_{in_{i}})$ and are terminated by the observed event time $T_{i}=\mbox{min}(C_{i},T_{i}^{*})$ with $C_{i}$ and $T_{i}^{*}$ being the right censoring time and true event time, respectively. Thus, observing a longitudinal measurement $Y_{ij}$ at  $s_{ij},$ $1\leq j\leq n_{i},$ implicitly implies $Y_{ij}|T_{i}^{*}>s_{ij}.$  We also denote $\delta_{i}=I(T_{i}^{*}<C_{i})$ as the corresponding event indicator, which takes value 1 if the true event time is observed and 0 otherwise, for subject $i.$ The observation time points $\bm s_{i}$ and the censoring process $C_{i}$ are assumed to be uninformative conditional on the baseline covariates.  

Suppose the longitudinal process for subject $i$ is specified by the following model:
\begin{eqnarray}
	Y_{ij}=\bm x_{ij}^{'}\bm\beta_{1}+\bm z_{ij}^{'}\bm b_{i}+\varepsilon_{ij},\mbox{  }i=1,...,n, \mbox{   }j=1,...,n_{i},\label{longsub}
\end{eqnarray}
where  $\bm x_{ij}=\bm x_{i}(s_{ij})$ is a $p\times1$ covariate vector with fixed effects $\bm\beta_{1},$ $\bm z_{ij}=\bm z_{i}(s_{ij})$ is a $r\times1$ covariate vector for random effects  $\bm b_{i}\sim N_{r}(\bm0,\bm D),$  and the unexplained variation  $\displaystyle\varepsilon_{ij}=\varepsilon_{i}(s_{ij})\sim N(0,\sigma^{2}).$  

The corresponding survival process for this subject is given by a proportional hazard model (Cox, 1972\cite{cox72}) with frailty terms as:
 \begin{eqnarray}
 	h_{i}(t)=h_{0}(t)\mbox{exp}\left\{\bm w_{i}^{'}\bm\beta_{2}+\gamma\bm z_{i}(t)^{'}\bm b_{i}\right\},\label{sursub}
 \end{eqnarray}
where $\bm w_{i}$ is a $q\times1$ vector of baseline explanatory variables for the survival process with associated regression parameter vector $\bm\beta_{2},$ the parameter $\gamma$ characterises the dependency between the two sub-models. Common covariates are allowed for $\bm x_{i}$ and $\bm w_{i}$. The baseline hazard function is common for all subjects and assumed to be a piecewise-constant function having $K-1$ equally spaced internal knots (Rizopoulos, 2010\cite{riz10}) as:
\[h_{0}(t)=\sum_{k=1}^{K}\lambda_{k}I(v_{k-1}<t\leq v_{k}),\]
where $0=v_{0}<v_{1}<\cdots<v_{K}=t_{max}=\mbox{max}\left\{t_{i},\mbox{ }i=1,...,n\right\},$ such that $[0,t_{max}]$ is split into $K$ intervals, each with a constant baseline hazard $\lambda_{k}$.

According to the recording time point $\bm s_{i}$ of the longitudinal process, the information contributed by subject $i$ can be described in a progressive way as follows:

i. At the origin of the two processes, i.e. at $t=0,$ some baseline covariates are taken. Does the event occur before the first scheduled longitudinal measurement planned at $t=s_{i1}?$ If not, the two processes continue; 

ii. At $t=s_{ij},$ $j=1,...,n_{i}-1,$ a longitudinal measurement is observed and we monitor if the event is censored or observed before the next scheduled longitudinal measurement planned at $t=s_{i(j+1)}.$ If not, the two processes continue; 

iii. At $t=s_{in_{i}},$ a longitudinal measurement is observed, then the event is censored or observed at $T_{i}=t_{i}$ before the next longitudinal measurement can be recorded. The two processes are terminated at $T_{i}=t_{i}$ with an associated event indicator $\delta_{i}.$

\noindent The above steps decompose the observed likelihood of subject $i$ as:
\begin{eqnarray}
	\nonumber	L_{i}&=&f_{T_{i}^{*},\bm Y_{i}}(t_{i}, \bm y_{i})\\
	\nonumber	&=&P_{T_{i}^{*}}(T_{i}^{*}>s_{i1})\times f_{T_{i}^{*},Y_{i1}}(T_{i}^{*}>s_{i2}, y_{i1}| T_{i}^{*}>s_{i1})\times f_{T_{i}^{*},Y_{i2}}(T_{i}^{*}>s_{i3}, y_{i2}| T_{i}^{*}>s_{i2}, y_{i1})\times\cdots\\
	\nonumber	&\times& f_{T_{i}^{*},Y_{i(n_{i}-1)}}(T_{i}^{*}>s_{in_{i}}, y_{i(n_{i}-1)}| T_{i}^{*}>s_{i(n_{i}-1)}, y_{i1},...,y_{i(n_{i}-2)})\\
	\nonumber	&\times&f_{T_{i}^{*},Y_{in_{i}}}(T_{i}^{*}>t_{i}, y_{in_{i}}| T_{i}^{*}>s_{in_{i}}, y_{i1},...,y_{i(n_{i}-1)})^{1-\delta_{i}}\\
	&\times&f_{T_{i}^{*},Y_{in_{i}}}(T_{i}^{*}=t_{i}, y_{in_{i}}| T_{i}^{*}>s_{in_{i}}, y_{i1},...,y_{i(n_{i}-1)})^{\delta_{i}}.\label{factorlik}
\end{eqnarray}
 Each term can be interpreted as the information updated between interval $[s_{ij},s_{i(j+1)})$ conditional on the previous information. Suppose the dependency between the intervals are introduced by the time-invariant subject-specific random effects $\bm b_{i},$ then (\ref{factorlik}) can be rewritten as:
	\begin{eqnarray}
	\nonumber	L_{i}&=&f_{T_{i}^{*},\bm Y_{i}}(t_{i}, \bm y_{i})=\int f_{T_{i}^{*},\bm Y_{i}}(t_{i}, \bm y_{i}|\bm b_{i})f_{\bm b_{i}}\left(\bm b_{i}\right) \, d\bm b_{i}\\
	\nonumber	&=&\int_{\bm b_{i}} P_{T_{i}^{*}}(T_{i}^{*}>s_{i1}| \bm b_{i})\times f_{T_{i}^{*},Y_{i1}}(T_{i}^{*}>s_{i2}, y_{i1}| T_{i}^{*}>s_{i1},\bm b_{i})\times f_{T_{i}^{*},Y_{i2}}(T_{i}^{*}>s_{i3}, y_{i2}| T_{i}^{*}>s_{i2}, \bm b_{i})\times\cdots\\
	\nonumber	&\times& f_{T_{i}^{*},Y_{i(n_{i}-1)}}(T_{i}^{*}>s_{in_{i}}, y_{i(n_{i}-1)}| T_{i}^{*}>s_{i(n_{i}-1)}, \bm b_{i})\times f_{T_{i}^{*},Y_{in_{i}}}(T_{i}^{*}>t_{i}, y_{in_{i}}| T_{i}^{*}>s_{in_{i}}, \bm b_{i})^{1-\delta_{i}}\\
	&\times&f_{T_{i}^{*},Y_{in_{i}}}(T_{i}^{*}=t_{i}, y_{in_{i}}| T_{i}^{*}>s_{in_{i}}, \bm b_{i})^{\delta_{i}}f_{\bm b_{i}}\left(\bm b_{i}\right) \, d\bm b_{i}.\label{factorlikbi}
\end{eqnarray}
Model (\ref{factorlikbi}) implies $\varepsilon_{ij}$ includes both the measurement error and  local biological variation. While the local biological variations could be regraded as independent if the gaps of the two adjacent visit times are large (Tsiatis and Davidian, 2004\cite{tsi04}), it may still have impact on the survival process. Unlike the permanent effect imposed by the random effects, the local biological variation only has local influence on the survival process, thus conditional independence between intervals is assumed.

According to the two sub-models in (\ref{longsub}) and (\ref{sursub}), the distribution of the two processes, conditional on the random effects $\bm b_{i}$ and subject $i$ being alive up to $t=s_{ij}$, are given by:
\begin{eqnarray*}
	F_{T_{i}^{*}}(t|\bm b_{i}, T_{i}^{*}>s_{ij})=1-\mbox{exp}\left\{-\int_{s_{ij}}^{t}h_{i}(u)du\right\},	
\end{eqnarray*}
and
\begin{eqnarray*}
	F_{Y_{ij}}(y_{ij}|\bm b_{i}, T_{i}^{*}>s_{ij})=\Phi\left(\frac{y_{ij}-\bm x_{ij}^{'}\bm\beta_{1}-\bm z_{ij}^{'}\bm b_{i}}{\sigma}\right).	
\end{eqnarray*}

In the remaining paper, we denote $\displaystyle U_{t|\bm b_{i},s_{ij}}=F_{T_{i}^{*}}(t|\bm b_{i}, T_{i}^{*}>s_{ij})$ and $\displaystyle  U_{y_{ij}|\bm b_{i},s_{ij}}=F_{Y_{ij}}(y_{ij}|\bm b_{i}, T_{i}^{*}>s_{ij})$ for convenience. The joint distribution of $U_{T_{i}^{*}|\bm b_{i},s_{ij}}$ and $U_{Y_{ij}|\bm b_{i},s_{ij}}$ at each $s_{ij}$ is modelled by the bivariate copulas with a Kendall's correlation $\tau_{ij}$ capturing their residual dependency arising from local biological variation. The bivariate Gaussian and $t_{\nu}$ copulas are considered here since they are comprehensive copulas which have  an unrestricted Kendall's $\tau$ between -1 and 1.  Although the bivariate Clayton and Frank copulas also claim to be comprehensive, their formulas are actually defined separately on the copula parameters to take full range and numerical problems could occur when performing estimation (Yuan, 2007\cite{yua07}). Nevertheless, their joint likelihood under these two copulas are also provided in Appendix A for completeness.

 Let $\phi(\cdot),$ $\Phi(\cdot)$ and $\psi(\cdot;\nu),$ $\Psi(\cdot;\nu)$ be the pdfs and cdfs of the standard normal distribution and Student's $t$ distribution with $\nu$ degrees of freedom, respectively.

	\subsection{Bivariate Gaussian copula joint model} 
	Suppose the bivariate Gaussian copula is applied to characterise the joint distribution of $\displaystyle U_{T_{i}^{*}|\bm b_{i},s_{ij}}$ and $\displaystyle U_{Y_{ij}|\bm b_{i},s_{ij}}.$ Let $\displaystyle Z_{t|\bm b_{i},s_{ij}}=\Phi^{-1}(U_{t|\bm b_{i},s_{ij}})$ and $\displaystyle Z_{y_{ij}|\bm b_{i},s_{ij}}=\Phi^{-1}(U_{y_{ij}|\bm b_{i},s_{ij}})=(y_{ij}-\bm x_{ij}^{'}\bm\beta_{1}-\bm z_{ij}^{'}\bm b_{i})/\sigma.$ 
	Let $\Phi_{2}(\cdot; \alpha)$ and $\phi_{2}(\cdot; \alpha)$  denote the joint CDF and pdf of a bivariate standardised normal random vector with mean $\bm 0$ and the Pearson's correlation $\alpha$. The joint CDF of $T_{i}^{*}, Y_{ij}|T_{i}^{*}>s_{ij},\bm b_{i},$ $j=1,...,n_{i},$ is given by:
		\begin{eqnarray}
		\displaystyle
	\nonumber	F_{T_{i}^{*},Y_{ij}}(t, y_{ij}|\bm b_{i},T_{i}^{*}>s_{ij})=\Phi_{2}\left(Z_{t|\bm b_{i},s_{ij}},Z_{y_{ij}|\bm b_{i},s_{ij}};\alpha_{ij}\right),
	\end{eqnarray}
	Therefore its likelihood, depending on censored or not, can be derived as:
		\begin{eqnarray}
		\displaystyle
		f_{T_{i}^{*},Y_{ij}}(t, y_{ij}|\bm b_{i}, T_{i}^{*}>s_{in_{i}})=\sigma^{-1}\phi_{2}\left( Z_{t|\bm b_{i},s_{ij}}, Z_{y_{ij}|\bm b_{i},s_{ij}}; \alpha_{ij}\right)\frac{dU_{t|\bm b_{i},s_{ij}}/dt}{\phi\left(Z_{t|\bm b_{i},s_{ij}}\right)},\label{bivGaulikobs}
	\end{eqnarray}
or
	\begin{eqnarray}
	\displaystyle
	f_{T_{i}^{*},Y_{ij}}(T_{i}^{*}>t, y_{ij}|\bm b_{i},T_{i}^{*}>s_{ij})=\Phi\left(-\frac{Z_{t|\bm b_{i},s_{ij}}-\alpha_{ij}Z_{y_{ij}|\bm b_{i},s_{ij}}}{\sqrt{1-\alpha_{ij}^2}}\right)\frac{dU_{y_{ij}|\bm b_{i},s_{ij}}}{dy_{ij}},\label{bivGaulikcen}
\end{eqnarray}
where $-1<\alpha_{ij}<1$ controls the strength of dependency and it is a  function of Kendall's correlation as $\displaystyle\alpha_{ij}=\mbox{sin}(\pi\tau_{ij}/2).$ 
The complete likelihood under the bivariate Gaussian copula joint model can be obtained by substituting (\ref{bivGaulikcen}) and (\ref{bivGaulikobs}) back into (\ref{factorlikbi}). 

	\subsection{Bivariate $t_{\nu}$  copula joint model} 
	Suppose the bivariate $t_{\nu}$ copula is used to characterise the joint distribution of $\displaystyle U_{T_{i}^{*}|\bm b_{i},s_{ij}}$ and $\displaystyle U_{Y_{ij}|\bm b_{i},s_{ij}}.$ Let $\displaystyle W_{t|\bm b_{i},s_{ij}}^{\nu}=\Psi^{-1}\left(U_{t|\bm b_{i},s_{ij}};\nu\right)$ and $\displaystyle W_{y_{ij}|\bm b_{i},s_{ij}}^{\nu}=\Psi^{-1}\left(U_{y_{ij}|\bm b_{i},s_{ij}};\nu\right).$ 
	Let $\Psi_{2}(\cdot; \alpha,\nu)$ and $\psi_{2}(\cdot; \alpha,\nu)$ denote the joint CDF and pdf of a bivariate $t_{\nu}$ random vector with mean $\bm 0$ and Pearson's correlation $\alpha$. The joint CDF of  $T_{i}^{*}, Y_{ij}|T_{i}^{*}>s_{ij},\bm b_{i},$ $j=1,...,n_{i},$ is given by:
		\begin{eqnarray}
		\displaystyle
		\nonumber	F_{T_{i}^{*},Y_{ij}}(t, y_{ij}|\bm b_{i},T_{i}^{*}>s_{ij})=\Psi_{2}\left(W_{t|\bm b_{i},s_{ij}}^{\nu},W_{y_{ij}|\bm b_{i},s_{ij}}^{\nu};\alpha_{ij},\nu\right)
	\end{eqnarray}
	Therefore its likelihood, depending on censored or not, can be derived as:
		\begin{eqnarray}
		\displaystyle
		f_{T_{i}^{*},Y_{ij}}(t, y_{ij}|\bm b_{i}, T_{i}^{*}>s_{ij})=\psi_{2}\left(W_{t|\bm b_{i},s_{ij}}^{\nu}, W_{y_{ij}|\bm b_{i},s_{ij}}^{\nu}; \alpha_{ij}, \nu\right)\frac{dU_{t|\bm b_{i},s_{ij}}/dt}{\psi\left(W_{t|\bm b_{i},s_{ij}}^{\nu};\nu\right)}
			\frac{dU_{y_{ij}|\bm b_{i},s_{ij}}/dy_{ij}}{\psi\left(W_{y_{ij}|\bm b_{i},s_{ij}}^{\nu};\nu\right)},\label{bivtlikobs}
	\end{eqnarray}
or
	\begin{eqnarray}
	\displaystyle
	f_{T_{i}^{*},y_{ij}}(T_{i}^{*}>t, y_{ij}|\bm b_{i},T_{i}^{*}>s_{ij})=\Psi\left\{-\frac{W_{t|\bm b_{i},s_{ij}}^{\nu}-\alpha_{ij}W_{y_{ij}|\bm b_{i},s_{ij}}^{\nu}}{\sigma(s_{ij}|\bm b_{i},y_{ij})};\nu+1\right\}\frac{dU_{y_{ij}|\bm b_{i},s_{ij}}}{dy_{ij}},\label{bivtlikcen}
\end{eqnarray}
where $\displaystyle\sigma(s_{ij}|\bm b_{i},y_{ij})^{2}=[\{\nu+(W_{y_{ij}|\bm b_{i},s_{ij}}^{\nu})^{2}\}\left(1-\alpha_{ij}^{2}\right)]/(\nu+1)$. The $-1<\alpha_{ij}<1$ controls the strength of dependency and it is a  function of Kendall's correlation as $\displaystyle\alpha_{ij}=\mbox{sin}(\pi\tau_{ij}/2).$ The complete likelihood under the bivariate  $t_{\nu}$ copula joint model can be obtained by substituting (\ref{bivtlikobs}) and (\ref{bivtlikcen}) back into (\ref{factorlikbi}).

\subsection{Likelihood maximisation}
The integrals in terms of random effects in (\ref{factorlikbi}) do not have a closed form expression under either the bivariate Gaussian or $t_{\nu}$ copula joint model, thus a multivariate Gaussian quadrature technique (J\"{a}ckel, 2005\cite{jac05}) is applied to approximate the integration. However, the common weighting kernel $f_{\bm b_{i}}(\bm b_{i})$ and its quadrature points do not consider subject level information, thus they always concentrate around $\bm 0,$ while the main mass of the integrand in (\ref{factorlikbi}) is more likely to locate around the subject-specific random effects.  The inconsistency of the quadrature points and main mass of the integrand can result in an inaccurate numerical approximation even with a large number of quadrature points. For faster and more accurate calculation, we rearrange (\ref{factorlikbi}) as:
\begin{eqnarray}
	\displaystyle
\nonumber f_{T_{i}^{*},\bm Y_{i}}(t_{i},\bm y_{i})&=&\int_{\bm b_{i}}f_{T_{i}^{*},\bm Y_{i},\bm b_{i}}(t_{i},\bm y_{i},\bm b_{i})\, d\bm b_{i}=\int_{\bm b_{i}}f_{\bm Y_{i}}(\bm y_{i})f_{T_{i}^{*}}(t_{i}|\bm y_{i},\bm b_{i})f_{\bm b_{i}}(\bm b_{i}|\bm y_{i})\,d\bm b_{i}\\
	&=&\bm\phi_{n_{i}}\left(y_{i1}-\bm x_{i1}^{'}\bm \beta_{1},\dots,y_{in_{i}}-\bm x_{in_{i}}^{'}\bm \beta_{1};\bm V_{\bm y_{i}}\right)\int_{\bm b_{i}}f_{T_{i}^{*}}(t_{i}|\bm y_{i},\bm b_{i})f_{\bm b_{i}}(\bm b_{i}|\bm y_{i})\,d\bm b_{i},\label{factornewlikbi}
\end{eqnarray}
where $\bm\phi_{n_{i}}\left(\cdot;\bm V_{\bm y_{i}}\right)$ is the pdf of multivariate normal distribution with mean vector $\bm0$ and covariance matrix $\displaystyle\bm V_{\bm y_{i}}=(\bm z_{i1},\cdots,\bm z_{in_{i}})^{'}\bm D(\bm z_{i1},\cdots,\bm z_{in_{i}})+\sigma^2\bm I_{n_{i}},$ the $f_{\bm b_{i}}(\bm b_{i}|\bm y_{i})$ can be derived by the joint normality of $\bm Y_{i}$ and $\bm b_{i}$ (Wulfsohn and Tsiatis, 1997\cite{wul97}),  while 
$f_{T_{i}^{*}}(t_{i}|\bm y_{i},\bm b_{i})=f_{T_{i}^{*},\bm Y_{i}}(t_{i}, \bm y_{i}|\bm b_{i})/f_{\bm Y_{i}}(\bm y_{i}|\bm b_{i}).$ The expressions for $f_{T_{i}^{*}}(t_{i}|\bm y_{i},\bm b_{i})$ under the four bivariate copula joint models are provided in Appendix B.

The quadrature points of the weighting kernel $f_{\bm b_{i}}(\bm b_{i}|\bm y_{i})$ in (\ref{factornewlikbi}) are subject adaptive by including the information from the longitudinal process and expected to be closer to the subject-specific random effects as well as the main mass of the integrand $f_{T_{i}}(t_{i}|\bm y_{i},\bm b_{i}),$ which normally concentrates around subject-specific random effects. Thus higher accuracy of numerical approximation can be achieved by using fewer nodes in (\ref{factornewlikbi}) compared to the original parameterisation in (\ref{factorlikbi}). 

Although the Kendall's correlation function can be modelled continuously by cubic B-spline basis functions, the direct estimation of $\tau(t)$ is restricted by its range of [-1,1]. Therefore we consider $\displaystyle r(t)=\log\left[\left\{1+\tau(t)\right\}/\left\{1-\tau(t)\right\}\right]$ in the estimating process, then transform $\hat{r}(t)$ back to  $\hat{\tau}(t)$. The modelling of $r(t)$ is performed by a linear combination of $l$ cubic B-spline basis functions, $\bm B_{l}(t)=\left[B_{1}(t),...,B_{l}(t)\right]^{'},$  with $l\times1$ coefficient vector $\bm\eta,$ such that $r(t)=\bm B_{l}(t)^{'}\bm\eta$. Once the estimations of the coefficient vector $\bm\eta$ are obtained, the estimation of the $r(t)$ can be constructed as $\hat{r}(t)=\bm B_{l}(t)^{'}\hat{\bm\eta}.$ 

The maximisation of (\ref{factornewlikbi}) can be carried out numerically through a Newton-type algorithm (Dennis, \textit{et al,} 1983\cite{den83}) or the approach in Nelder and Mead (1965)\cite{nel65}, which are implemented by the \verb|nlm| and \verb|optim| functions, respectively, in \verb|R|, and the standard errors can be estimated from the inverse Hessian matrix as a by-product from the two functions. The initial values for the longitudinal and survival sub-models can be obtained by fitting a regular joint model through  \verb|jointModel| or  \verb|joint| functions from \verb|JM| or \verb|joineR| packages, respectively, while the correlation parameters in the bivariate copula functions can be initialised as 0s.

\subsection{Copula misspecification}
Since our model assume a parametric bivariate copula, we would like to investigate the impacts of misspecifying copula on parameter estimation. Assume the two sub-models and random effects distribution are correctly specified. Denote $\bm\theta=(\bm\theta_{y}^{'},\bm\theta_{t}^{'},\bm\theta_{b}^{'},\bm\theta_{\alpha}^{'})^{'}$ as the parameters of interest, where $\bm\theta_{y}=(\bm\beta_{1}^{'},\sigma)^{'}$, $\bm\theta_{t}=(\bm\beta_{2}^{'},\gamma,\lambda_{1},\dots,\lambda_{K})^{'}$, $\bm\theta_{b}=\mbox{vech}(\bm D)$ and $\bm\theta_{\alpha}$ is the parameters in the bivariate copulas. According to the likelihood function in (\ref{factornewlikbi}), its log-likelihood is:
\begin{eqnarray}
\nonumber \mbox{log}\left\{f_{T_{i}^{*},\bm Y_{i}}(t_{i},\bm y_{i})\right\}&=&\mbox{log}\left\{\bm\phi_{n_{i}}\left(y_{i1}-\bm x_{i1}^{'}\bm \beta_{1},\dots,y_{in_{i}}-\bm x_{in_{i}}^{'}\bm \beta_{1};\bm V_{\bm y_{i}}\right)\right\}\\
&+&\mbox{log}\left\{\int_{\bm b_{i}}f_{T_{i}^{*}}(t_{i}|\bm y_{i},\bm b_{i})f_{\bm b_{i}}(\bm b_{i}|\bm y_{i})\,d\bm b_{i}\right\}
	\end{eqnarray}
Denote $\mbox{log}\{\tilde{f}_{T_{i}^{*},\bm Y_{i}}(t_{i},\bm y_{i})\}$ as the log-likelihood under the misspecified copula $\tilde{c}(\cdot),$ we have
\begin{eqnarray}
\nonumber	\mbox{log}\left\{\tilde{f}_{T_{i}^{*},\bm Y_{i}}(t_{i},\bm y_{i})\right\}&=&\mbox{log}\left\{\bm\phi_{n_{i}}\left(y_{i1}-\bm x_{i1}^{'}\bm \beta_{1},\dots,y_{in_{i}}-\bm x_{in_{i}}^{'}\bm \beta_{1};\bm V_{\bm y_{i}}\right)\right\}\\
	&+&\mbox{log}\left\{\int_{\bm b_{i}}\tilde{f}_{T_{i}^{*}}(t_{i}|\bm y_{i},\bm b_{i})f_{\bm b_{i}}(\bm b_{i}|\bm y_{i})\,d\bm b_{i}\right\}
\end{eqnarray}
Denote $\hat{\bm\theta}$ and $\tilde{\bm\theta}$ as the maximum likelihood estimators under the correct and misspecified log-likelihoods. The difference between $f_{T_{i}^{*}}(t_{i}|\bm y_{i},\bm b_{i})$ and $\tilde{f}_{T_{i}^{*}}(t_{i}|\bm y_{i},\bm b_{i})$ would result in biases in $\tilde{\bm\theta}$.

We then investigate the performance of  $\tilde{\bm\theta}$ as $n_{i}\to\infty$. Note that $\mbox{log}\left\{f_{\bm b_{i}}\left(\bm b_{i}|\bm y_{i}\right)\right\}=\mbox{Constant}+\mbox{log}\left\{f_{\bm Y_{i}}(\bm y_{i}|\bm b_{i})\right\}+\mbox{log}\left\{f_{\bm b_{i}}(\bm b_{i})\right\}$. Since $\mbox{log}\left\{f_{\bm Y_{i}}(\bm y_{i}|\bm b_{i})\right\}=\sum_{j=1}^{n_{i}}\mbox{log}\left\{f_{Y_{ij}}(y_{ij}|\bm b_{i})\right\}$ is $O(n_{i})$, while $\mbox{log}\left\{f_{\bm b_{i}}(\bm b_{i})\right\}$ is $O(1)$ and smooth functions with regard to $\bm b_{i}$. The asymptotic Bayesian theory  (Cox and Hinkley, 1974\cite{cox74}) states that $f_{\bm b_{i}}\left(\bm b_{i}|\bm y_{i}\right)$
converges to a multivariate normal density, with mean vector $\tilde{\bm b}_{i}=\operatorname{arg\,max}_{\bm b_{i}}\mbox{log}\left\{f_{\bm Y_{i}}(\bm y_{i}|\bm b_{i})\right\}$ and variance-covariance matrix $\tilde{D}_{i}=[-\partial^{2}\mbox{log}\left\{f_{\bm Y_{i}}(\bm y_{i}|\bm b_{i})\right\}/\partial\bm b_{i}\partial\bm b_{i}^{\top}|_{\bm b_{i}=\tilde{\bm b}_{i}}]^{-1}$ as $n_{i}$ increases. Moreover, this distribution is going to degenerate to its mean $\tilde{\bm b}_{i}$ as $n_{i}\to\infty$. 

Note that $\mbox{log}\left\{\bm\phi_{n_{i}}\left(y_{i1}-\bm x_{i1}^{'}\bm \beta_{1},\dots,y_{in_{i}}-\bm x_{in_{i}}^{'}\bm \beta_{1};\bm V_{\bm y_{i}}\right)\right\}$ is $O_{p}(n_{i})$, while the second terms of (10) and (11) converge to 
$\mbox{log}\{f_{T_{i}^{*}}(t_{i}|\bm y_{i},\tilde{\bm b}_{i})\}$ and $\mbox{log}\{\tilde{f}_{T_{i}^{*}}(t_{i}|\bm y_{i},\tilde{\bm b}_{i})\}$, respectively, and they are both $O_{p}(1)$. For fixed sample size $n$, let  $m=\mbox{min}\left\{n_{i}, i=1,\dots,n\right\}\to\infty$, the correct and misspecified log-likelihood, after normalising by $m$, are
\begin{eqnarray}
\nonumber \frac{1}{m}l=\frac{1}{m}\sum_{i=1}^{n}\mbox{log}\left\{\bm\phi_{n_{i}}\left(y_{i1}-\bm x_{i1}^{'}\bm \beta_{1},\dots,y_{in_{i}}-\bm x_{in_{i}}^{'}\bm \beta_{1};\bm V_{\bm y_{i}}\right)\right\}+\frac{1}{m}\sum_{i=1}^{n}\mbox{log}\left\{f_{T_{i}^{*}}(t_{i}|\bm y_{i},\tilde{\bm b}_{i})\right\}
\end{eqnarray}
\begin{eqnarray}
\nonumber \frac{1}{m}\tilde{l}=\frac{1}{m}\sum_{i=1}^{n}\mbox{log}\left\{\bm\phi_{n_{i}}\left(y_{i1}-\bm x_{i1}^{'}\bm \beta_{1},\dots,y_{in_{i}}-\bm x_{in_{i}}^{'}\bm \beta_{1};\bm V_{\bm y_{i}}\right)\right\}+\frac{1}{m}\sum_{i=1}^{n}\mbox{log}\left\{\tilde{f}_{T_{i}^{*}}(t_{i}|\bm y_{i},\tilde{\bm b}_{i})\right\}.
\end{eqnarray}
The second terms of $l/m$ and $\tilde{l}/m$ are both $o_{p}(1)$, while their first term are the same and $O_{p}(1).$ Therefore,  $l/m\overset{p}{\to} \tilde{l}/m$ and the maximum likelihood estimators of $\bm\theta_{y}$ and $\bm\theta_{b}$ under $l$ and $\tilde{l}$ both converge in probability to the maximiser of $1/m\sum_{i=1}^{n}\mbox{log}\left\{\bm\phi_{n_{i}}\left(y_{i1}-\bm x_{i1}^{'}\bm \beta_{1},\dots,y_{in_{i}}-\bm x_{in_{i}}^{'}\bm \beta_{1};\bm V_{\bm y_{i}}\right)\right\}$, which means $\tilde{\bm\theta}_{y}\overset{p}{\to} \hat{\bm\theta}_{y}$ and $\tilde{\bm\theta}_{b}\overset{p}{\to} \hat{\bm\theta}_{b}$ as $m\to\infty$.

	\section{Simulation studies}
Simulation studies are conducted to investigate the finite sample performance of the proposed model. The following sub-models for the two processes are applied throughout the simulation studies. The longitudinal process is specified as:
		\begin{eqnarray}
			Y_{ij}=\beta_{10}+\beta_{11}s_{ij}+\beta_{12}x_{i1}+\beta_{13}x_{i2}+\beta_{14}I(cat_{i}=x_{3})+\beta_{15}I(cat_{i}=x_{4})+b_{i0}+b_{i1}s_{ij}+\varepsilon_{ij},\label{longsubsim}
		\end{eqnarray} 
		and the survival process is taken to be:
		\begin{eqnarray}
			h_{i}(t)=h_{0}(t)\mbox{exp}\left\{\beta_{21}x_{i1}+\beta_{22}x_{i2}+\beta_{23}I(cat_{i}=x_{3})+\beta_{24}I(cat_{i}=x_{4})+\gamma(b_{i0}+b_{i1}t)\right\},\label{sursubsim}  
		\end{eqnarray}
		where $x_{i1}$ and $x_{i2}$ have probability 0.5 taking value 1 or 0, while $cat_{i}$ is a factor following a categorical distribution with probability $0.3,0.5$ and 0.2 being one of the categories of $x_{3},$ $x_{4}$ and $x_{5},$ emulating a covariate having three levels with $x_{5}$ as the reference level. The measurement error $\varepsilon_{ij}\sim N\left(0,\sigma^{2}\right)$ is independent of the random effects $(b_{i0},b_{i1})\sim N_{2}\left(0,\bm D\right)$ with vech$(\bm D)=\left(D_{11}, D_{12}, D_{22}\right)$. A constant baseline hazard function, $h_{0}(t)=1$, is selected for simplicity. 
		
		The four bivariate copulas discussed in Section 2 are used to link the two sub-models and three cases of Kendall's tau correlations are considered, namely case 1: constant positive correlation with $\tau(t)=0.5,$ case 2: time-varying correlation with $\displaystyle\tau(t)=\left[\mbox{exp}\left\{r(t)\right\}-1\right]/\left[\mbox{exp}\left\{r(t)\right\}-1\right],$ where $r(t)$ are consist of 6 cubic B-spline basis functions with equally spaced knots over $[0, 10.2]$ and coefficient vector $\bm\eta=(0.1,-1,2,-4,4,-0.1),$ and case 3:  constant negative correlation with $\tau(t)=-0.5.$ 
		
		The above setup are used for simulating data in the two scenarios of $n_{i}$ with scenario 1: the measurement times are scheduled at $t=0, 1,...,9, 10 $ with up to $\mbox{max}(n_{i})=11$ measurements per subject and scenario 2: the measurements are planned at $t=0, 2,...,8, 10 $ with up to $\mbox{max}(n_{i})=6$ measurements per subject. Except the origin $t=0,$ the time points are subjected to a uniform distribution error between $[-0.2,0.2].$ An independent censoring process following an exponential distribution of rate 0.011 is considered with the event process finally terminated at $t=11,$ resulting in around $50\%$ censoring rate for the event times.
		
		In each scenario, $N=500$ Monte Carlo samples each with sample size $n=200$ subjects are generated under all the combinations of copulas and Kendall's tau functions. The following candidate models are used to fit the simulated datasets.
		\begin{itemize}
			\item  \textbf{Bivariate Gaussian copula joint model (GJM-k)}: the marginals are correctly specified as (\ref{longsubsim}) and (\ref{sursubsim}), while the correlation between the two sub-models after conditioning on the random effects is introduced by the bivariate Gaussian copula function with $r(t)$ modelled by $k$ cubic B-spline basis functions with equally spaced knots over [0,10.2].
			\item  \textbf{Bivariate $t_{4}$ copula joint model (T4JM-k)}: the marginals are correctly specified as (\ref{longsubsim}) and (\ref{sursubsim}), while the correlation between the two sub-models after conditioning on the random effects is introduced by the bivariate $t_{4}$ copula function  with $r(t)$  modelled by $k$ cubic B-spline basis functions with equally spaced knots over [0,10.2].
			\item  \textbf{Regular joint model (RJM)}: the marginals are correctly specified as (\ref{longsubsim}) and (\ref{sursubsim}) and assumed to be conditionally independent given the random effects.
		\end{itemize}
	
		\subsection{Fitted results}
	In each scenario, twelve types of datasets are generated under different combinations of copulas and Kendall's tau functions while four types of datasets under the four types of bivariate copulas are simulated in each case. Denote them as Gaussian, $t_{4}$, Frank and Clayton datasets for simplicity. Each type of dataset are fitted by the three candidate models. We present and discuss some of the results from scenario 1. In Tables 1 and 2, SE denotes the model based standard calculated from the observed Fisher information matrix, SD is the standard deviation calculated from the estimates based on the 500 Monte Carlo samples and CP are coverage probability for the 95\% confidence intervals based on SE. The remaining outputs for the full simulation studies are available in the supplementary materials.

\noindent\textbf{Case 1: positive constant Kendall's tau correlation: $\tau(t)=0.5$}

 Table 1 displays the results fitted by GJM-4, T4JM-4 and RJM  for the four types of data  in case 1 of scenario 1.   In this case, except the random effects, there is a strong positive constant correlation introduced by the bivariate copula functions between the two processes in the simulated data. The RJM completely ignore the correlation in the copulas, thus obvious biases is observed in some of the parameter estimations. Specifically, the association parameter $\gamma$ is severely overestimated to compensate for the extra dependency from the bivariate copula functions. Moderate biases can be observed in the regression parameters $\bm\beta_{2}$ of the survival sub-model and $\beta_{11}$ of the longitudinal sub-model, although the biases in the remaining parameters of $\bm\beta_{1}$ is small. If comparing the fitted outputs by RJM across the four types of dataset, the most significant biases can be found for the dataset from the bivariate Clayton copula joint model, while the least biases are observed when fitting the Frank datasets. The biases of fitting RJM to the Gaussian and $t_{4}$ datasets are roughly the same and between that of the previous two datasets. 

On the other hand, even the max$(n_{i})$ is just 11, the parameters estimated by GJM-4 and T4JM-4 are almost unbiased across the  four types datasets no matter the bivariate copula functions are misspecified or not. Denote  $f^{*}_{T_{i}^{*},\bm Y_{i}}(t_{i}, \bm y_{i})$ as the likelihood of the data under the exact joint model (EJM), which is the correct model with the true parameter values substituted in. White (1982)\cite{whi82} states that the maximum likelihood estimator $\tilde{\bm\theta}$ will converge in probability to the value that minimise the Kullback-Leibler divergence $D_{KL}(f^{*}||\tilde{f})=\int f^{*}_{T_{i}^{*},\bm Y_{i}}(t_{i}, \bm y_{i})\mbox{log}\{f^{*}_{T_{i}^{*},\bm Y_{i}}(t_{i}, \bm y_{i})/\tilde{f}_{T_{i}^{*},\bm Y_{i}}(t_{i}, \bm y_{i})\}d\bm y_{i}dt_{i}$. According to the log likelihood in (10) and (11), the minimisation can be achieved by setting  $\tilde{\bm\theta}_{y}$ and $\tilde{\bm\theta}_{b}$ around their true values then select the values of $\tilde{\bm\theta}_{t}$ and $\tilde{\bm\theta}_{\alpha}$ that minimised  $|\mbox{log}\{\tilde{f}_{T_{i}^{*}}(t_{i}|\bm b_{i},\bm y_{i})\}-\mbox{log}\{f^{*}_{T_{i}^{*}}(t_{i}|\bm b_{i},\bm y_{i})\}|$. In fact, for GJM-4 and T4JM-4, the simulation results further suggest that their $\tilde{\bm\theta}_{t}$ are also close to its true value and their $|\mbox{log}\{\tilde{f}_{T_{i}^{*}}(t_{i}|\bm b_{i},\bm y_{i})\}-\mbox{log}\{f^{*}_{T_{i}^{*}}(t_{i}|\bm b_{i},\bm y_{i})\}|$ can be effectively reduce by adjusting the copula parameters $\bm\theta_{\alpha}$, while the RJM lacks this flexibility and results in biased $\tilde{\bm\theta}_{t}$.

\begin{figure}[H]
		\centering
		\begin{minipage}{0.49\textwidth}
			\includegraphics[width=\linewidth]{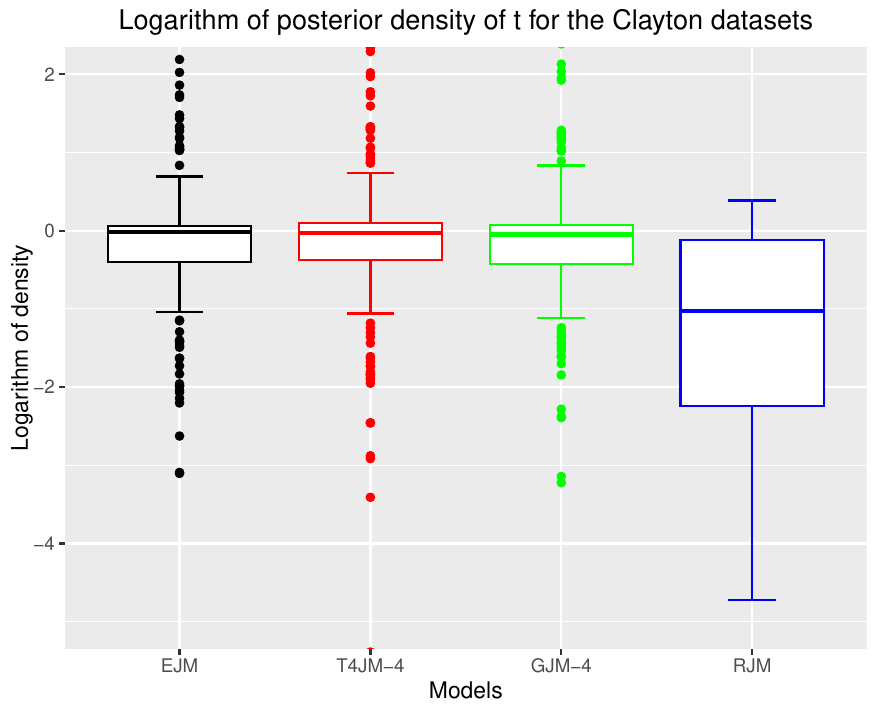}
		\end{minipage}
		\begin{minipage}{0.49\textwidth}
			\includegraphics[width=\linewidth]{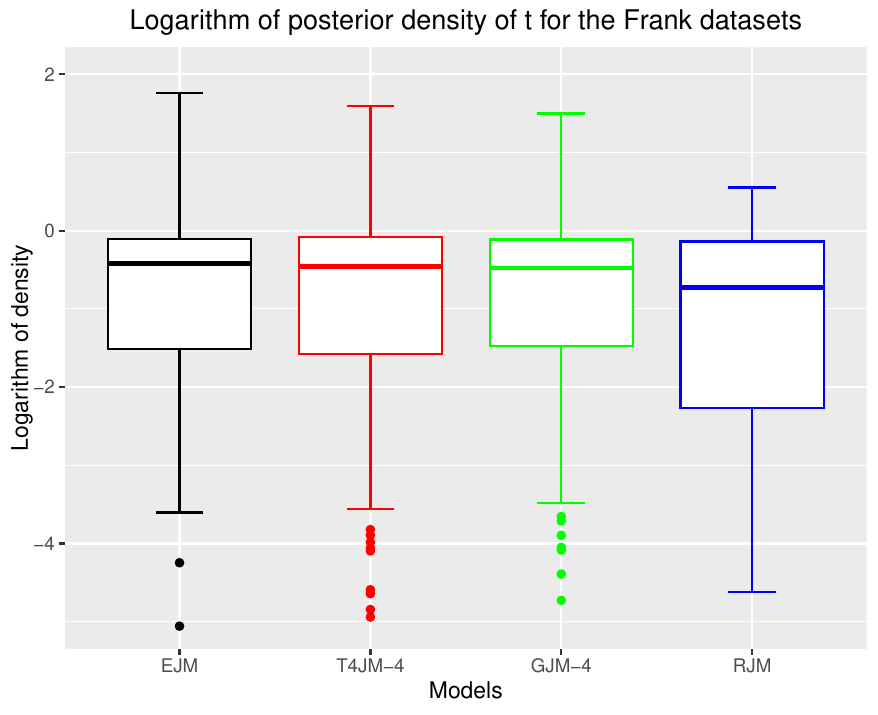}
		\end{minipage}
		\caption{The logarithm of ${f}_{T_{i}}(t_{i}|\bm b_{i},\bm y_{i})$ by the EJM, GJM-4, T4JM-4 and RJM with the true value of $\bm\theta_{y}$, $\bm\theta_{t}$, $\bm\theta_{b}$ and the estimated $\hat{\tau}(t)$ in Figure \ref{estClayandFratau} substituted in under the Clayton and Frank datasets from case 1 of scenario 1.}
		\label{ftipos_c0.5_200plot}
	\end{figure}

\begin{figure}[H]
		\centering
		\begin{minipage}{0.49\textwidth}
			\includegraphics[width=\linewidth]{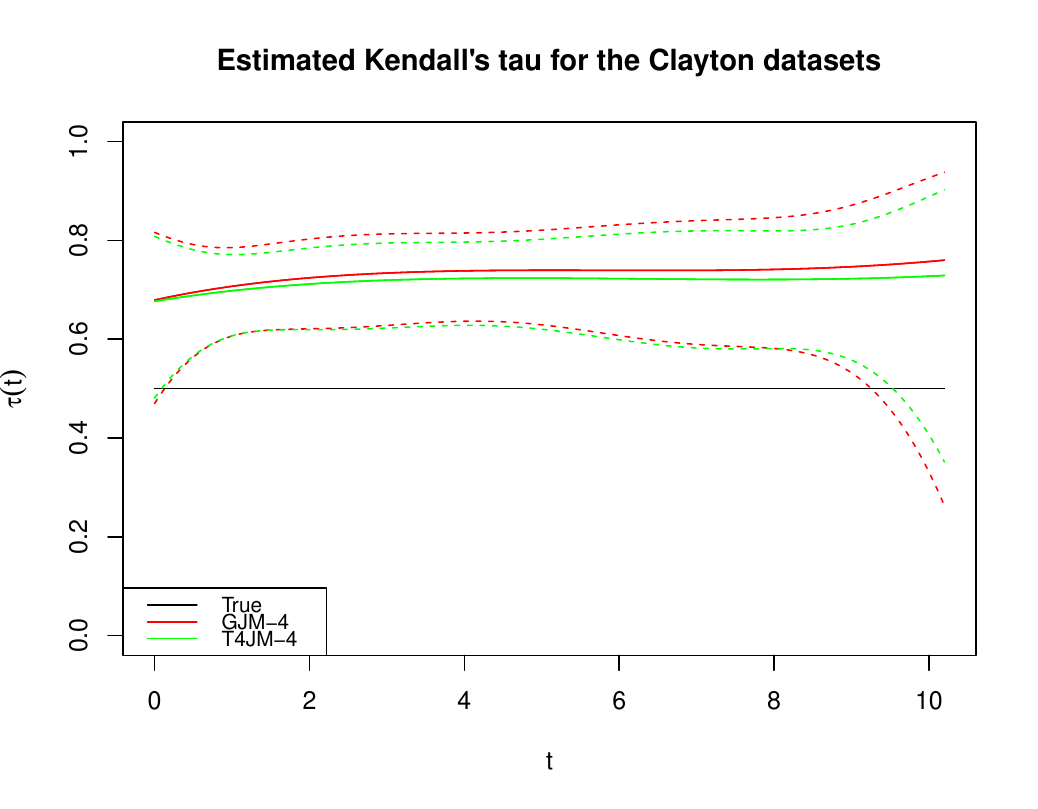}
		\end{minipage}
		\begin{minipage}{0.49\textwidth}
			\includegraphics[width=\linewidth]{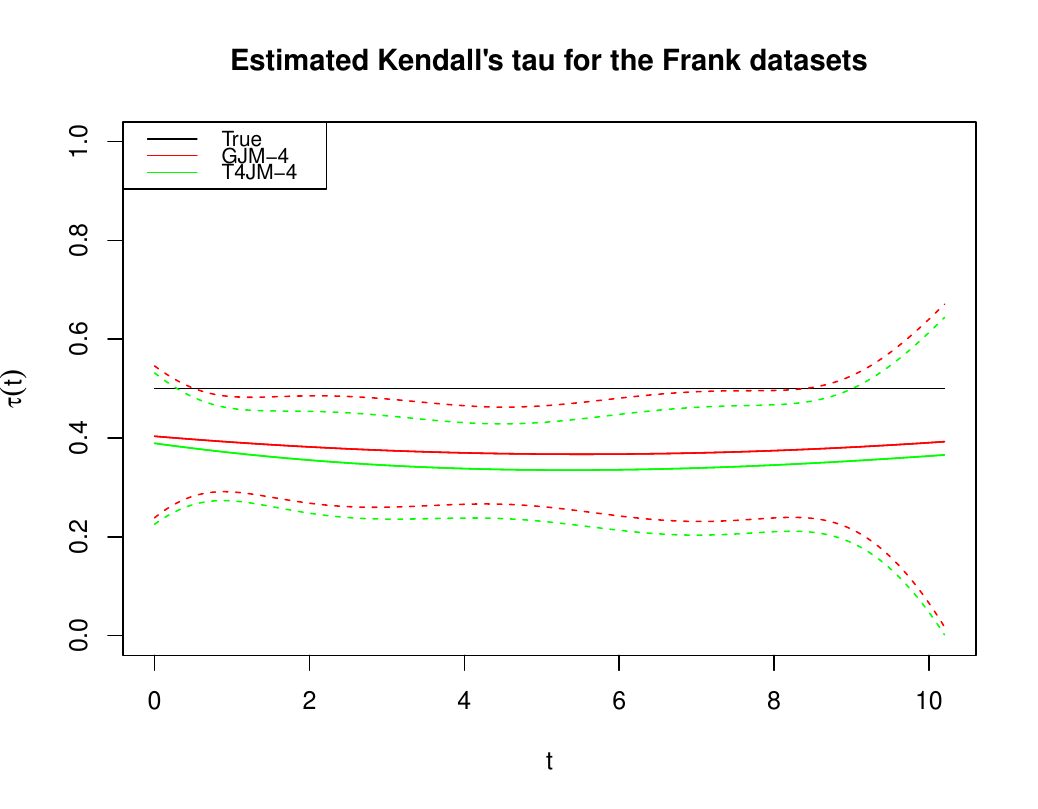}
		\end{minipage}
		\caption{The fitted Kendall's correlation functions $\hat{\tau}(t)$ (solid lines) with their corresponding 95\% confidence interval functions (dashed lines) by the GJM-4 and T4JM-4 for the Clayton and Frank datasets from case 1 of scenario 1.}
		\label{estClayandFratau}
	\end{figure}

We calculate values of $\mbox{log}\left\{f_{T_{i}}(t_{i}|\bm b_{i},\bm y_{i})\right\}$ under different models for a Monte Carlo sample of sample size 200 under the Clayton and Frank datasets. Figure \ref{ftipos_c0.5_200plot} presents the boxplot of $\mbox{log}\{f_{T_{i}}^{*}(t_{i}|\bm b_{i},\bm y_{i})\}$ under the EJM, and the $\mbox{log}\{\tilde{f}_{T_{i}}(t_{i}|\bm b_{i},\bm y_{i})\}$ under the GJM-4, T4JM-4 and RJM with the true values of $\bm\theta_{y}$, $\bm\theta_{t}$, $\bm\theta_{b}$ and the estimated $\hat{\tau}(t)$ in Figure \ref{estClayandFratau} substituted in. This posterior density under the GJM-4 and T4JM-4 are very close to that of EJM, while the density under the RJM are quite different from that of them and their differences are more prominent under the Clayton dataset, which validates our aforementioned arguments. 

Therefore, only the copula parameters $\bm\theta_{\alpha}$ are biasedly estimated in GJM-4 and T4JM-4. Figure \ref{estClayandFratau} presents the $\hat{\tau}(t)$ estimated by GJM-4 and T4JM-4 for the Clayton and Frank datasets. Under the same strength of Kendall's tau correlation, the bivariate Clayton joint model introduce stronger dependency between the sub-models than the bivariate Gaussian and $t_{4}$ copula joint models, while the dependency among the bivariate Frank one is the weakest among the four models. These conclusions are consistent with the discussions regarding  biases of parameter estimations fitted by RJM under the four datasets in Table 1.

Although, there is also moderate biases in $\tilde{\bm\theta}_{b}$ by fitting the RJM to the simulated datasets, these biases has been decreasing as $n_{i}$ increase by comparing to the fitted results in scenario 2 with $\mbox{max}(n_{i})=6$. This tendency agrees with the discussion in Section 2.4.

Overall, apart from misusing the independence copula, i.e. RJM, the estimations for $\bm\theta_{y}, \bm\theta_{t}$ and $\bm\theta_{b}$ are very robust against to the copula misspecification under the proposed joint models even with a moderate number of repeated measurements in each subject.

In the case of $\tau(t)=-0.5$, the conclusions are very similar to that of $\tau(t)=0.5$, except the biases of parameter estimations are in the opposite direction, thus it is not discussed here. The results are provided in the supplementary materials.

\noindent\textbf{Case 2: time varying Kendall's tau correlation}

 In this case, the Kendall's tau correlation function is time-varying and designed in the way that the duration and magnitude of positive and negative parts are roughly equivalent. Table 2 lists the fitted outputs by the GJM-6, T4JM-6 and RJM  for the four types  datasets in scenario 1. As discussed in case 1, the estimators for $\bm\theta_{y}, \bm\theta_{t}$ and $\bm\theta_{b}$ are robust against copula misspecification, thus the results  are almost unbiased  by fitting GJM-6 and T4JM-6. 
 
 In contrast to case 1,  the same level of robustness is also observed for the parameter estimations by RJM here.  Note that the estimators of $\bm\theta_{y},$ $\bm\theta_{t}$ and $\bm\theta_{b}$ fitted by RJM are biased in opposite direction for datasets between cases 1 and 3, thus it is possible to shrink the biases by selecting some correlation functions with almost equal impact from the negative and positive sides.
 
 	\begingroup
 \setlength{\tabcolsep}{6pt} % Default value: 6pt
 \renewcommand{\arraystretch}{1.18} 
 \begin{table}[H]
 	\tbl{{ Average AIC and BIC for the candidate models fitted on the Gaussian datasets from case 2 in scenario 1.}}
 	{\begin{tabular}{lcccccccccc}
 			\toprule
 			& GJM-4    & GJM-5    &GJM-6  &GJM-7 & GJM-8 
 			\\
 			\hline
 			AIC          &  12802.43  & 12806.35    & 12770.16 & 12772.97  & 12771.02
 			\\            
 			BIC          &  12868.39  &  12875.61  & 12842.73 & 12848.83  & 12850.18
 			\\
 			\hline
 	\end{tabular}}
 	\label{AICBIC}
 \end{table}
 \endgroup
  
    Unlike case 1, there is also a risk of misspecifying correlation function in case 2. Following Yao (2007)\cite{yao07} and Wang \textit{et al.} (2024a\cite{wan24a} and 2024b\cite{wan24b}), we select the optimal numbers and locations of knots by AIC and BIC criteria. Taking the Gaussian datasets for example, given the two sub-models,  random effects distribution and copula function are correctly specified. As shown in Table \ref{AICBIC},  the two criteria are able to select the true model among the candidate models with 4, 5, 6, 7 and 8 cubic B-spline basis functions with equally space knots over [0, 10.2]. Figure 3 presents the fitted correlation functions with 4, 6 and 8 B-spline basis functions. The correlation function is best-fitted by 6 cubic B-spline basis functions, as this is the correct model, but the fitted curved by 8 basis functions is quite close. On the other hand, 4 cubic B-spline basis functions are no enough to provide adequate fit for the correlation function. However, if the interest is only on estimating $\bm\theta_{y}, \bm\theta_{t}$ and $\bm\theta_{b}$, then the specification for the correlation function does not matter for the datasets from case 2, since all the candidates can provide accurate estimations on these parameters for the reasons discussed above.

  \begin{figure}[H]
  	\centering
  	\begin{minipage}{0.326\textwidth}
  		\includegraphics[width=\linewidth]{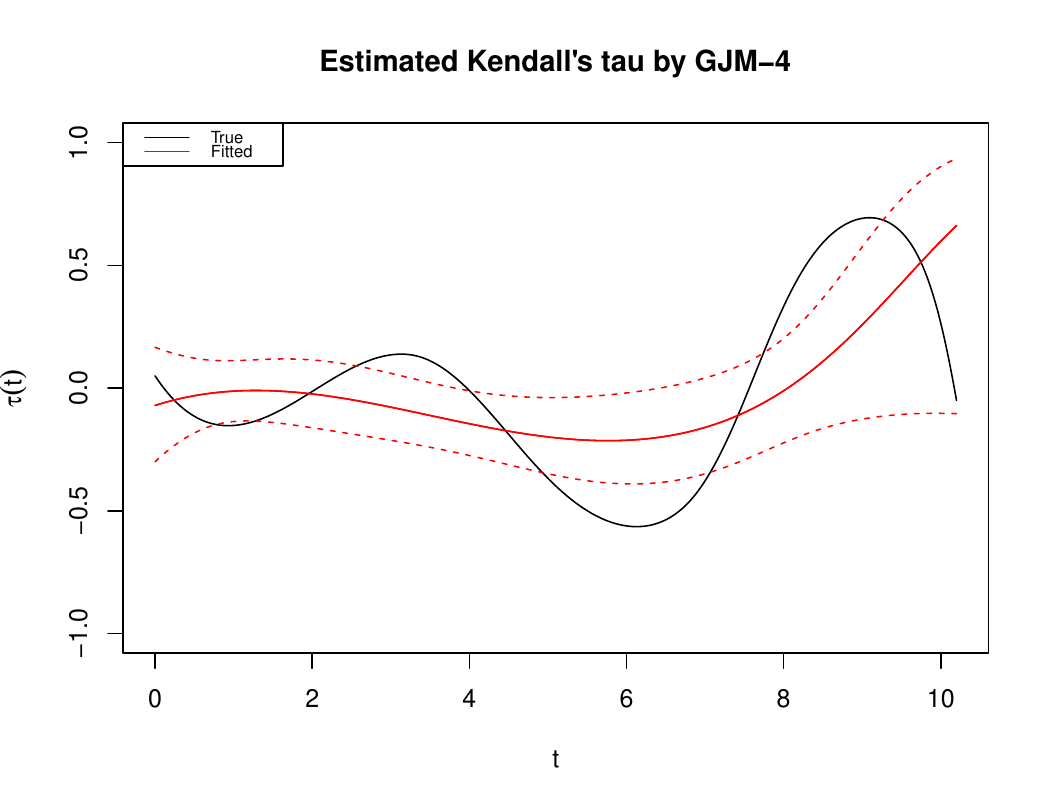}
  	\end{minipage}
  	\begin{minipage}{0.326\textwidth}
  		\includegraphics[width=\linewidth]{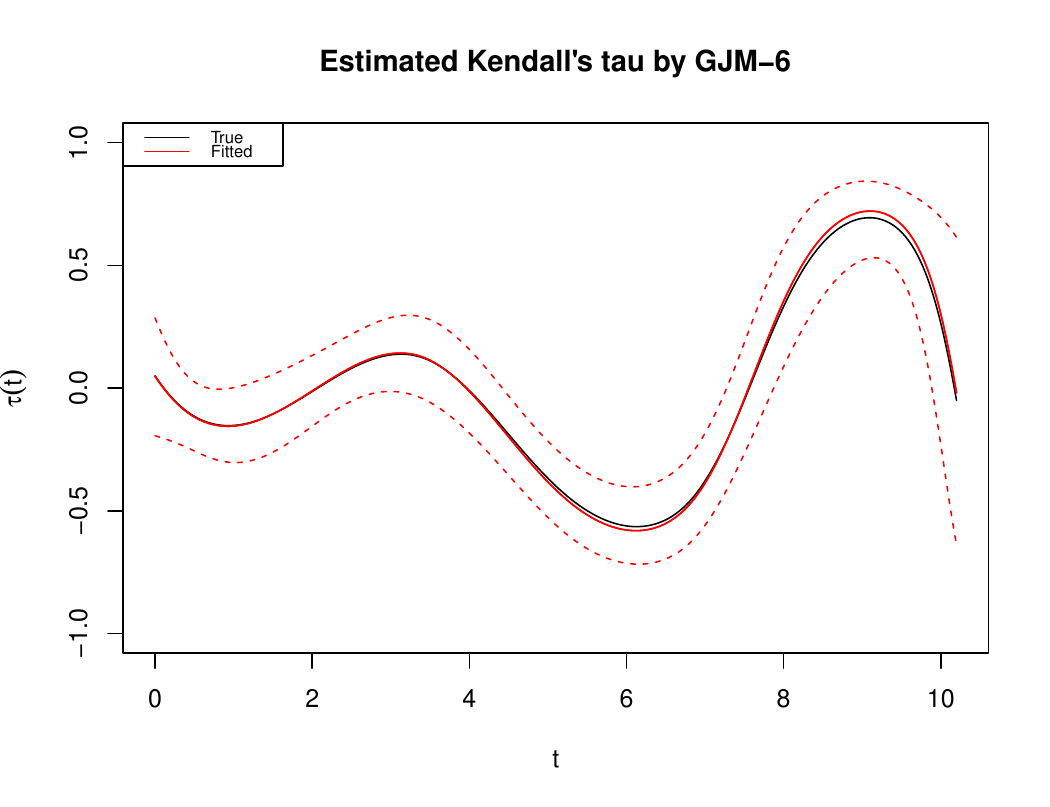}
  	\end{minipage}
  	\begin{minipage}{0.326\textwidth}
  	\includegraphics[width=\linewidth]{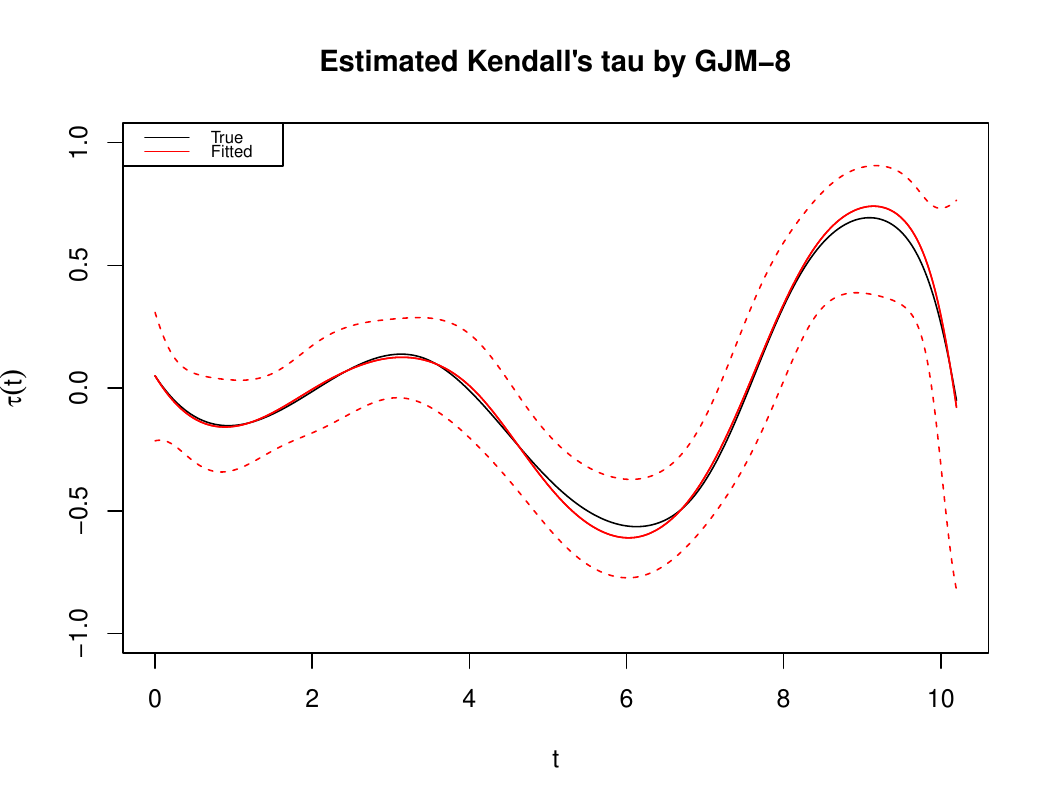}
  \end{minipage}
  	\caption{The fitted Kendall's correlation functions $\hat{\tau}(t)$ (solid lines) with their corresponding 95\% confidence interval functions (dashed lines) by the GJM-4, GJM-6 and GJM-8  when the true datasets are the Gaussian datasets from case 2 of scenario 1.}
  	\label{estGautau}
  \end{figure}

		\begingroup
	\setlength{\tabcolsep}{6pt} % Default value: 6pt
	\renewcommand{\arraystretch}{1.09} 
	\begin{table}[H]
		\tbl{ Estimation of the parameters  by GJM-4, T4JM-4 and RJM for simulated data from case 1 of scenario 1.}
		{\begin{tabular}{lccccccccccccccccccc}
				\toprule
				True  &$\beta_{10}$ &$\beta_{11}$ &$\beta_{12}$&$\beta_{13}$&$\beta_{14}$&$\beta_{15}$&$\beta_{21}$&$\beta_{22}$&$\beta_{23}$&$\beta_{24}$&$D_{11}$&$D_{22}$&$D_{12}$&$\sigma$&$\gamma$
				\\
				value     & 10       &   -0.5   &   1       &   0.5    &    0.5  &   1      &    -2    &    -1    &   -1.5    &    -2    &    2      &    0.2   &   -0.1   &    2     & -0.5       
				\\
				\hline
				&\multicolumn{15} {c}{Data generated by the bivariate Gaussian copula joint model (Gaussian dataset)}
				\\
				\cline{1-16}
			   GJM-4      &            &            &           &            &           &           &           &            &             &            &            &           &            &           &
				\\
				\cdashline{1-16}
				Est.      & 10.010 & -0.498 & 1.009 & 0.503  & 0.496 & 0.986 &-2.026 & -1.020 & -1.508  & -1.999 & 1.907  & 0.198 & -0.096 & 2.000 & -0.509
				\\
				SE        & 0.336   & 0.036  & 0.235 & 0.234  &  0.351 &0.323 & 0.251  & 0.225   & 0.300   & 0.284  & 0.300  & 0.030 & 0.072  & 0.029 & 0.056
				\\
				SD        & 0.350   & 0.036  & 0.231 & 0.243  & 0.356 & 0.331 & 0.242  & 0.236   & 0.309   & 0.289  & 0.325  & 0.030 & 0.072  & 0.027 & 0.054
				\\
				CP	      & 0.950   & 0.944  & 0.944 & 0.934 & 0.952  & 0.954 & 0.970 & 0.930   & 0.944   & 0.944  & 0.900  & 0.932 & 0.936  & 0.964  & 0.966
				\\
				\cdashline{1-16}
				T4JM-4   &&& &  &   &   &    &   &  &  &  &  &
				\\
				\cdashline{1-16}
				Est.     & 10.017 & -0.496 & 1.006 & 0.501 & 0.495 & 0.985 & -1.982 & -0.998 &-1.480 & -1.962 & 1.915  & 0.198 & -0.098 & 2.004 & -0.492
				\\
				SE       & 0.336  & 0.036   & 0.236 & 0.235 &  0.352 & 0.324 & 0.243 & 0.219 & 0.294 & 0.278 & 0.299 & 0.030 & 0.073  &0.029  & 0.054
				\\
				SD       & 0.354  & 0.036   & 0.232 & 0.244 & 0.357 & 0.333 & 0.239 & 0.232 & 0.306 & 0.287 & 0.325 & 0.030  & 0.072  & 0.027 & 0.053
				\\
				CP	    & 0.948  &  0.936  & 0.942 & 0.938 & 0.950 & 0.950 & 0.958 & 0.926 & 0.946 & 0.932  & 0.898  & 0.932 & 0.940 & 0.966  & 0.940
				\\
				\cdashline{1-16}
				RJM  && & &  &   &   &    &   &  &  &  &  &
				\\
				\cdashline{1-16}
				Est.     & 9.916  & -0.561  & 1.075  & 0.534  & 0.548  & 1.052  & -2.387 & -1.198 & -1.743 & -2.314 & 2.246  & 0.249 & -0.130 & 1.983  & -0.766
				\\
				SE       & 0.348  & 0.038   & 0.247  & 0.246 &  0.365 &  0.335 & 0.337  & 0.297  & 0.393  & 0.375  & 0.339   & 0.039 & 0.089  & 0.029 & 0.086
				\\
				SD       & 0.368  & 0.039   & 0.239  & 0.252 &  0.378 &  0.343 & 0.326  & 0.303  & 0.384  & 0.356  & 0.367  & 0.041  & 0.087 & 0.027  & 0.084
				\\
				CP	     & 0.940  & 0.658  & 0.938  & 0.944  & 0.940  & 0.950  & 0.838  & 0.908  & 0.926  & 0.914  & 0.906  & 0.850  & 0.950 & 0.922 & 0.058
				\\
				\hline
					&\multicolumn{15} {c}{Data generated by the bivariate $t_{4}$ copula joint model ($t_{4}$ dataset)}
				\\
				\cline{1-16}
				GJM-4       &            &            &           &            &            &            &            &            &            &            &             &           &             &           &
				\\
				\cdashline{1-16}
				Est.       &  9.979 & -0.500  & 1.008  & 0.498  & 0.511   & 1.014   & -2.043 & -1.032 & -1.519 & -2.046 & 1.942   & 0.200 & -0.095  & 1.999 & -0.513
				\\
				SE         & 0.339  &   0.035  & 0.237 & 0.236  &  0.354 & 0.326  & 0.251  & 0.224  & 0.299  & 0.286   & 0.305  & 0.030 & 0.074   & 0.029  & 0.055
				\\
				SD         & 0.327  & 0.036   & 0.235  & 0.235 & 0.347   & 0.319  & 0.252  & 0.223  & 0.305  & 0.274   & 0.315  & 0.032  & 0.074   & 0.029 & 0.055
				\\
				CP	       & 0.972  & 0.940   & 0.940  &  0.948 & 0.954  &  0.958 & 0.962 &  0.944 &  0.942 &  0.952  &  0.924 &  0.922 & 0.954  &0.940  &  0.958
				\\
				\cdashline{1-16}
				T4JM-4  &&& &  &   &   &    &   &  &  &  &  &
				\\
				\cdashline{1-16}
				Est.      & 9.981  & -0.498 & 1.005  & 0.498 & 0.513  & 1.013  & -2.024 & -1.022 &-1.511  &-2.032 & 1.916 & 0.197 & -0.094 & 2.001 & -0.508
				\\
				SE        & 0.336  & 0.035   & 0.236 & 0.235 & 0.352 & 0.324 & 0.241  & 0.217   & 0.293  & 0.278 & 0.299 & 0.030 & 0.072 & 0.029  & 0.052
				\\
				SD        & 0.324  & 0.035  & 0.235 & 0.233 & 0.346  & 0.316 & 0.248  & 0.219  & 0.299  & 0.268  & 0.308 & 0.032 & 0.073 & 0.029  & 0.051
				\\
				CP	     &  0.974  &0.934   & 0.934 & 0.948 & 0.950  &0.962  &  0.964 & 0.948  & 0.946 & 0.954  & 0.912  & 0.916 & 0.946 & 0.944  & 0.962
				\\
				\cdashline{1-16}
				RJM  &&& &  &   &   &    &   &  &  &  &  &
				\\
				\cdashline{1-16}
				Est.     & 9.892 & -0.562 & 1.066  & 0.530 & 0.560 & 1.077 & -2.398 & -1.213 & -1.737 & -2.366 & 2.266 & 0.252  & -0.128 & 1.981   & -0.770
				\\
				SE       & 0.350 & 0.038   & 0.248 & 0.246 & 0.366 & 0.337 & 0.339  & 0.299  & 0.394  & 0.380 & 0.342  & 0.040 & 0.090  & 0.029   & 0.086
				\\
				SD       & 0.345 & 0.039   & 0.241 & 0.246 & 0.366 & 0.337 & 0.322 & 0.278   & 0.375  & 0.343  & 0.377  & 0.042  & 0.092  & 0.029   & 0.083
				\\
				CP	    & 0.956  & 0.628   & 0.944 & 0.944 & 0.942 & 0.956 & 0.818 & 0.918   & 0.920  & 0.886 &  0.910 &  0.808  & 0.938  & 0.880   & 0.034
				\\
					\hline
				&\multicolumn{15} {c}{Data generated by the bivariate Clayton copula joint model (Clayton dataset)}
				\\
				\cline{1-16}
				GJM-4       &            &            &           &            &            &            &            &            &            &            &             &           &             &           &
				\\
				\cdashline{1-16}
				Est.       & 9.989  & -0.498  & 1.003  & 0.494  & 0.520  & 1.011  & -2.065 & -1.021 & -1.538 & -2.042 & 1.952   & 0.199 & -0.100  & 2.001  & -0.515
				\\
				SE         & 0.337  &   0.032  & 0.236 & 0.235  &  0.353 & 0.325  & 0.220  & 0.197  & 0.272  & 0.258   & 0.300  & 0.030 & 0.073   & 0.029 & 0.043
				\\
				SD         & 0.333  &  0.034   & 0.230 & 0.235  & 0.375  & 0.326  & 0.231  & 0.209  &0.300  & 0.289   & 0.307  & 0.029  & 0.074   & 0.028 & 0.048
				\\
				CP	       & 0.952  &  0.924   & 0.952 &  0.954 &  0.926 &  0.954 & 0.942 &  0.936 &  0.934 &  0.918  &  0.926 &  0.942 & 0.954  & 0.960 &  0.914 
				\\
				\cdashline{1-16}
				T4JM-4  &&& &  &   &   &    &   &  &  &  &  &
				\\
				\cdashline{1-16}
				Est.      & 9.985 & -0.497  & 1.007 & 0.494  & 0.528 & 1.016   & -2.047 & -1.010 &-1.532  &-2.032 & 1.940 & 0.199 & -0.100 & 1.999  & -0.510
				\\
				SE        & 0.335 & 0.032   & 0.235 & 0.234 &   0.351& 0.323  &  0.216  &  0.195 &  0.270 &  0.256&  0.295&  0.030&  0.072 & 0.028 & 0.041
				\\
				SD        & 0.331 & 0.033   &  0.224& 0.233 &  0.372 & 0.321  &  0.220  & 0.203 &  0.294 & 0.279 &  0.300 & 0.029 & 0.073 & 0.028  & 0.044
				\\
				CP	     & 0.954  & 0.940  &  0.956 & 0.952 &  0.930& 0.962  &  0.950 &  0.934 &  0.930 & 0.924 &  0.912 & 0.942 & 0.958 & 0.958  &  0.922
				\\
				\cdashline{1-16}
				RJM  &&& &  &   &   &    &   &  &  &  &  &
				\\
				\cdashline{1-16}
				Est.     & 9.866 &-0.572  &1.077   & 0.530  & 0.600 & 1.108 &-2.545 &-1.255  &-1.854  & -2.472  & 2.419  & 0.275 & -0.153 &  1.973  &-0.851
				\\
				SE       & 0.358 & 0.038  &  0.253 & 0.251 &  0.374 & 0.344 &0.366 &  0.319  & 0.425  &  0.408  & 0.359 &  0.044 &  0.098 & 0.028 & 0.094
				\\
				SD       & 0.364 & 0.039  &  0.249 & 0.249 & 0.405 &  0.349& 0.341 & 0.305  &  0.420 &   0.410  & 0.381 & 0.046  & 0.102  &  0.027 &   0.102
				\\
				CP	    & 0.926  & 0.526  & 0.936 & 0.948  & 0.928 &  0.936& 0.734 &  0.912 &  0.886 &  0.810   & 0.822 & 0.690  & 0.954 & 0.850  & 0.004
				\\
					\hline
				&\multicolumn{15} {c}{Data generated by the bivariate Frank copula joint model (Frank dataset)}
				\\
				\cline{1-16}
				GJM-4       &            &            &           &            &            &            &            &            &            &            &             &           &             &           &
				\\
				\cdashline{1-16}
				Est.       & 9.990  & -0.496  & 1.008 & 0.509  & 0.494  & 1.005   & -2.002 & -1.030 &-1.511  & -2.013  & 1.929  & 0.199  & -0.096 & 2.004  &-0.509 
				\\
				SE         & 0.340  & 0.037   & 0.237  &  0.236 & 0.355 &  0.327  & 0.262  &  0.236 &  0.314 &  0.298  &  0.305 & 0.030 & 0.073   & 0.029  &  0.061
				\\
				SD         & 0.341  & 0.035   &  0.246 & 0.236  & 0.351 &  0.315  &  0.257  &  0.235 & 0.319  &  0.293  & 0.306  & 0.031 &  0.075  & 0.029  &   0.061
				\\
				CP	       & 0.958  & 0.956  &  0.936 &  0.948 & 0.948 &  0.954  & 0.954  & 0.952  &  0.946 &  0.968  &  0.912 & 0.932 & 0.942   & 0.952  &  0.968
				\\
				\cdashline{1-16}
				T4JM-4  &&& &  &   &   &    &   &  &  &  &  &
				\\
				\cdashline{1-16}
				Est.      & 9.999  & -0.493  & 1.001 & 0.506 & 0.490 & 0.998 & -1.952 &-1.003  & -1.475 &-1.964 & 1.942 & 0.199 &-0.098 & 2.015  &-0.493
				\\
				SE        & 0.341  & 0.037   & 0.238 & 0.237 &  0.357 & 0.329 &  0.251 &  0.227 &  0.304 & 0.288 & 0.306 & 0.030 & 0.074 & 0.029 &   0.058
				\\
				SD        & 0.343  & 0.036  & 0.248 & 0.239 &  0.353 & 0.318 &  0.256 &  0.233 &  0.317 &  0.290 & 0.308 & 0.031 & 0.075 & 0.029 & 0.063
				\\
				CP	     & 0.956   & 0.950  &  0.938& 0.948 &  0.946 & 0.954 & 0.930  & 0.948 &  0.940 & 0.952 &  0.922 & 0.932 & 0.944 & 0.928 & 0.924
				\\
				\cdashline{1-16}
				RJM  &&& &  &   &   &    &   &  &  &  &  &
				\\
				\cdashline{1-16}
				Est.     & 9.909 &-0.545  & 1.060 & 0.536  & 0.532  & 1.059 & -2.297 & -1.174  & -1.697 &-2.278  & 2.167   &0.234  & -0.118  & 1.993  &-0.704 
				\\
				SE       & 0.347  & 0.038  & 0.244 & 0.243 &  0.363 & 0.334 & 0.320  &  0.284 &  0.375 &  0.359 &  0.330  & 0.036 &  0.084 & 0.029  & 0.080
				\\
				SD       & 0.349 & 0.037  &  0.255 &0.244  &  0.363 &  0.319 & 0.314  &  0.276 & 0.367  & 0.338  & 0.332   & 0.038 &  0.088 & 0.029  & 0.083
				\\
				CP	    & 0.956  & 0.790  & 0.924  & 0.954 & 0.942  &  0.952 & 0.892 &  0.934 & 0.942 &  0.898 &  0.950  &  0.904 &  0.948 & 0.938  &0.242
				\\
				\hline
		\end{tabular}}
	\end{table}
	\endgroup

\begingroup
\setlength{\tabcolsep}{6pt} % Default value: 6pt
\renewcommand{\arraystretch}{1.09} 
\begin{table}[H]
	\tbl{ Estimation of the parameters  by GJM-6, T4JM-6 and RJM for simulated data from case 2 of scenario 1.}
	{\begin{tabular}{lccccccccccccccccccc}
			\toprule
			True  &$\beta_{10}$ &$\beta_{11}$ &$\beta_{12}$&$\beta_{13}$&$\beta_{14}$&$\beta_{15}$&$\beta_{21}$&$\beta_{22}$&$\beta_{23}$&$\beta_{24}$&$D_{11}$&$D_{22}$&$D_{12}$&$\sigma$&$\gamma$
			\\
			value     & 10       &   -0.5   &   1       &   0.5    &    0.5  &   1      &    -2    &    -1    &   -1.5    &    -2    &    2      &    0.2   &   -0.1   &    2     & -0.5       
			\\
			\hline
			&\multicolumn{15} {c}{Data generated by the bivariate Gaussian copula joint model (Gaussian dataset)}
			\\
			\cline{1-16}
			GJM-6      &            &            &           &            &           &           &           &            &             &            &            &           &            &           &
			\\
			\cdashline{1-16}
			Est.      & 9.979   & -0.497 & 0.988 & 0.510  & 0.515  & 1.018 & -2.044 & -1.035 & -1.534  & -2.035 & 1.925 & 0.200  &-0.105 & 2.002 &-0.517
			\\
			SE        & 0.335   & 0.037  & 0.234 & 0.234  & 0.348 & 0.320 & 0.270  & 0.241   &  0.320  & 0.305  & 0.299 &  0.030 & 0.072 & 0.029 & 0.062
			\\
			SD        & 0.347   &  0.037 & 0.236 & 0.230  & 0.357 & 0.320 &  0.286 & 0.254  &  0.333  &  0.291  & 0.298 &  0.029 & 0.069 & 0.028 & 0.064
			\\
			CP	      & 0.942   & 0.964 & 0.952 &  0.952 & 0.936 & 0.960 &  0.932  & 0.948  & 0.936  &  0.960  & 0.924 & 0.938  & 0.970 & 0.958  & 0.950
			\\
			\cdashline{1-16}
			T4JM-6   &&& &  &   &   &    &   &  &  &  &  &
			\\
			\cdashline{1-16}
			Est.     &9.981   & -0.501  & 0.989 & 0.511  & 0.514  & 1.018  & -2.008& -1.015&-1.508 &-1.999 & 1.935  & 0.201 & -0.107  & 2.013 &-0.512
			\\
			SE       & 0.336 &  0.037   & 0.235 & 0.235 & 0.350 & 0.322 &  0.261 & 0.234 &  0.312 & 0.297 & 0.299  & 0.030 & 0.072 & 0.029 & 0.061
			\\
			SD       & 0.347 & 0.037    & 0.238 & 0.233 & 0.356 & 0.323 & 0.284 & 0.253 &  0.330 & 0.288 & 0.300  &  0.029 &0.069 & 0.029 & 0.065
			\\
			CP	    & 0.942  & 0.956   & 0.956 &  0.948 & 0.936 & 0.952 &  0.930& 0.932 &  0.946 & 0.962 & 0.922 &  0.952 & 0.972 & 0.934 & 0.934
			\\
			\cdashline{1-16}
			RJM  && & &  &   &   &    &   &  &  &  &  &
			\\
			\cdashline{1-16}
			Est.     & 9.973  &-0.487  & 0.984  & 0.509  & 0.515  & 1.015   & -2.029 & -1.029& -1.527 &-2.029 & 1.928  &0.198  &-0.106  &1.999   &-0.500
			\\
			SE       & 0.334  &  0.038 & 0.234  & 0.234  & 0.349  & 0.320  & 0.275  & 0.247  & 0.324  & 0.308 & 0.299  & 0.030 & 0.072 & 0.029  & 0.066 
			\\
			SD       & 0.348  & 0.037  &  0.234 & 0.231  &  0.358 &  0.320 &  0.285 &  0.261 &  0.335  & 0.292 & 0.300  & 0.029 & 0.071 & 0.028  &  0.067
			\\
			CP	     & 0.942  & 0.946  & 0.954  & 0.950 &  0.930 &  0.956 &  0.954 &  0.938 &  0.946  & 0.966 &  0.910 & 0.940 & 0.970 & 0.952  & 0.948
			\\
			\hline
			&\multicolumn{15} {c}{Data generated by the bivariate $t_{4}$ copula joint model ($t_{4}$ dataset)}
			\\
			\cline{1-16}
			GJM-6       &            &            &           &            &            &            &            &            &            &            &             &           &             &           &
			\\
			\cdashline{1-16}
			Est.       & 9.986  & -0.497  & 1.018  & 0.504  & 0.495  & 0.996  & -2.057 & -1.022 & -1.557 & -2.055  & 1.961   & 0.204 & -0.098 & 1.995  & -0.518 
			\\
			SE         & 0.338  &  0.037  &  0.236 & 0.235  & 0.352  & 0.325  &  0.270 &  0.241 &  0.320 &  0.306  & 0.304  &  0.030 & 0.073  & 0.029  & 0.061
			\\
			SD         & 0.353  & 0.037   & 0.247  & 0.244  &  0.373 & 0.354  &  0.273 &  0.248 &  0.337 &   0.317  & 0.316   & 0.030  & 0.079 & 0.029  & 0.064
			\\
			CP	       & 0.930  & 0.952   & 0.934 &  0.936  &  0.950 & 0.930  &  0.966 &  0.940 & 0.934 & 0.940   & 0.926   & 0.952 &  0.936 & 0.940  & 0.936
			\\
			\cdashline{1-16}
			T4JM-6  &&& &  &   &   &    &   &  &  &  &  &
			\\
			\cdashline{1-16}
			Est.      & 9.986  & -0.495 & 1.014  & 0.501  & 0.500  & 0.998 & -2.044& -1.014 & -1.556 & -2.043&1.927  & 0.200 & -0.097 & 2.000  & -0.514
			\\
			SE        & 0.335  & 0.036  &  0.235 & 0.234 &  0.350 & 0.323 & 0.263 &  0.235 &  0.313 &  0.299 & 0.297 & 0.029 & 0.072  & 0.028  &  0.059
			\\
			SD        & 0.349  & 0.037  &  0.247 & 0.247  &  0.369 & 0.350 & 0.271 &  0.242 &  0.327 &  0.309 & 0.309 & 0.029 & 0.076  & 0.029  & 0.062
			\\
			CP	      & 0.940  &  0.952 & 0.936 &  0.934 &  0.950 & 0.934 & 0.960 &  0.930 &  0.952 &  0.950 & 0.914 & 0.938 & 0.942  & 0.950  & 0.936
			\\
			\cdashline{1-16}
			RJM  &&& &  &   &   &    &   &  &  &  &  &
			\\
			\cdashline{1-16}
			Est.     & 9.985 &-0.486  & 1.011  & 0.503 & 0.491  & 0.986 & -2.024 &-1.008 & -1.541 & -2.033 & 1.970  &  0.201  &-0.099 & 1.993   &-0.494
			\\
			SE       & 0.337 & 0.038   & 0.237 & 0.236 & 0.353 & 0.325 &  0.275 &  0.248 &  0.324 & 0.308  &  0.305 & 0.030  &  0.073 & 0.029  & 0.065
			\\
			SD       & 0.350 & 0.038  &  0.245 & 0.250 & 0.374 & 0.352 &  0.280 &  0.259 & 0.337  & 0.317  & 0.323  & 0.030   &  0.078 & 0.029  & 0.063
			\\
			CP	    & 0.942  & 0.946  &  0.936 & 0.930 & 0.952 & 0.934 & 0.952  & 0.946  & 0.950 &  0.946 &  0.922 & 0.948   &  0.938 & 0.930 & 0.956
			\\
			\hline
			&\multicolumn{15} {c}{Data generated by the bivariate Clayton copula joint model (Clayton dataset)}
			\\
			\cline{1-16}
			GJM-6       &            &            &           &            &            &            &            &            &            &            &             &           &             &           &
			\\
			\cdashline{1-16}
			Est.       & 9.990  &-0.497   & 0.989 & 0.499  & 0.519   & 1.004  & -2.053 & -1.026 & -1.567 & -2.069 & 1.922  & 0.200 & -0.101  &2.000  &-0.527
			\\
			SE         & 0.334  & 0.035   & 0.234 &  0.233 &  0.348  & 0.321  &  0.256  &  0.229 & 0.308 & 0.295  &  0.297  & 0.029 & 0.072  & 0.028  & 0.054
			\\
			SD         & 0.329  &  0.035  & 0.224 &   0.241 & 0.353  &  0.325 &  0.276  &  0.242 & 0.329  &  0.310 & 0.309  & 0.029  & 0.073 & 0.030  & 0.060
			\\
			CP	       & 0.952  & 0.944   & 0.968 &   0.950 & 0.948  & 0.952  &  0.922 &  0.942 &  0.938 &  0.944 &  0.910 & 0.942  & 0.952 & 0.928  &  0.894
			\\
			\cdashline{1-16}
			T4JM-6  &&& &  &   &   &    &   &  &  &  &  &
			\\
			\cdashline{1-16}
			Est.      & 9.987  &-0.504  & 0.990 & 0.501  & 0.520 & 1.005 & -2.018 & -1.006 & -1.542 &-2.032 & 1.938 & 0.201 & -0.102 & 2.010  &-0.522
			\\
			SE        & 0.335  &  0.035 &  0.235& 0.234 & 0.350 & 0.323 & 0.247  & 0.221  & 0.300  &  0.287 & 0.298 & 0.029 & 0.072 & 0.028  & 0.052
			\\
			SD        & 0.334  & 0.036  &  0.225& 0.243 & 0.353 &  0.327 & 0.276 & 0.237  & 0.320  & 0.304 &  0.310 & 0.030 & 0.074 & 0.030  & 0.061
			\\
			CP	     & 0.950   & 0.934  & 0.968 &  0.954 & 0.944 & 0.946 &  0.916 & 0.930 & 0.930  & 0.944 & 0.914  & 0.946 & 0.952 & 0.918  & 0.900
			\\
			\cdashline{1-16}
			RJM  &&& &  &   &   &    &   &  &  &  &  &
			\\
			\cdashline{1-16}
			Est.     & 9.980 & -0.488 & 0.988 & 0.495 & 0.523 & 1.008 & -2.026 & -1.014 & -1.547 & -2.051  & 1.938  & 0.198  & -0.105 & 1.996  &-0.505
			\\
			SE       & 0.335 & 0.038  &  0.234 & 0.234 & 0.350 & 0.322& 0.276   & 0.249 & 0.325  &  0.310   &  0.300 & 0.030 & 0.072  & 0.029 & 0.066
			\\
			SD       & 0.325 & 0.037  &  0.228 & 0.241 &  0.349 & 0.317 & 0.289  & 0.255 &   0.328 & 0.321   & 0.312   & 0.030 & 0.072 &  0.031 & 0.066
			\\
			CP	    & 0.952  & 0.942  & 0.966  & 0.952 & 0.954 & 0.958 & 0.950  & 0.944 &  0.940 &  0.944  & 0.910   & 0.942 & 0.956 & 0.930  & 0.952
			\\
			\hline
			&\multicolumn{15} {c}{Data generated by the bivariate Frank copula joint model (Frank dataset)}
			\\
			\cline{1-16}
			GJM-6       &            &            &           &            &            &            &            &            &            &            &             &           &             &           &
			\\
			\cdashline{1-16}
			Est.       & 9.991   & -0.499  & 1.010  & 0.507  & 0.504  & 1.002  & -2.026 & -0.999 &-1.554 &-2.043  & 1.929  & 0.200 & -0.096  & 2.000 &-0.511
			\\
			SE         & 0.337   & 0.038   & 0.235 &  0.234 &  0.351  &  0.323 & 0.274  & 0.245  & 0.324 &  0.309  & 0.299 &  0.030 &  0.072  & 0.029 &  0.065
			\\
			SD         & 0.348   & 0.039  &  0.240 &  0.236 & 0.348  &  0.332 & 0.266  &  0.247 & 0.321  &  0.315  &  0.303 & 0.029 & 0.072  & 0.029  &  0.068
			\\
			CP	       & 0.938   & 0.940  &  0.952 &  0.946 &  0.954 & 0.938  &  0.958 &  0.948 & 0.948  &  0.950 & 0.914  & 0.956 &  0.960 & 0.948   & 0.946
			\\
			\cdashline{1-16}
			T4JM-6  &&& &  &   &   &    &   &  &  &  &  &
			\\
			\cdashline{1-16}
			Est.      & 9.991  & -0.503 & 1.012  & 0.508  &0.505  &1.002   &-1.983  & -0.977 & -1.521 & -1.999 & 1.942 & 0.202 & -0.099 & 2.014 & -0.506
			\\
			SE        & 0.338  & 0.038  &  0.236 & 0.236 &  0.353 & 0.325 & 0.264  & 0.236 &  0.315 &  0.300 &  0.301 & 0.030 & 0.073  & 0.029 &  0.063
			\\
			SD       & 0.351   & 0.039  &  0.242 & 0.237 &   0.350 & 0.336 & 0.263 &  0.245 & 0.324 & 0.316   & 0.307 & 0.030 &  0.073 & 0.029 &  0.070
			\\
			CP	     & 0.940  & 0.936  & 0.946  &  0.944 &  0.954 & 0.938 & 0.956 &  0.930 &  0.944 & 0.942  &  0.912 & 0.950 & 0.954 & 0.938 &  0.934
			\\
			\cdashline{1-16}
			RJM  &&& &  &   &   &    &   &  &  &  &  &
			\\
			\cdashline{1-16}
			Est.     & 9.988 &-0.492  & 1.008  & 0.505 & 0.502 & 0.998 &-2.023 & -1.002 & -1.552 & -2.041 & 1.931 & 0.198  & -0.097 & 1.999 & -0.505
			\\
			SE       & 0.336 & 0.038  &  0.235 & 0.234 &  0.351 & 0.323 & 0.275 &  0.247 &  0.325 & 0.309  & 0.299 & 0.030 & 0.072  & 0.029 & 0.065
			\\
			SD       & 0.344 & 0.038  & 0.241  &  0.236 & 0.346 & 0.330 & 0.265 & 0.249 & 0.324  &  0.315  & 0.303 & 0.030 & 0.071  & 0.029 & 0.068
			\\
			CP	     & 0.952 & 0.942  & 0.950 &  0.954 & 0.950 &  0.942 & 0.960 & 0.932 & 0.948  & 0.954  & 0.916  & 0.936 & 0.958 & 0.942 & 0.948
			\\
			\hline
	\end{tabular}}
\end{table}
\endgroup

	\subsection{Dynamic prediction}
		After a joint model is fitted, we also focus on predictions of the subject-specific survival probabilities based on some baseline covariates and updated longitudinal information. The quality of the predictions is of interest as well and can be assessed by some metrics like the area under the receiver operating characteristic curve (AUC) and prediction error (PE).
		\subsubsection{Formula for prediction }
		Suppose a time-varying bivariate copula joint model is fitted based on a random sample of $n$ subjects $\mathcal{\bm D}_{n}=\left\{T_{i},\delta_{i},\bm y_{i};i=1,...,n\right\}.$  Predictions of survival probabilities at time $u>t$ for a new subject $i,$ which has $j,$ $1\leq j\leq n_{i},$ longitudinal measurements $\mathcal{\bm Y}_{i}(t)=\left\{y_{i}(s);0\leq s<t\right\}$ up to $t$ and a vector of baseline covariates $\bm w_{i},$ can be derived as:
		\begin{eqnarray}
			\nonumber	\pi_{i}(u|t)&=&P(T_{i}^{*}>u|T_{i}^{*}>t,\mathcal{\bm Y}_{i}(t),\bm w_{i},\mathcal{\bm D}_{n};\bm\theta)=P(T_{i}^{*}>u|T_{i}^{*}>t,\mathcal{\bm Y}_{i}(t),\bm w_{i};\bm\theta)\\
			\nonumber	&=&\int_{\bm b_{i}}P(T_{i}^{*}>u|T_{i}^{*}>t,\mathcal{\bm Y}_{i}(t),\bm w_{i},\bm b_{i};\bm\theta)f_{\bm b_{i}}(\bm b_{i}|T_{i}^{*}>t,\mathcal{\bm Y}_{i}(t),\bm w_{i};\bm\theta)d\bm b_{i}\\  
			&=&\int_{\bm b_{i}}\frac{P_{T_{i}^{*}}\left(T_{i}^{*}>u|T_{i}^{*}>s_{ij},\mathcal{\bm Y}_{i}(t),\bm w_{i},\bm b_{i};\bm\theta\right)}{P_{T_{i}^{*}}\left(T_{i}^{*}>t|T_{i}^{*}>s_{ij},\mathcal{\bm Y}_{i}(t),\bm w_{i},\bm b_{i};\bm\theta\right)}f_{\bm b_{i}}(\bm b_{i}|T_{i}^{*}>t,\mathcal{\bm Y}_{i}(t),\bm w_{i};\bm\theta)d\bm b_{i}.\label{intgelpredt}
		\end{eqnarray}
		The integral in (\ref{intgelpredt}) can be approximated by its first-order estimate (Rizopoulos, 2011\cite{riz11}) using the empirical Bayes estimate for $\bm b_{i}:$
		\begin{eqnarray}
			\hat{\pi}_{i}(u|t)\approx\frac{P_{T_{i}^{*}}\left(T_{i}^{*}>u|T_{i}^{*}>s_{ij},\mathcal{\bm Y}_{i}(t),\bm w_{i},\hat{\bm b}_{i};\hat{\bm\theta}\right)}{P_{T_{i}^{*}}\left(T_{i}^{*}>t|T_{i}^{*}>s_{ij},\mathcal{\bm Y}_{i}(t),\bm w_{i},\hat{\bm b}_{i};\hat{\bm\theta}\right)},\label{predt}
		\end{eqnarray}
		where $\hat{\bm b}_{i}$ denotes the mode of the conditional distribution $f_{\bm b_{i}}(\bm b_{i}|T_{i}^{*}>t,\mathcal{\bm Y}_{i}(t),\bm w_{i};\hat{\bm\theta}).$
		
		According to (\ref{bivGaulikcen}),  the bivariate Gaussian copula joint model has the following expression for (\ref{predt}):
		\begin{eqnarray}
			\hat{\pi}_{i}(u|t)\approx\Phi\left(-\frac{Z_{u|\hat{\bm b}_{i},s_{ij}}-\hat{\alpha}(s_{ij})Z_{y_{ij}|\hat{\bm b}_{i},s_{ij}}}{\sqrt{1-\hat{\alpha}(s_{ij})^{2}}}\right)\div\Phi\left(-\frac{Z_{t|\hat{\bm b}_{i},s_{ij}}-\hat{\alpha}(s_{ij})Z_{y_{ij}|\hat{\bm b}_{i},s_{ij}}}{\sqrt{1-\hat{\alpha}(s_{ij})^{2}}}\right),\label{Gaucopredt}
		\end{eqnarray}
		which reduces to $P_{T_{i}^{*}}(T_{i}^{*}>u|\bm w_{i},\hat{\bm b}_{i};\hat{\bm\theta})/P_{T_{i}^{*}}(T_{i}^{*}>t|\bm w_{i},\hat{\bm b}_{i};\hat{\bm\theta})=\displaystyle\mbox{exp}\{-\int_{t}^{u}h_{i}(s|\bm w_{i},\hat{\bm b}_{i};\hat{\bm\theta})ds\}$ for the regular joint model given $\alpha(t)$ is a constant function of 0. 
		
	According to (\ref{bivtlikcen}),   the bivariate $t_{\nu}$ copula joint model has the following expression for (\ref{predt}):
		\begin{eqnarray}
		\hat{\pi}_{i}(u|t)\approx\Psi\left(-\frac{W_{u|\hat{\bm b}_{i},s_{ij}}-\hat{\alpha}(s_{ij})W_{y_{ij}|\hat{\bm b}_{i},s_{ij}}}{\hat{\sigma}(s_{ij}|\hat{\bm b}_{i},s_{ij})};\nu+1\right)\div\Psi\left(-\frac{W_{t|\hat{\bm b}_{i},s_{ij}}-\hat{\alpha}(s_{ij})W_{y_{ij}|\hat{\bm b}_{i},s_{ij}}}{\hat{\sigma}(s_{ij}|\hat{\bm b}_{i},y_{ij})};\nu+1\right).
			\label{tcopredt}
		\end{eqnarray}

	\subsubsection{Assessments of prediction performance}
	The overall performance of the model in predicting survival probabilities are evaluated in terms of discrimination by AUC and calibration by PE. Consider a pair of randomly selected subjects at risk by $t$, denoted as $i_{1}$ and $i_{2}$, from the population. Suppose subject $i_{1}$ experiences the event in the interval $(t,u]$ whereas subject $i_{2}$ does not.  A good predictive model is supposed to assign higher predicted survival probabilities to subject $i_{2}$ than subject $i_{1}$ (Garre \textit{et al.,} 2008\cite{Gar08}). The AUC defined as:
	\[\mbox{AUC}(u|t)=P\left\{\pi_{i_{2}}(u|t)>\pi_{i_{1}}(u|t) | T_{i_{2}}^{*}>u, t<T_{i_{1}}^{*}\leq u\right\}\]
	is a popular tool to assess the discriminative performance of the model. An AUC value closer to 1 implies a better discriminative capability of the model. On the other hand, if a subject $i$ is event free up to time $u,$ an accurate predicting model is expected to provide  $\pi_{i}(u|t)$ close to 1 and close to 0 otherwise. The PE, or the Brier score, defined as:
	\[\mbox{PE}(u|t)=E\left[\left\{I(T_{i}^{*}>u)-\pi_{i}(u|t)\right\}^{2}| T_{i}^{*}>t\right]\]
	is a common approach to calibrate the predictive accuracy of the model. A model with a smaller PE value (closer to 0) is preferable. To account for censoring in the population, the weighted AUC and PE estimators from Andrinopoulou  \textit{et al.} (2018)\cite{and18} are adopted.

		\subsubsection{Predicted results}
	We simulate $N=100$ new Monte Carlo datasets each with sample size $n=200$ under the same parameter settings as cases 1 and 2 in simulation studies with longitudinal measurements scheduled at time points $t=0,1,\dots,9,10.$ 
	
	\noindent\textbf{Case 1: prediction under copula misspecification }

	In case 1 of the simulation study, the fitted results suggest the misspecification on copula function has minimal impact on the estimators for $\bm\theta_{y}, \bm\theta_{t}$ and $\bm\theta_{b}$ at the price of distorting the estimators of $\bm\theta_{\alpha}$ in the bivariate copula functions. We investigate its impact on the predictions of survival probabilities. Parameter estimations from Table 1 and $\hat{\tau}(t)$ from Figure \ref{estClayandFratau} are substitute into (\ref{Gaucopredt}) and (\ref{tcopredt}) to make predictions under GJM-4, T4JM-4 and RJM. In addition, the results predicted by the EJM is also provided for comparison.
	
			\begin{figure}[H]
		\centering
		\begin{minipage}{0.241\textwidth}
			\includegraphics[width=\linewidth]{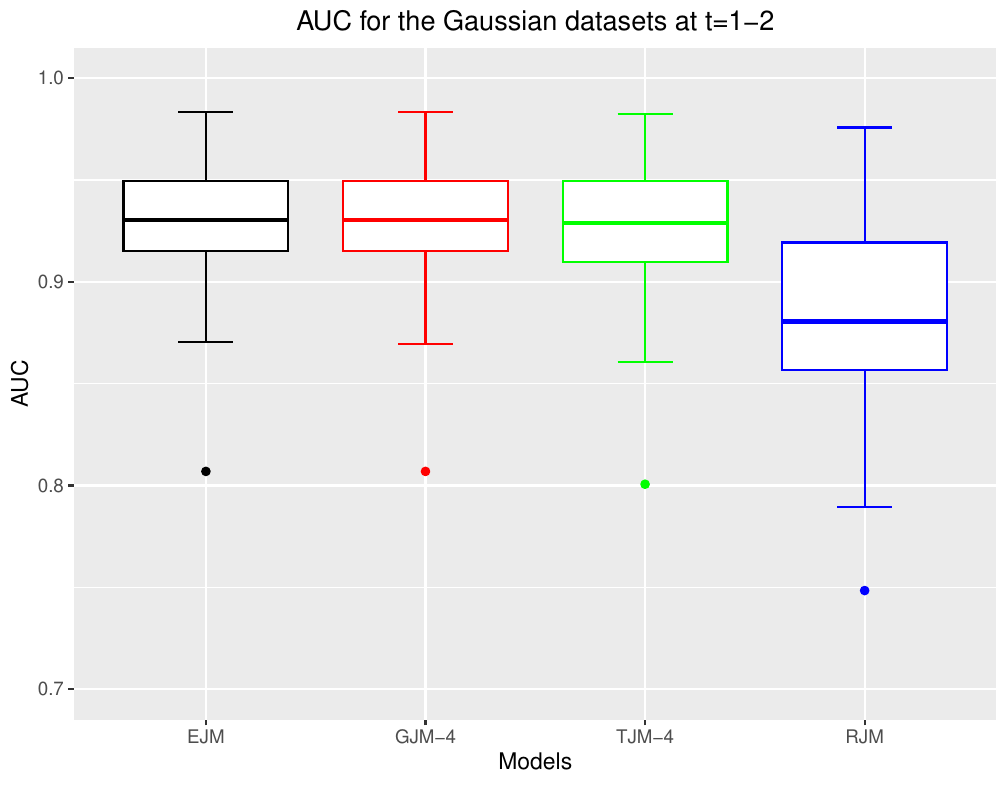}
		\end{minipage}
		\begin{minipage}{0.241\textwidth}
			\includegraphics[width=\linewidth]{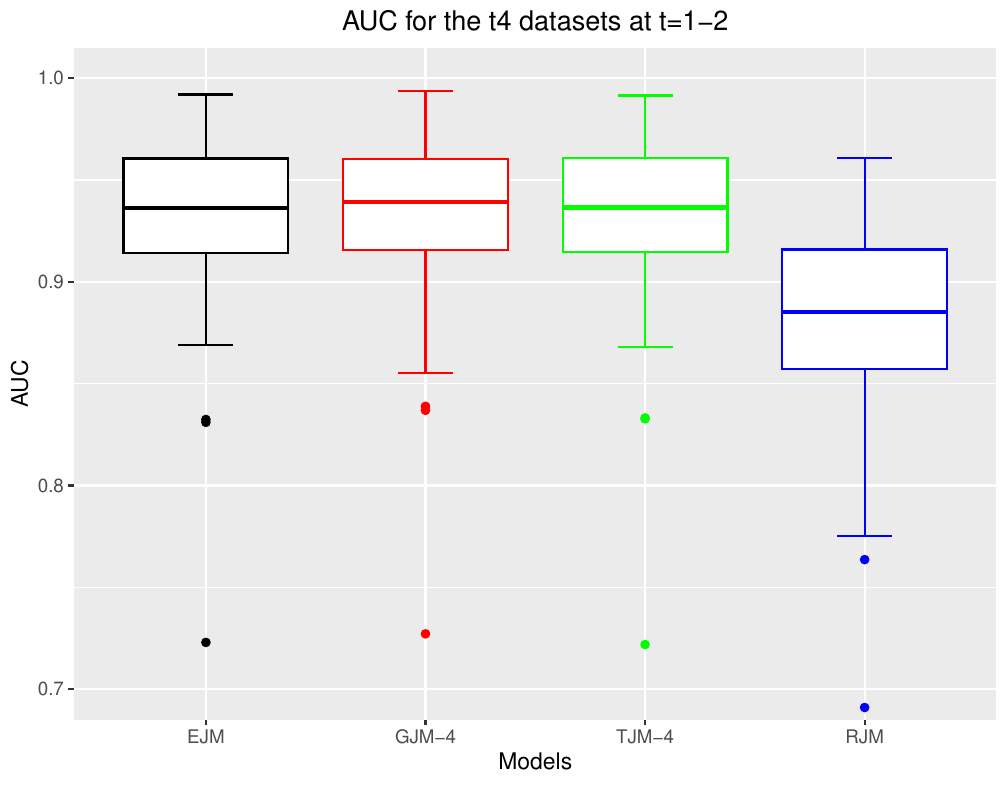}
		\end{minipage}
		\begin{minipage}{0.241\textwidth}
		\includegraphics[width=\linewidth]{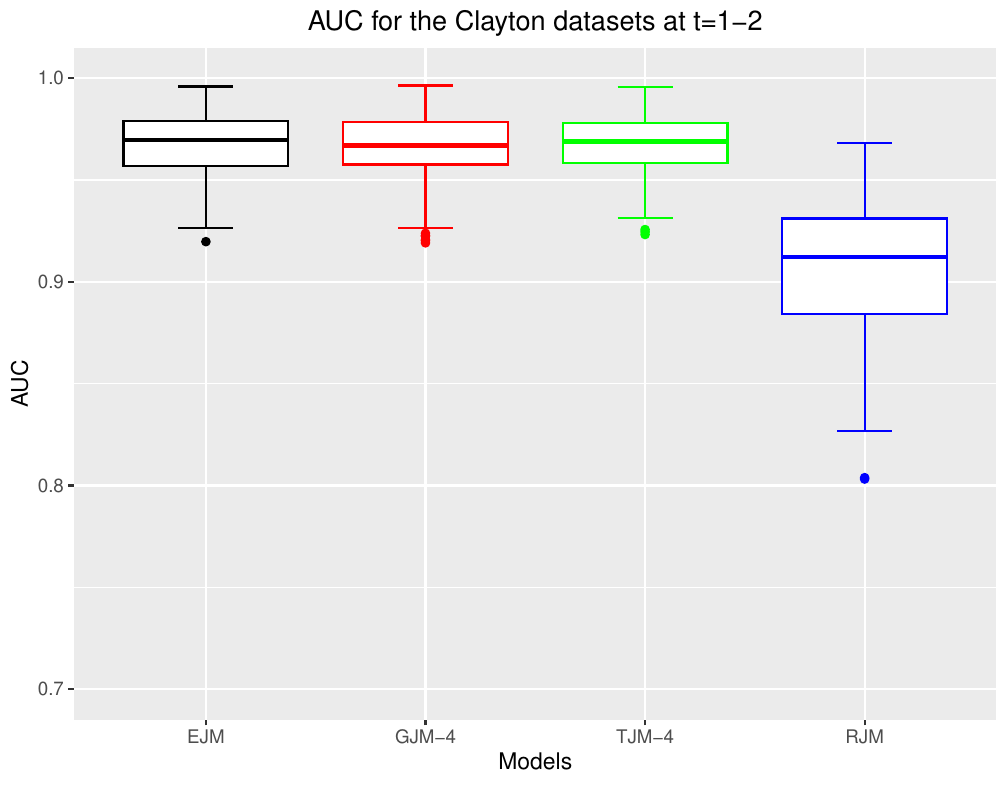}
	\end{minipage}
	\begin{minipage}{0.241\textwidth}
	\includegraphics[width=\linewidth]{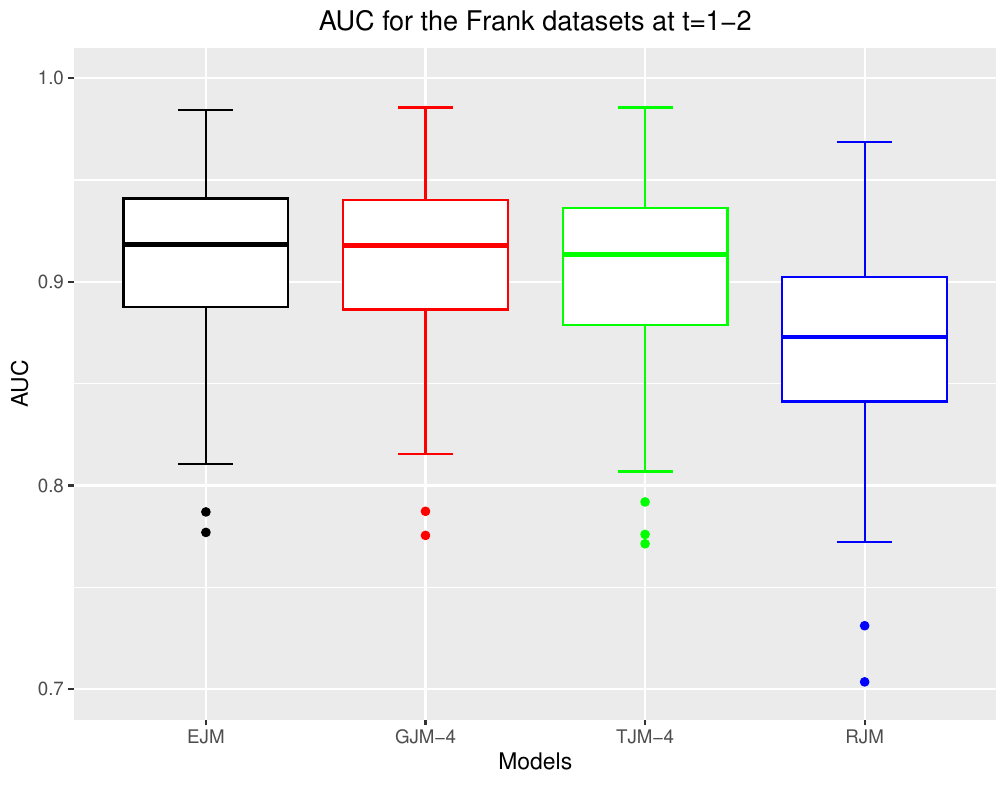}
\end{minipage}
	\begin{minipage}{0.241\textwidth}
		\includegraphics[width=\linewidth]{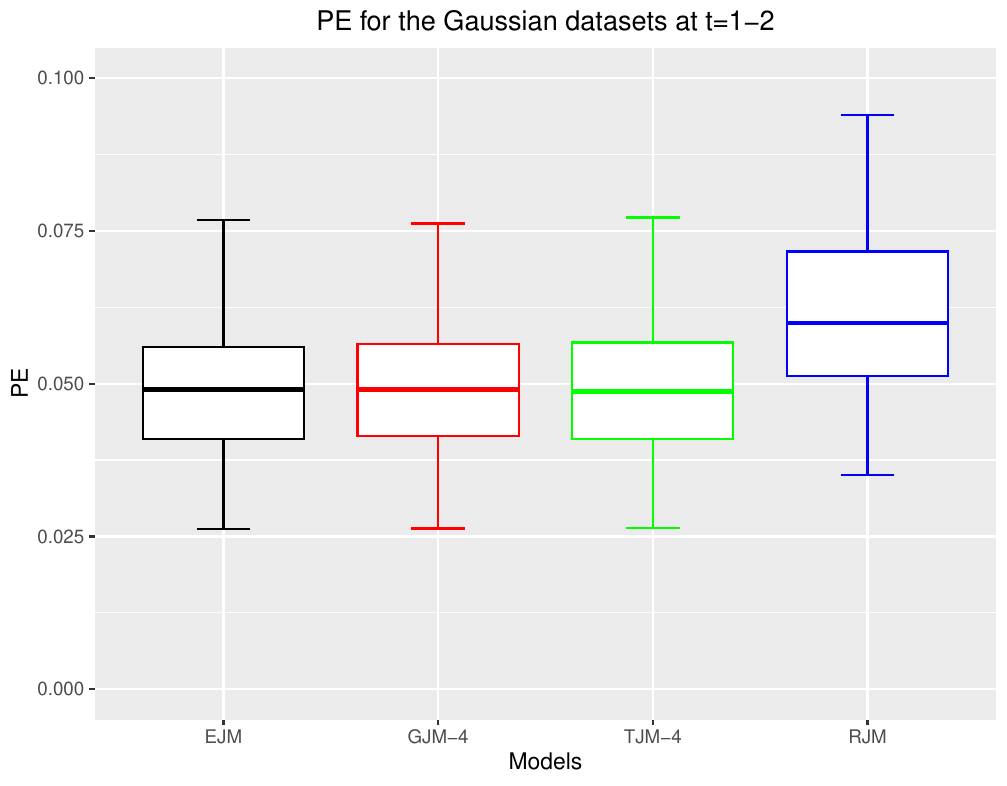}
	\end{minipage}
	\begin{minipage}{0.241\textwidth}
		\includegraphics[width=\linewidth]{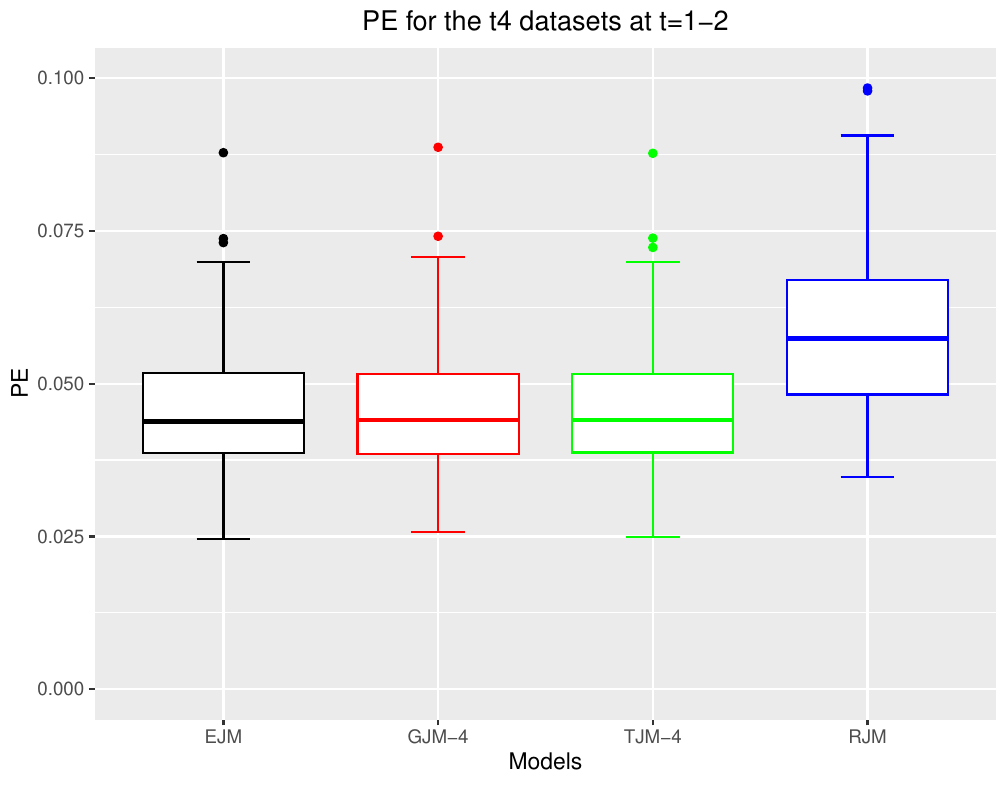}
	\end{minipage}
	\begin{minipage}{0.241\textwidth}
		\includegraphics[width=\linewidth]{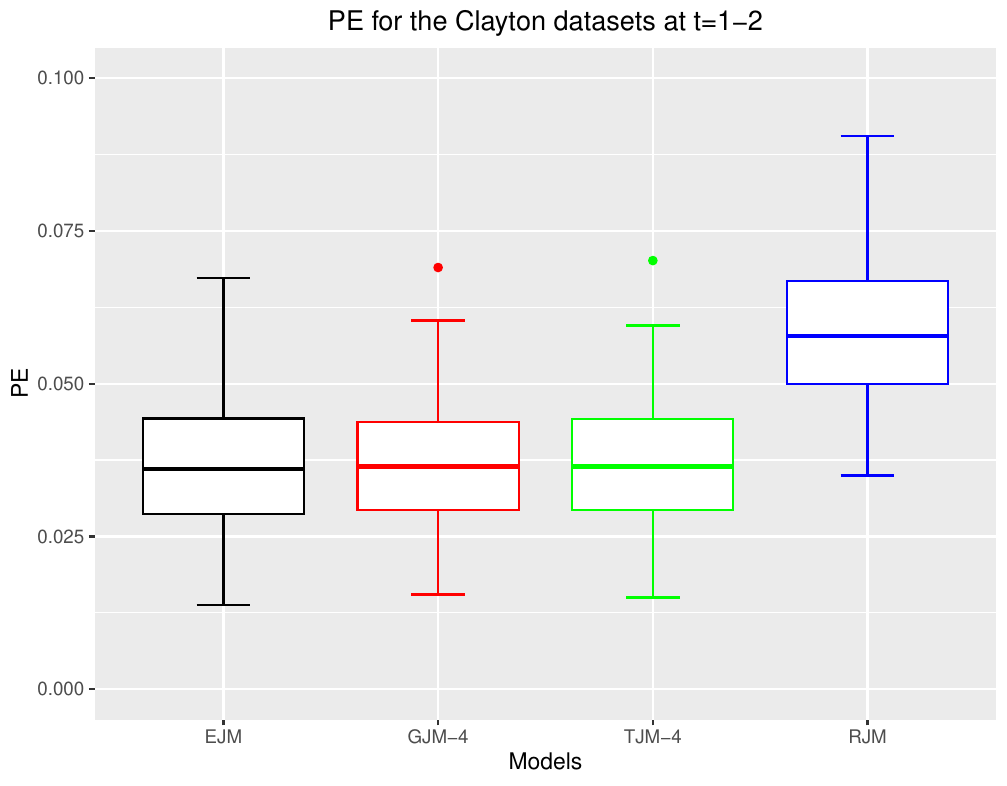}
	\end{minipage}
	\begin{minipage}{0.241\textwidth}
		\includegraphics[width=\linewidth]{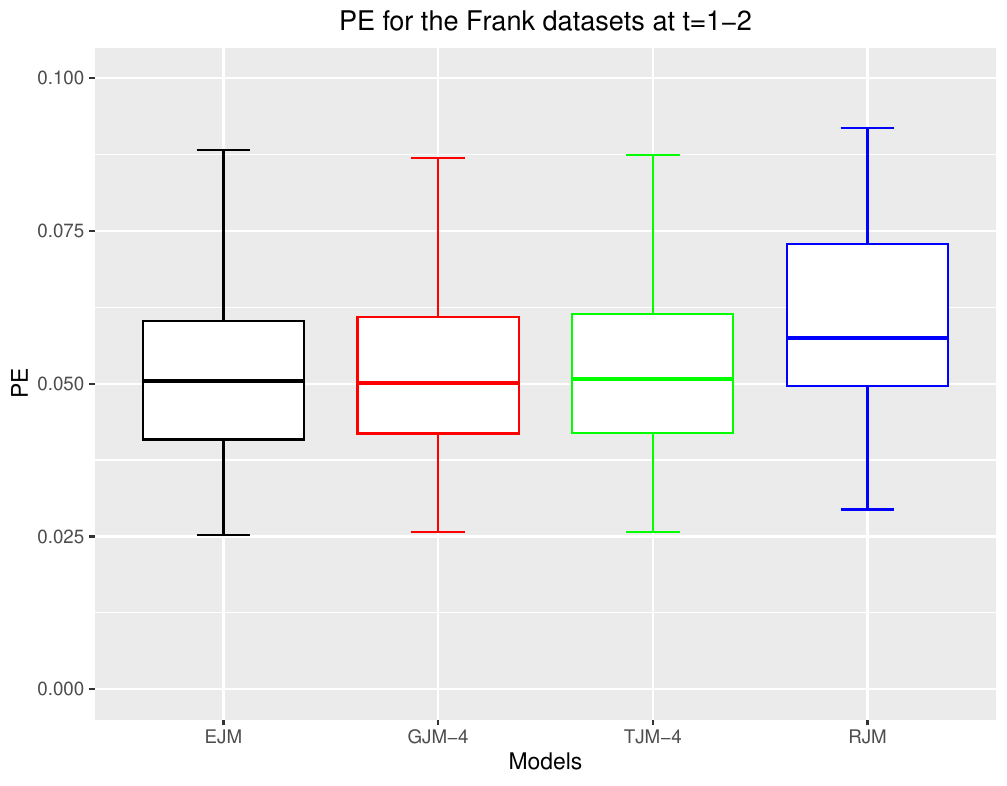}
	\end{minipage}
	\caption{The AUC and PE by the  EJM, GJM-4, T4JM-4 and RJM  from 100 Monte Carlo samples with sample size $200$ from case 1 in scenario 1.}
	\label{AUCPE_1to2}
\end{figure}

Figure \ref{AUCPE_1to2} presents the boxplots of AUC and PE calculated from the 100 Monte Carlo samples under the four models at $t=1$ with $\Delta t=u-t=1$. The Kendall's tau correlation of 0.5 is introduced between the two-submodels by the bivariate copula functions for the simulated datasets, but the RJM fails to capture all the copula association and its performance on survival prediction is inferior to the other models. The performance of GJM-4 and T4JM-4 are almost identical and they are both as good as EJM, which suggests that copula misspecification has minimal impact on predicting survival probability provided that the sub-models and structure of the copula correlation function are correctly specified.

Note that $\hat{\pi}(u|t)$ can also be calculated by $P_{T_{i}^{*}}\{t_{i}>u|\hat{\bm b}_{i},\mathcal{\bm Y}_{i}(t)\}/P_{T_{i}^{*}}\{t_{i}>t|\hat{\bm b}_{i},\mathcal{\bm Y}_{i}(t)\}$. As explained in Section 3.1,  $|f_{T_{i}^{*}}^{*}(t_{i}|\bm b_{i},\bm y_{i})-\tilde{f}_{T_{i}^{*}}(t_{i}|\bm b_{i},\bm y_{i})|$ is significantly decreased by fitting GJM-4 and T4JM-4, while it cannot be effectively reduced in RJM. Thus except misspecifying as the independence copula, the predicted survival probability is robust under copula misspecification.

	\noindent\textbf{Case 2: prediction under correlation function misspecification }	

In this case, we investigate the impact of misspecifying the correlation function in the bivariate copula given the copula is correctly chosen. Taking the Gaussian datasets from case 2 as an example, Figure \ref{estGautau} indicates the lack of fit of the correlation functions by GJM-4, while GJM-8 provides similar fitting to the correct model GJM-6. In fact, underfitting is more of an issue than overfitting here, as there is no measurement error in the correlation function. Nevertheless, the most appropriate number and locations of knots could be selected by AIC and BIC. 

\begin{figure}[H]
	\centering
	\begin{minipage}{0.241\textwidth}
		\includegraphics[width=\linewidth]{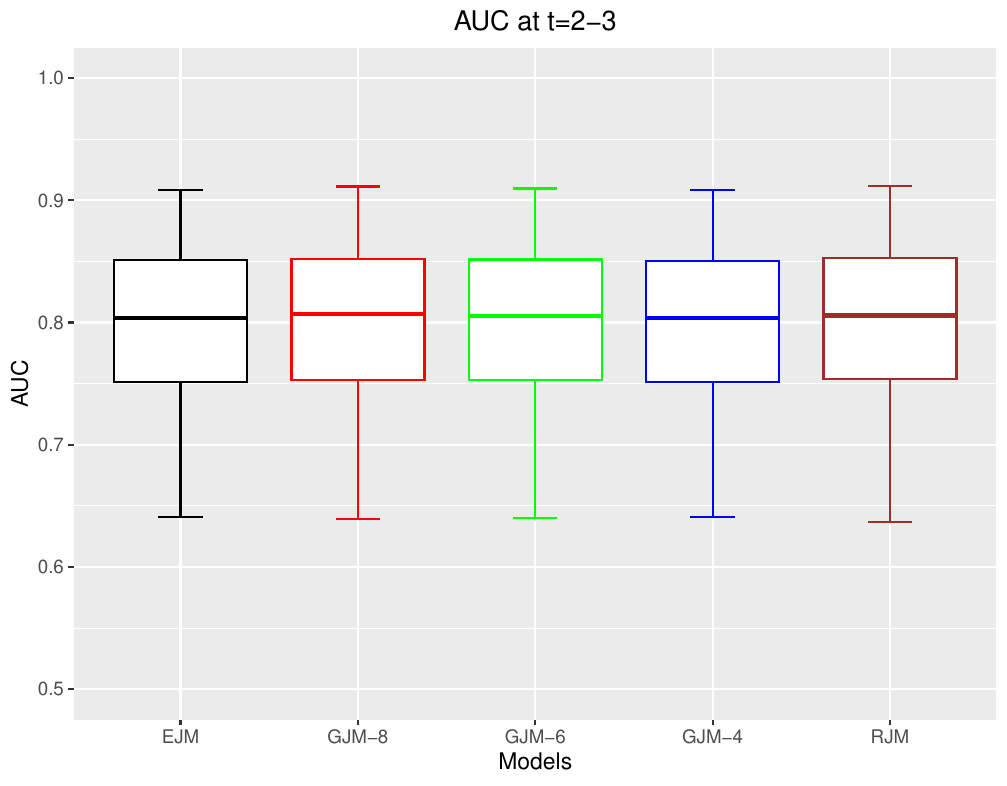}
	\end{minipage}
	\begin{minipage}{0.241\textwidth}
		\includegraphics[width=\linewidth]{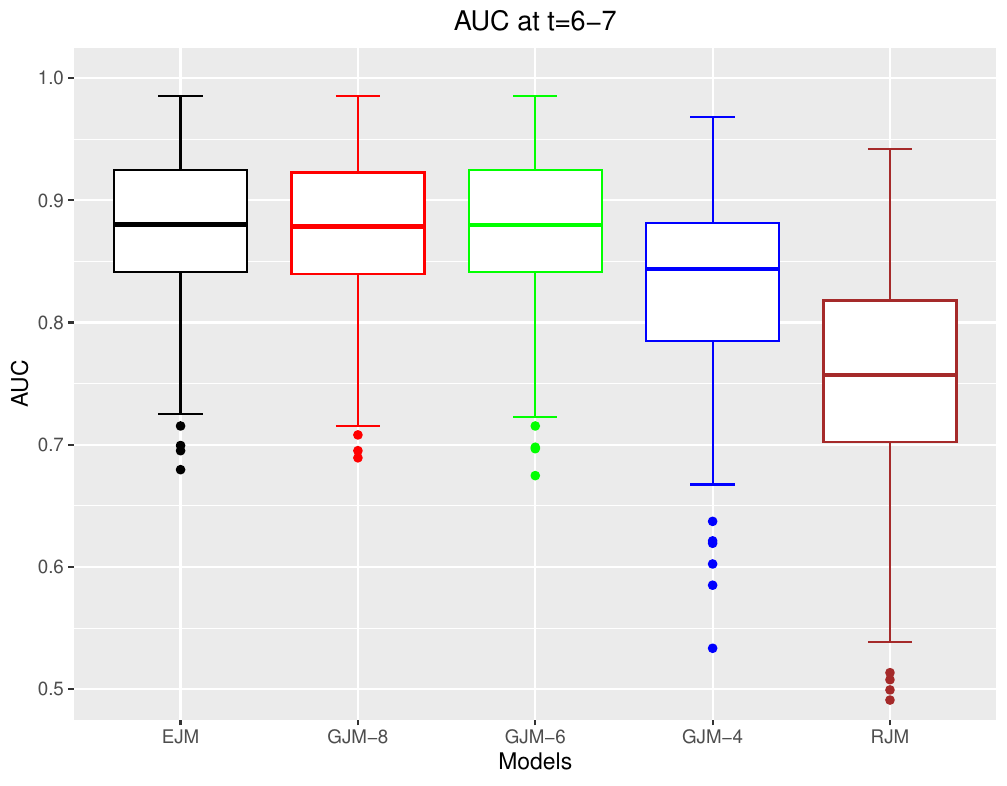}
	\end{minipage}
	\begin{minipage}{0.241\textwidth}
		\includegraphics[width=\linewidth]{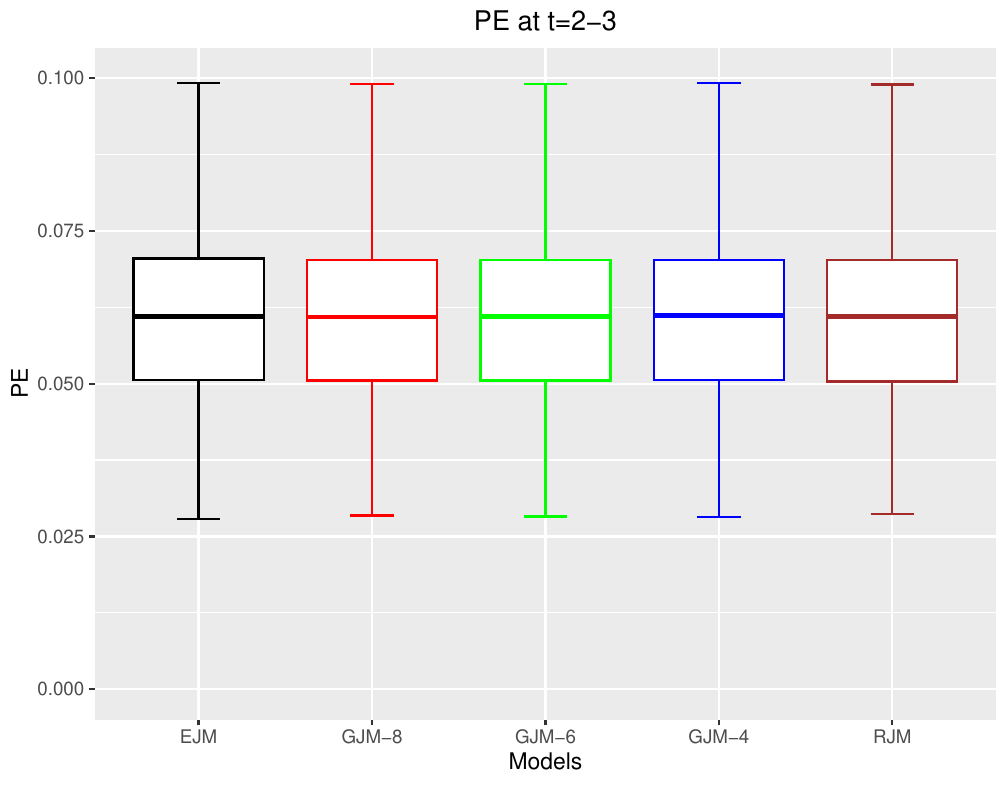}
	\end{minipage}
	\begin{minipage}{0.241\textwidth}
		\includegraphics[width=\linewidth]{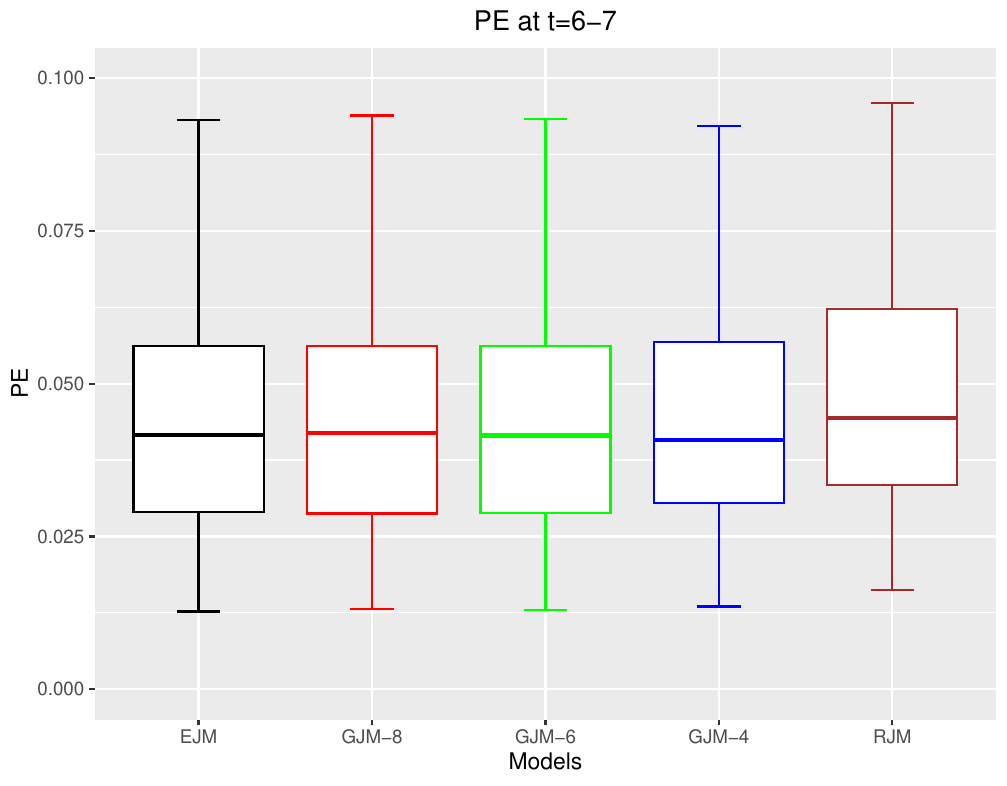}
	\end{minipage}
	\caption{The  AUC  and PE by the EJM, GJM-8, GJM-6, GJM-4 and RJM from 100 Monte Carlo samples with sample size $200$ for Gaussian datasets from case 2 in scenario 1.}
	\label{AUCPE_sin}
\end{figure}

Figure \ref{AUCPE_sin} shows the dynamic AUC and PE for the EJM, GJM-8, GJM-6, GJM-4 and RJM at $t=2$ and 6 with $\Delta t=u-t=1$. At $t=2,$ the five models provide almost equivalent prediction, and this is expected, as the correlation introduced by copula is about 0 at this time point and the five models have similar estimates of parameters. However, the prediction by EJM, GJM-6 and GJM-8 are superior than that of GJM-4 and RJM at $t=6.$ This is because the simper correlation structures in the latter two models fail to capture the strong negative local correlation  and are unable to use all the information from the longitudinal process for predicting at this time point. 

Since the estimations on parameters and predictions on survival probabilities are insensitive to copula misspecification, the bivariate Gaussian copula is recommended in practice due to its computational simplicity. But the structure of the correlation function in the copula should be treated with cautiousness, especially avoiding lack of fit.

\section{Application to the PBC data}
The PBC data (Murtaugh, \textit{et al,} 1994\cite{mur94}) records the information of 312 patients with primary biliary cirrhosis from 1974 to 1984. The main task of this study is to evaluate the treatment effect of D-penicillamine (158 randomised patients) compared with placebo (154 randomised patients). Baseline covariates such as age at the entry of study, gender and treatment are recorded. In the meanwhile, several biomarkers, such as serum bilirubin level (mg/dl), the presence of spiders and hepatomegaly (indicator variables), are monitored as follow-up. Among them, the serum bilirubin level is considered as a stronger indicator for the progression of the disease. After the baseline measurement, each patient was scheduled for visits at six months, one year and yearly afterwards, but not all patients attended the appointments at the scheduled time points and some even missed appointments, leading to an unbalanced dataset. Also due to death and censoring, only 1945 measurements of serum bilirubin levels are recorded, with the maximum measurement time up to 14.106 years and the number of measurements for each patient vary from 1 to 16. By the end of study, 149 patients died, 29 had a transplantation and 143 were still alive. 

Due to the relatively long gap between successive measurements and the high variability presented in the longitudinal trajectories of this dataset, we would like to investigate if there is local residual correlation between the two processes due to the unexplained local biological variation.

The proposed time-varying bivariate copula joint model is used to model the PBC data, with the longitudinal process specified as follows:
\begin{eqnarray}
	y_{ij}=\mu(s_{ij})+\beta_{11}drug_{i}+\beta_{12}sex_{i}+\beta_{13}age_{i}+b_{i0}+b_{i1}s_{ij}+\varepsilon_{ij},\label{pbclongsub}
\end{eqnarray}
where $\varepsilon_{ij}\sim N\left(0,\sigma^{2}\right),$ $(b_{i0},b_{i1})\sim N\left(0,\bm D\right)$ and $y_{ij}$ is the logarithm of serum bilirubin level for the $i$th subject at time $s_{ij}.$ Unlike in the simulation study, we model the population mean function $\mu(t)$ non-linearly by B-spline basis functions to allow more flexibility.
Death or transplantation is defined as the composite event. The time to event process is specified as:
\begin{eqnarray}
	h_{i}(t)=h_{0}(t)\mbox{exp}\left\{\beta_{21}drug_{i}+\beta_{22}sex_{i}+\beta_{23}age_{i}+\gamma\left(b_{i0}+b_{i1}t\right)\right\},\label{pbcsursub}
\end{eqnarray}
where $drug_{i}=1$ for D-penicillamine, $gender_{i}=1$ for female and $h_{0}(t)$ is a piecewise-constant function with equally spaced knots between 0 and the maximum observed event time at 14.306.

\begin{figure}[H]
	\centering
	\begin{minipage}{0.328\textwidth}
		\includegraphics[width=\linewidth]{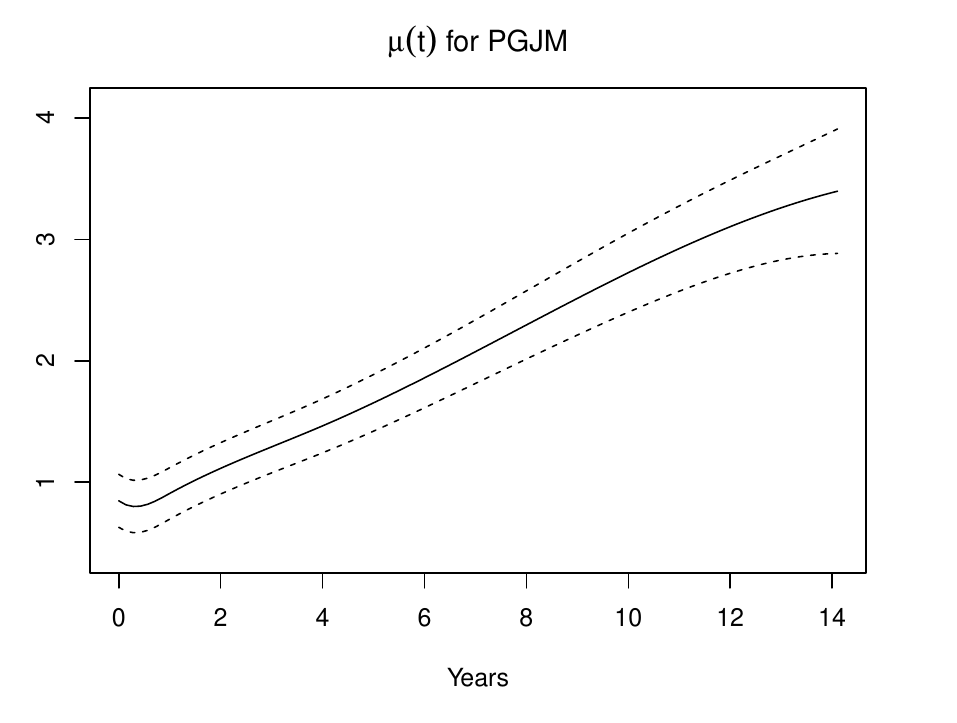}
	\end{minipage}
	\begin{minipage}{0.328\textwidth}
		\includegraphics[width=\linewidth]{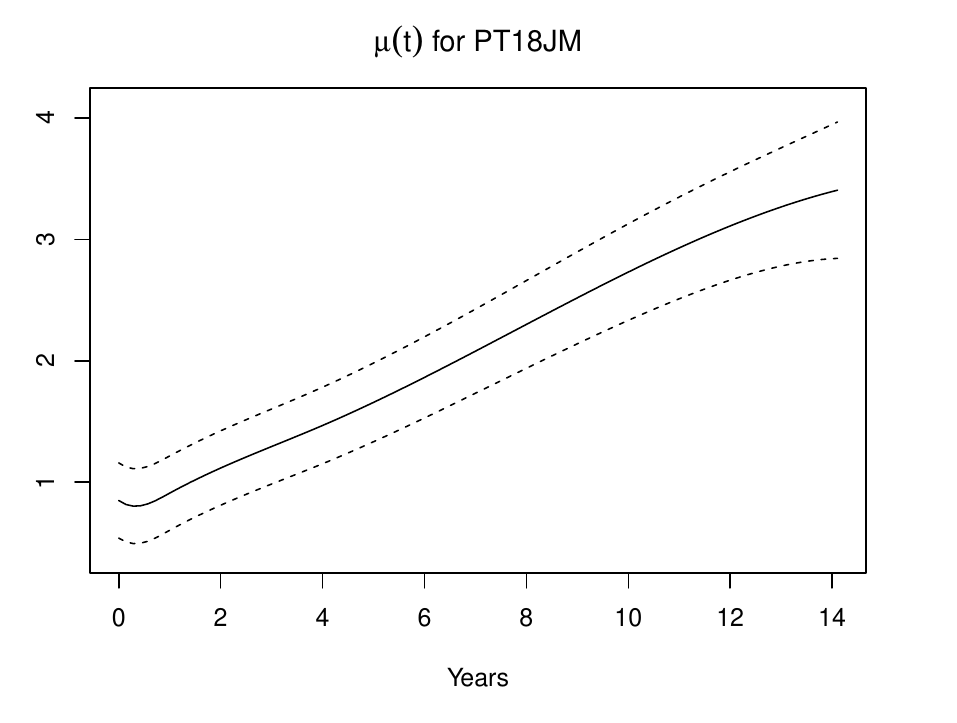}
	\end{minipage}
	\begin{minipage}{0.328\textwidth}
		\includegraphics[width=\linewidth]{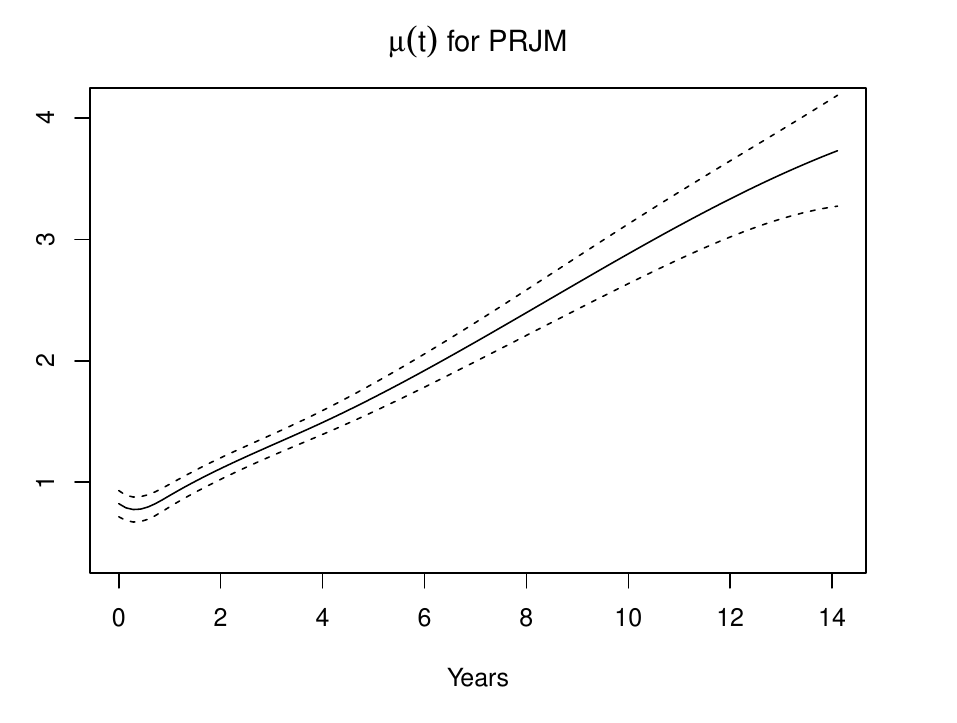}
	\end{minipage}
	\caption{The fitted mean functions $\hat{\mu}(t)$ for PBC dataset by the PGJM, PT18JM  and PRJM with the corresponding 95\% confidence band (dashed lines).}
	\label{mupbcbigplot}
\end{figure}

\begin{figure}[H]
	\centering
	\begin{minipage}{0.49\textwidth}
		\includegraphics[width=\linewidth]{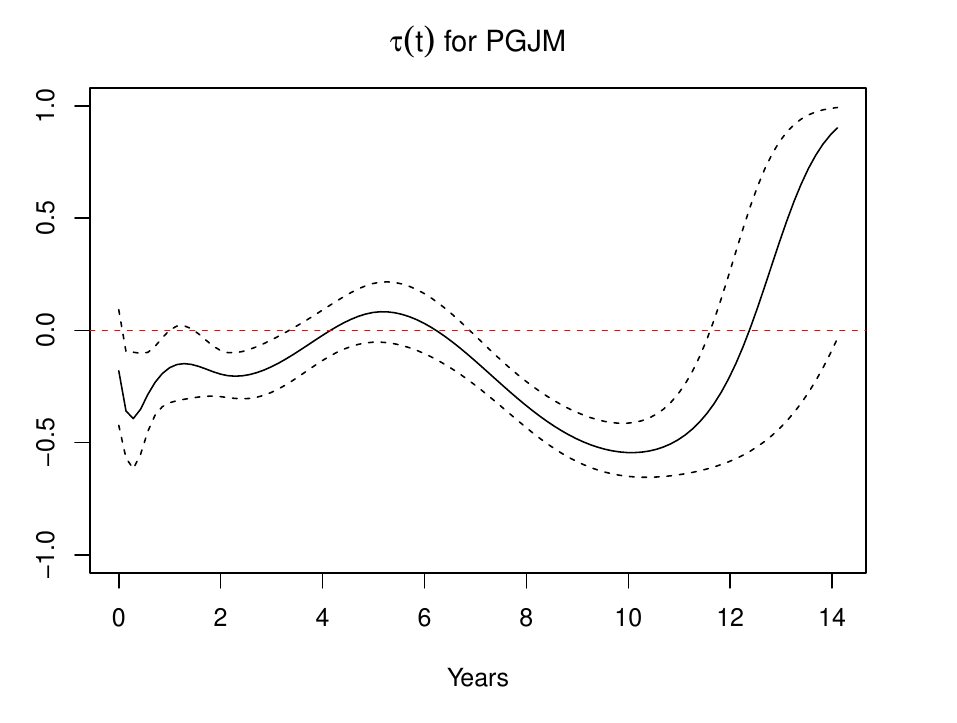}
	\end{minipage}
	\begin{minipage}{0.49\textwidth}
		\includegraphics[width=\linewidth]{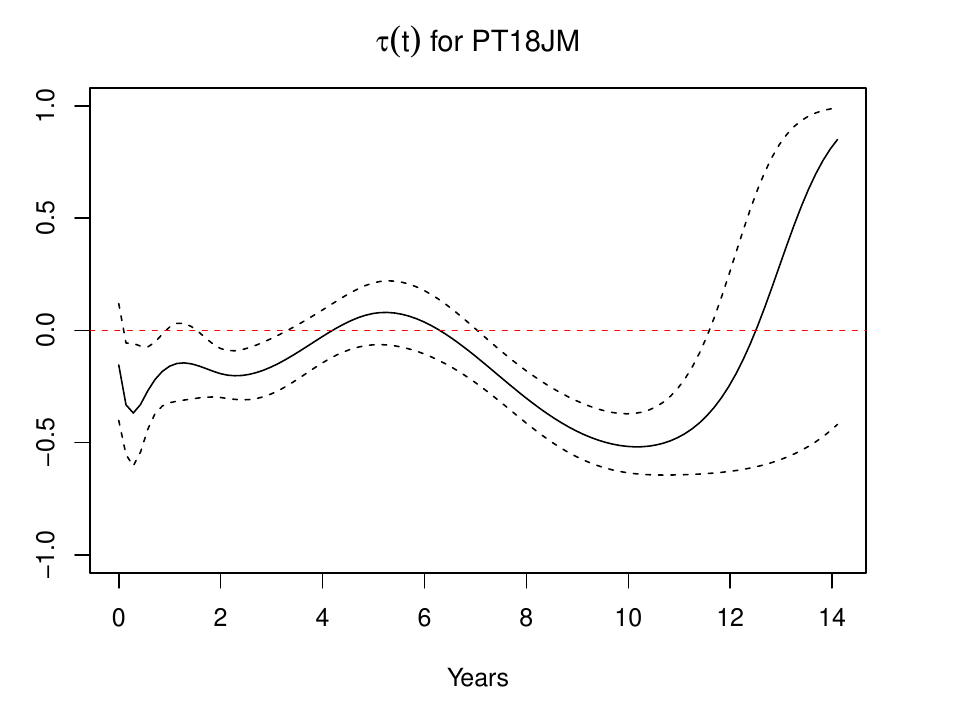}
	\end{minipage}
	\caption{The fitted Kendall's tau correlation functions $\hat{\tau}(t)$ for PBC dataset by the PGJM model and PT18JM with the corresponding 95\% confidence band (dashed lines) and reference lines (horizontal dashed lines) at zero.}
	\label{tautypbcbigplot}
\end{figure}

Three candidate models are applied for fitting the PBC dataset:
\begin{itemize}
	\item  \textbf{The bivariate Gaussian copula joint model (PGJM)}: the two sub-models are specified as (\ref{pbclongsub})  and (\ref{pbcsursub}), while the correlation between them is introduced by the bivariate Gaussian copula with $\mu(t)$ and $r(t)$ modelled by $k$ and $l$  cubic B-spline basis functions with knots located at the quantiles of the longitudinal measurement time points.
	\item  \textbf{The bivariate $t_{\nu}$ copula joint model (PT$\nu$JM)}: the two sub-models are specified as (\ref{pbclongsub})  and (\ref{pbcsursub}), while the correlation between them is introduced by the bivariate $t_{\nu}$ copula function with $\mu(t)$ and $r(t)$ modelled by $k$ and $l$ cubic B-spline basis functions with knots located at the quantiles of the longitudinal measurement time points. 
	\item  \textbf{The regular joint model (PRJM)}: the two sub-models are specified as (\ref{pbclongsub})  and (\ref{pbcsursub}) and assumed to be conditionally independent given the random effects, which is equivalent to $\tau(t)=0$ in the bivariate Gaussian copula joint model.
\end{itemize}
We select the optimal number of piecewise-constant baseline hazard function and  knots of cubic B-spline basis functions by AIC and BIC. Unlike the simulation study, the knots of the basis functions are located at the sample quantiles of the longitudinal measurements time points. The optimal combination is 6 cubic B-spline basis functions  for $\mu(t),$ 7 cubic B-spline basis functions for $\tau(t)$  and a piecewise-constant baseline hazard function with 7 pieces. We also notice that the fitted trajectories of $\hat{\mu}(t)$ and $\hat{\tau}(t)$ remain approximately the same when increasing the numbers of basis function beyond the optimal selection, while there are some significant changes in the trajectories by decreasing the numbers. Although we are aware that the fitted results is not sensitive to the choice of copula function, we still try to find the optimal $\nu$ for the PT$\nu$JM. Under the same sub-models as in PGJM, its log-likelihood value is increasing before $\nu=18$ and gradually decreasing later on, thus the optimal value of $\nu$ is 18. The fitted results of the three candidate models are summarised in Table \ref{pbcfit}. The two bivariate copula joint models provide similar fits in terms of parameter estimation and they result in a  similar fitted population mean function $\hat{\mu}(t)$ and Kendall's tau correlation function $\hat{\tau}(t),$ according to Figures \ref{mupbcbigplot} and \ref{tautypbcbigplot}. These results are consistent with the discussions in Section 2.4. In addition, the two bivariate copula joint models both provide significantly better fitting than the PRJM in terms of AIC and BIC. The correlation between the two sub-models is close to 0 during $0\leq t\leq7$, while it is  stronger during $7\leq t\leq10$ and the confidence intervals are much wider, especially for PT18JM, after $t=11,$ as there are fewer data beyond this point. This correlation structure indicates a larger negative (positive) local longitudinal variations during 7 to 10 years after entry could implies a better (worse) condition for the subject at this period, while there is no such implication in the earlier stage.

Despite the big differences in AIC and BIC, we notice that the parameter estimates and fitted mean function $\hat{\mu}(t)$ of the PRJM are generally similar to that of the bivariate copula joint models. This is because the correlation functions stay around 0 for half of the time. In fact, the fitted mean function of PRJM is only slightly higher than that of the bivariate copula joint models at the later stage and this may be due to the stronger correlation after $t\geq7$. Therefore, the proposed model does not tell us much new information compared to the PRJM  in terms of the interpretations of the regression parameters. This is similar to the case 2 of the simulation study. However, we shall see the main improvements of the proposed model is in predicting survival probability.

\begingroup
\setlength{\tabcolsep}{6pt} % Default value: 6pt
\renewcommand{\arraystretch}{1.2} 
\begin{table}[H]
	\tbl{Parameter estimates for the PBC data by the three candidate models.}
	{\begin{tabular}{ccccccc}
			\toprule
			&\multicolumn{2} {c} {PRJM}&\multicolumn{2} {c} {PGJM}&\multicolumn{2} {c} {P18TJM}
			\\
			\hline
			& Est. & SE &  Est.  &  SE     &Est.  &SE
			\\
			\hline
			$\beta_{11}$                                            & -0.126    &  0.110    & -0.135    & 0.112    & -0.136   & 0.112  
			\\
			$\beta_{12}$                                            & -0.166    & 0.165    & -0.207     & 0.167  & -0.212   & 0.167
			\\
			$\beta_{13}$                                            & -0.001    & 0.005   & -0.0005   & 0.005  & -0.0004& 0.005  
			\\
			$\beta_{21}$                                            & -0.190   & 0.225    & -0.167     & 0.213   & -0.167  & 0.215
			\\
			$\beta_{22}$                                            & -0.173   & 0.282    & -0.199    & 0.271   & -0.185  & 0.270
			\\
			$\beta_{23}$                                            & 0.041   & 0.007     &  0.044    & 0.006  &  0.045   & 0.006
			\\
			$D_{11}$                                                  & 0.959    & 0.082    &  0.966    & 0.083 & 0.965    &0.083  
			\\
			$D_{22}$                                                 & 0.038   & 0.005     & 0.033     & 0.005 &  0.032   & 0.005
			\\
			$D_{12}$                                                 & 0.083   & 0.016     &  0.070      & 0.016  & 0.070    &0.015   
			\\
			$\sigma$                                                 & 0.342   & 0.007     & 0.342      & 0.007  & 0.342    & 0.007  
			\\
			$\gamma$                                                 & 1.318    & 0.088     & 1.243       & 0.083  & 1.243    & 0.084
			\\
			Loglik &\multicolumn{2} {c}{-1928.047}  &\multicolumn{2} {c}{-1889.185} &\multicolumn{2} {c}{-1887.934}
			\\
			AIC    &\multicolumn{2} {c}{3904.094}     &\multicolumn{2} {c}{3840.37}  &\multicolumn{2} {c}{3837.868}  
			\\
			BIC    &\multicolumn{2} {c}{3993.926}    &\multicolumn{2} {c}{3956.403}  &\multicolumn{2} {c}{3953.901}  
			\\
			\hline
	\end{tabular}}\label{pbcfit}
\end{table}
\endgroup

As the visit is scheduled yearly after the first year, the AUC and PE are calculated for the follow-up times $t=1,...,10$ with $\Delta t=1$ by leave-one-out cross-validation. The results are summarised in Table \ref{dynaAUCBSpbc}. When the correlation is weak between the two sub-models, such as when $3\leq t\leq 7,$ the discriminative performance of the three candidate models are similar, with PRJM sometimes even having slightly better performance, e.g. at $t=4$ and 5. When the correlation between the two sub-models is strong, for example at $t=8,$ 9 and 10, the two bivariate copula joint models have significantly better performance than the PRJM.   The calculation of AUC and PE values are terminated for $t\geq11$ as there are only three events beyond this time point. Generally, the bivariate copula joint model provides a better prediction by utilising the residual information in the local biological variation and the choice of copula function is of less importance. The bivariate Gaussian copula joint model is recommended in practice as it requires less computation.

	\begingroup
\setlength{\tabcolsep}{6pt} % Default value: 6pt
\renewcommand{\arraystretch}{1.18} 
\begin{table}[H]
	\tbl{{ AUC and PE for three candidate models at different timepoints with $\Delta t=$1 for the PBC dataset.}}
	{\begin{tabular}{lcccccccccc}
			\toprule
			& $t=1$   &$t=2$     &$t=3$  &$t=4$  & $t=5$& $t=6$  & $t=7$ & $t=8$& $t=9$  & $t=10$
			\\
			\hline
			AUC($t+\Delta t|t$)
			\\
			\cdashline{1-11}
			PGJM             &  0.772  & 0.932    & 0.853 & 0.843  & 0.757 & 0.869  & 0.799 & 0.782 & 0.883  & 0.929
			\\            
			PT18JM          &  0.774  &   0.931  & 0.848 & 0.843  & 0.760 & 0.875  & 0.799 & 0.778 & 0.878   & 0.928
			\\            
			PRJM             &  0.739  &  0.928   &  0.847 & 0.852  & 0.770 & 0.879  & 0.793 & 0.746 & 0.796   & 0.864
			\\
			\cdashline{1-11}
			PE($t+\Delta t|t$)
			\\
			\cdashline{1-11}
			PGJM            &  0.044  & 0.066    & 0.060 & 0.061  & 0.064 & 0.079   & 0.052 & 0.108 & 0.090  & 0.070
			\\            
			PT18JM         &  0.043  &   0.067  & 0.060 & 0.061   & 0.064 & 0.079  & 0.052 & 0.108 & 0.091   & 0.071
			\\            
			PRJM            &  0.043  &  0.068   &  0.059 & 0.062  & 0.064 & 0.077   & 0.054 & 0.109 & 0.107  & 0.097
			\\
			\hline
	\end{tabular}}
	\label{dynaAUCBSpbc}
\end{table}
\endgroup

\begin{figure}[H]
	\centering
	\begin{minipage}{0.325\textwidth}
		\includegraphics[width=\linewidth]{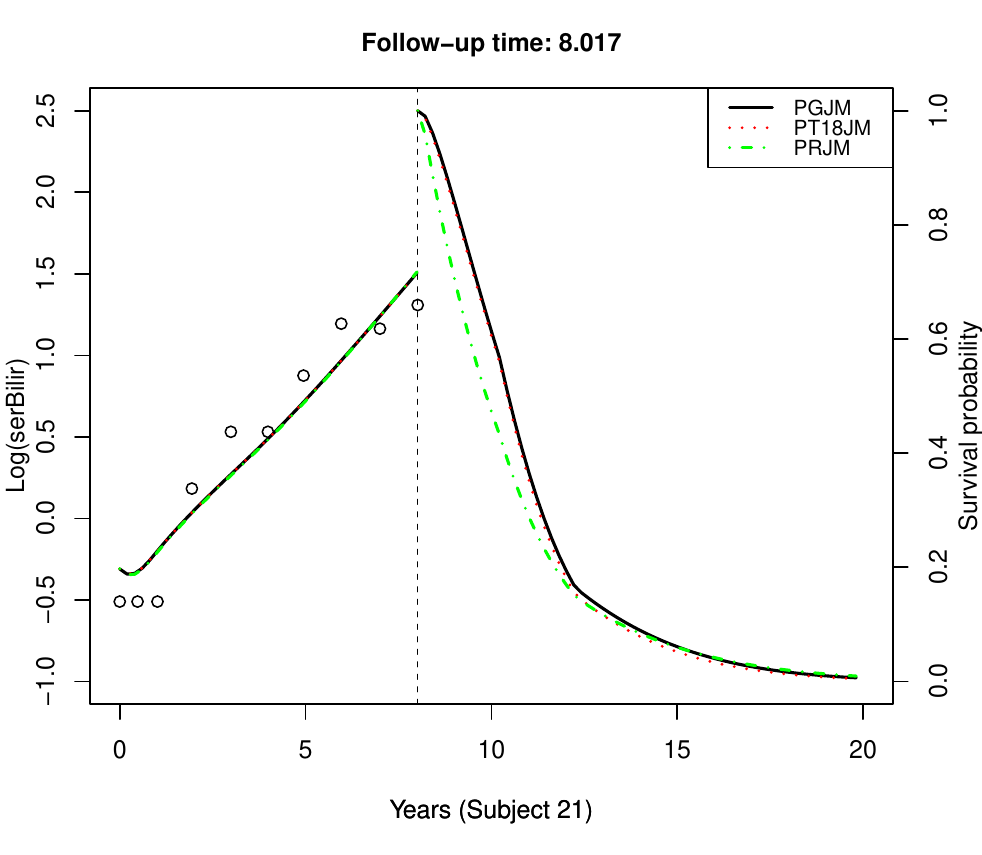}
	\end{minipage}
	\begin{minipage}{0.325\textwidth}
		\includegraphics[width=\linewidth]{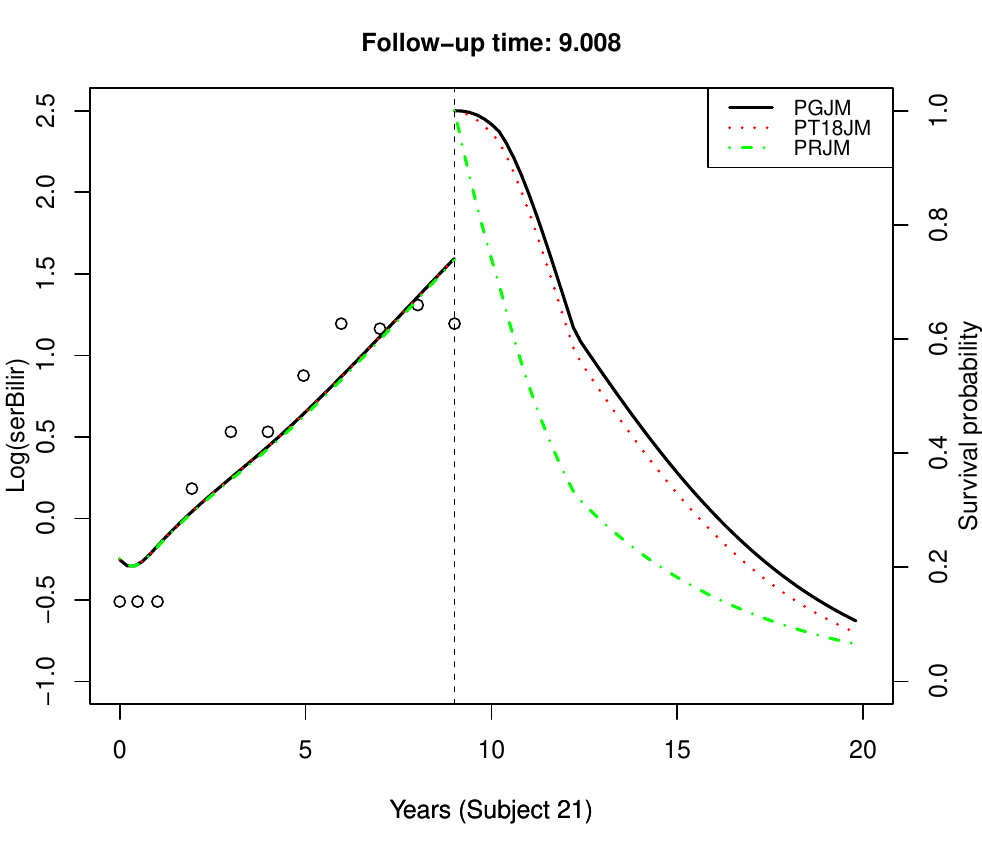}
	\end{minipage}
	\begin{minipage}{0.325\textwidth}
		\includegraphics[width=\linewidth]{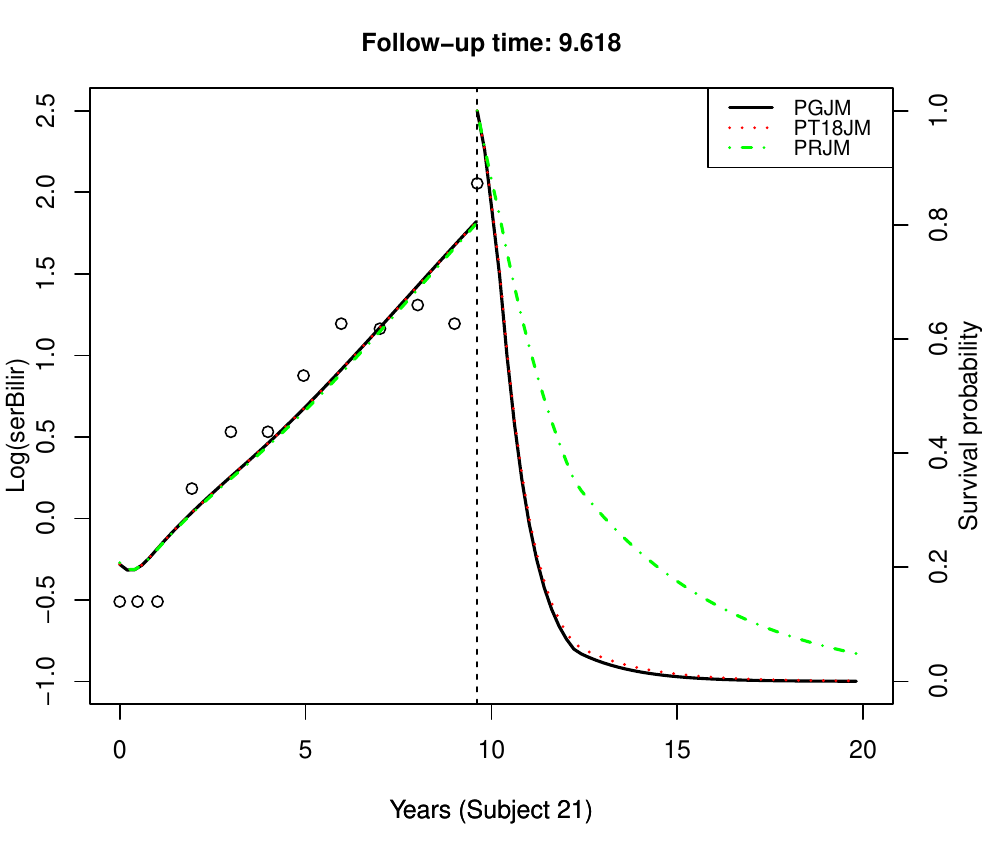}
	\end{minipage}
	\begin{minipage}{0.325\textwidth}
		\includegraphics[width=\linewidth]{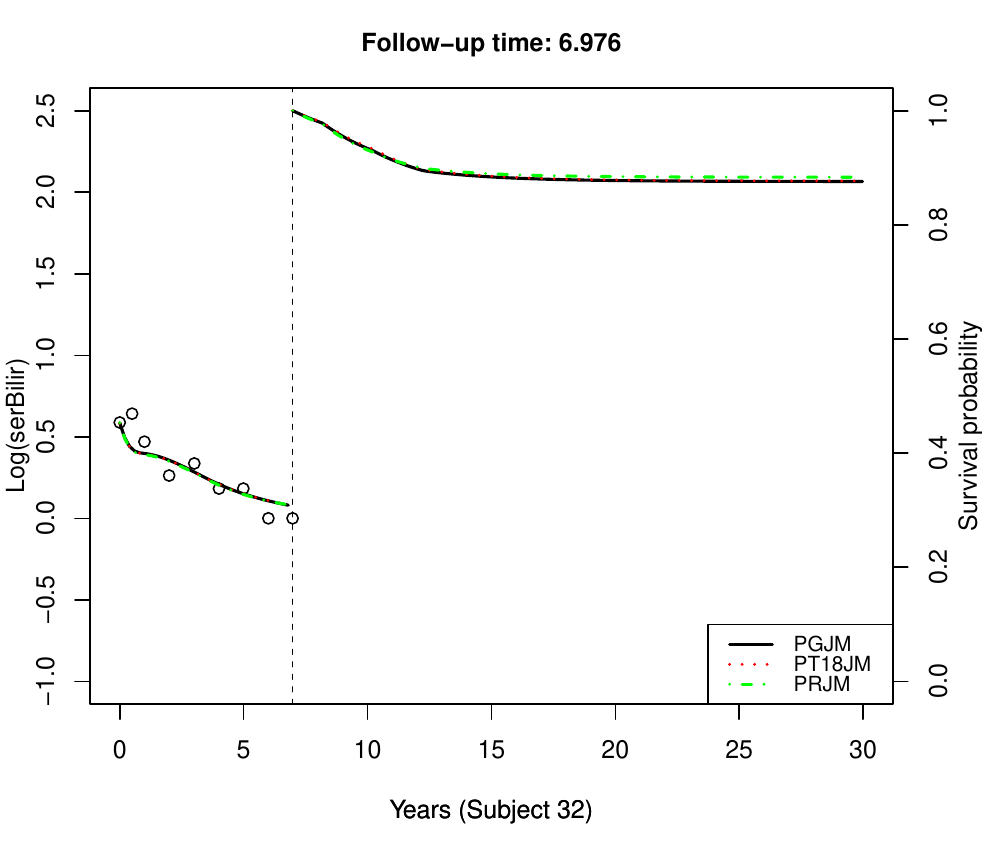}
	\end{minipage}
	\begin{minipage}{0.325\textwidth}
		\includegraphics[width=\linewidth]{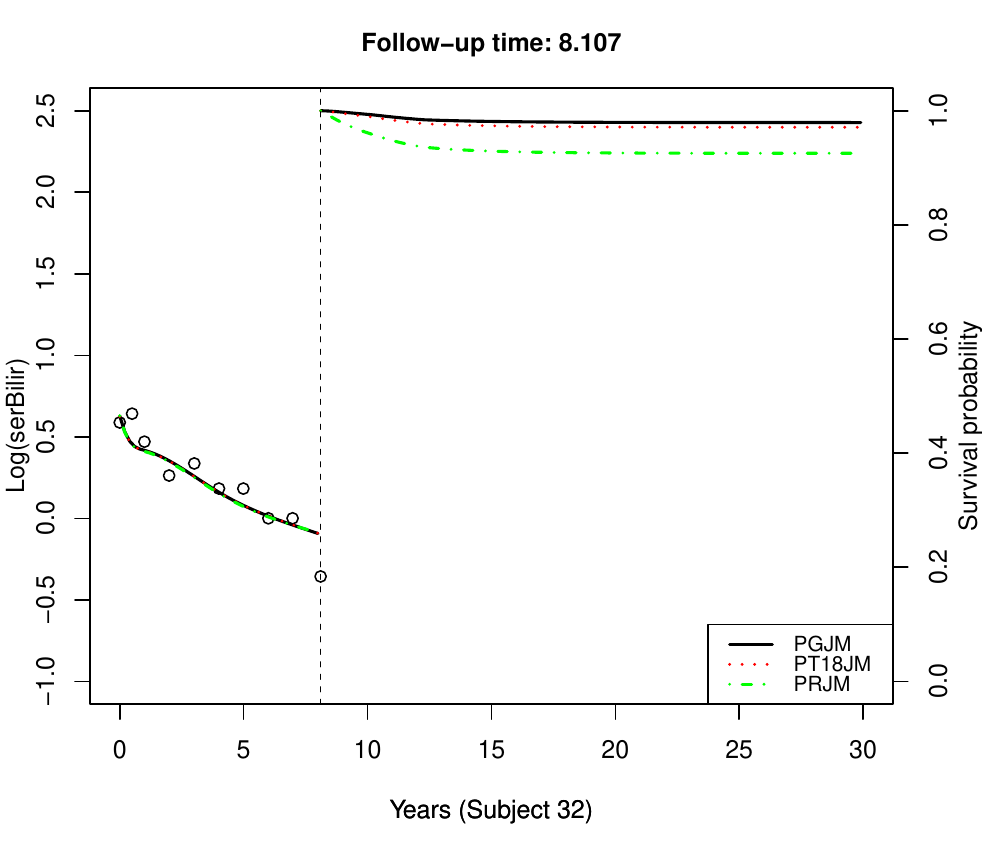}
	\end{minipage}
	\begin{minipage}{0.325\textwidth}
		\includegraphics[width=\linewidth]{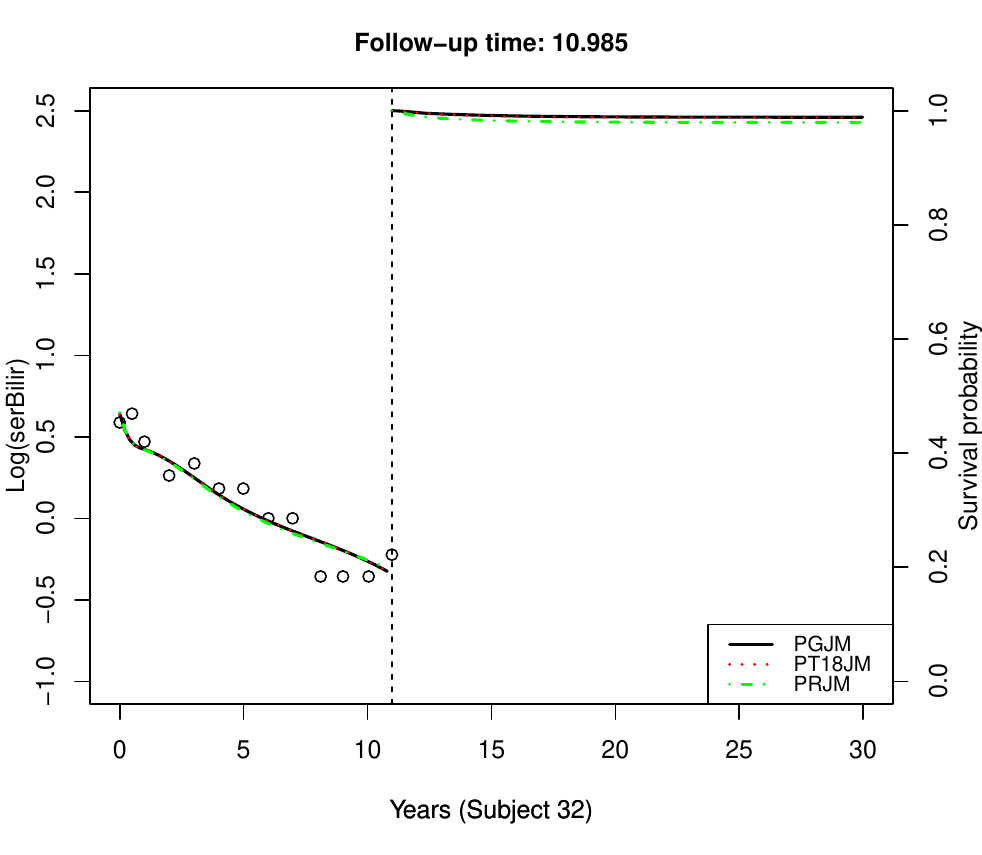}
	\end{minipage}
	\caption{Dynamic prediction of survival probabilities and fitted longitudinal trajectories for subjects 21 and 32 from the PBC dataset. The solid (black), dotted (red) and  dash-dotted (green) lines represent PGJM, PT18JM and PRJM, respectively, in Table \ref{pbcfit}.}
	\label{pbcbigplot}
\end{figure}

The first panel of Figure \ref{pbcbigplot} displays the predicted survival probabilities of subject 21 from the PBC dataset at $t=8.017,$ 9.008 and 9.618 by the PGJM, PT18JM and PRJM. It is noticeable that all the three models provide almost indistinguishable fitting for the longitudinal sub-model, while the predicted survival probabilities present obvious differences. To be more specific, all three models produce similar predictions in survival probabilities at $t=8.017,$ where the longitudinal observations scatter relatively evenly around the fitted curves. At $t=9.008,$ the new longitudinal observation is obviously lower than the fitted trajectories and the Kendall's tau correlation is quite strong at this time according to Figure \ref{tautypbcbigplot},  thus results in more optimistic predictions in survival probabilities by the two time-varying bivariate copula joint models than the PRJM. On the contrary, at $t=9.618,$ where the Kendall's correlation remains strong, an updated longitudinal measurement higher than the fitted curves makes the two time-varying bivariate  copula joint models produce lower predicted survival probabilities than the PRJM.  Given the subject is still alive at $t=9.618$, the PGJM and PT18JM  provide more accurate expected event times at 10.798 and 10.820 than that of PRJM at 12.627,  since the death time of this subject is at 10.013.

The second panel of Figure \ref{pbcbigplot} presents another example of subject 32, whose event time is censored at $t=14.215.$ Unlike subject 21, the longitudinal process of this subject experiences a downward trend. The fitted longitudinal trajectories of the three models are almost overlapped and their predicted survival probabilities  are generally very similar over time, with an exception at $t=8.107$, where a relatively lower longitudinal observation make the PGJM and PT18JM raise their predicted survival probability more than that of PRJM. 

In fact, the predictions by the  PRJM remain relatively stable across the time while the two time-varying bivariate copula joint models have more flexibility by accounting extra information from the large local variation if the correlation between the two sub-models is also strong at this moment.

\section{Discussion}
In the paper,  the regular joint model is generalised to have two layer of correlation, allowing us to relax the usually assumed but rarely checked assumption of conditional independence given the random effects in an old joint model. But the interpretations of the two dependencies are different. The predominant trend in longitudinal process acts on the event process directly by the commonly shared baseline covariates and time-invariant random effects,  whose impacts are permanent and directly related to the marginal distribution of the survival sub-model. On the other hand,
the unexplained variations have a dynamic but local influence on the event process via a time-varying bivariate copula and it does not affect the marginal distribution of the survival process.

In simulation studies, all the sub-models are correctly specified but fitted under different copulas. The estimators from the regular joint model may or may not present biases depending on the correlation structures in the bivariate copula for generating the data, while the estimators from the bivariate $t_{\nu}$ and Gaussian copula joint models are generally robust. The proposed models also have superior performance in predicting survival probability compared to the regular joint model  for having stronger discrimination capability and lower prediction error, and this is especially obvious when the correlation in the bivariate copulas, described by Kendall's tau function $\tau(t),$ is high.

The simulations also suggests both the parameter estimation and survival prediction are insensitive to the selection of the bivariate copula.
Due to computational simplicity, selecting the bivariate Gaussian copula joint model seems to be a reasonable choice in practice. More attention should be paid on how to select a correlation function that are complicated enough to capture the real one and  AIC and BIC criteria are applied to determine the optimal knot locations. Although even more computational expensive, the approaches of using larger number of knots by adding a roughness penalty term on the log-likelihood could also be explored in the future.

The real data application on the PBC dataset indicates our model provides significantly better fitting than the regular joint model. Despite that there are no obvious differences in the regression parameters between the proposed models and the regular joint model, the fitted correlation $\hat{\tau}(t)$  indicates there is strong extra correlation between the two processes arising from the local biological variation. The dynamic prediction of survival probabilities also shows the proposed models provide significant better predictions between $t=7$ and 11, where the correlation is strongest.

Although assuming conditional independence within the longitudinal process, conditional on the random effects, may not be unreasonable given the relatively large gap between the two successive measurement times, what happens under the violation of this assumption is still not clear and it might be interesting to investigate this issue. A more flexible non-parametric mean function is used to capture any non-linear trends in the longitudinal trajectories, but the fixed effect regression parameters and random effects are modelled as fixed in terms of $t.$ In future work, we could also consider modelling these components  functionally as in Brown \textit{et al.} (2005)\cite{bro05} and Yao (2007)\cite{yao07}. It could also be interesting to develop a similar score test for testing the conditional independence on the random effects like in Jacqmin-Gadda \textit{et al.} (2010)\cite{jac10}.  

\section*{Software}
R code is available at https://github.com/zhangzili0916/bivfun-copula-jointmodel-randomeffect on Github.

\section*{Appendix A}
\subsection*{Bivariate Clayton copula joint model} 
Suppose the bivariate Clayton copula is used to characterise the joint distribution of $\displaystyle U_{T_{i}^{*}|\bm b_{i},s_{ij}}$ and $\displaystyle U_{Y_{ij}|\bm b_{i},s_{ij}}.$ Then the joint CDF of $T_{i}^{*}, Y_{ij}|T_{i}^{*}>s_{ij},\bm b_{i},$ $j=1,...,n_{i}$  is given by:
\begin{eqnarray}
	\displaystyle
	\nonumber	F_{T_{i}^{*},Y_{ij}}(t, y_{ij}|\bm b_{i},T_{i}^{*}>s_{ij})=\left\{\left(U_{t|\bm b_{i},s_{ij}}\right)^{-\alpha_{ij}}+\left(U_{y_{ij}|\bm b_{i},s_{ij}}\right)^{-\alpha_{ij}}-1\right\}^{-\frac{1}{\alpha_{ij}}}.
\end{eqnarray}
Therefore its likelihood, depending on censored or not, can be derived as:
\begin{eqnarray}
	\displaystyle
	f_{T_{i}^{*},Y_{ij}}(t, y_{ij}|\bm b_{i},T_{i}^{*}>s_{ij})=\frac{\left(1+\alpha_{ij}\right)\left(U_{t|\bm b_{i},s_{ij}}\times U_{y_{ij}|\bm b_{i},s_{ij}}\right)^{-\alpha_{ij}-1}}{\left\{\left(U_{t|\bm b_{i},s_{ij}}\right)^{-\alpha_{ij}}+\left(U_{y_{ij}|\bm b_{i},s_{ij}}\right)^{-\alpha_{ij}}-1\right\}^{\frac{1}{\alpha_{ij}}+2}}\frac{dU_{t|\bm b_{i},s_{ij}}}{dt}\frac{dU_{y_{ij}|\bm b_{i},s_{ij}}}{dy_{ij}} \label{bivclaylikobs}
\end{eqnarray}
or
\begin{eqnarray}
	\displaystyle
	\nonumber	f_{T_{i}^{*},Y_{ij}}(T_{i}^{*}>t, y_{ij}|\bm b_{i}, T_{i}^{*}>s_{ij})&=&\left[1-\left(U_{y_{ij}|\bm b_{i},s_{ij}}\right)^{-\alpha_{ij}-1}\left\{\left(U_{t|\bm b_{i},s_{ij}}\right)^{-\alpha_{ij}}+\left(U_{y_{ij}|\bm b_{i},s_{ij}}\right)^{-\alpha_{ij}}-1\right\}^{-\frac{1}{\alpha_{ij}}-1}\right]\\
	&&\times\frac{dU_{y_{ij}|\bm b_{i},s_{ij}}}{dy_{ij}}, \label{bivclaylikcen}
\end{eqnarray}
where $\displaystyle\alpha_{ij}>0$ controls the strength of dependency and it is a  function of Kendall's correlation as $\displaystyle\alpha_{ij}=\frac{2\tau_{ij}}{1-\tau_{ij}}.$ The complete likelihood under the bivariate Clayton copula joint model can be obtained by substituting (\ref{bivclaylikobs}) and (\ref{bivclaylikcen}) back into (\ref{factorlikbi}). Note that the Clayton copula can be extended to allow $\displaystyle\alpha_{ij}\geq-1$ but the formulas for $-1\leq\displaystyle\alpha_{ij}\leq0$ are defined separately. Nevertheless, it can be extended to be a comprehensive copula in this way.

\subsection*{Bivariate Frank copula joint model} 
Suppose the bivariate Frank copula is used to characterise the joint distribution of $\displaystyle U_{T^{*}|\bm b_{i},s_{ij}}$ and $\displaystyle U_{Y_{ij}|\bm b_{i},s_{ij}}.$ Then the joint CDF of $T_{i}^{*}, Y_{ij}|T_{i}^{*}>s_{ij},\bm b_{i},$ $j=1,...,n_{i}$  is given by:
\begin{eqnarray}
	\displaystyle
	\nonumber	F_{T_{i}^{*},Y_{ij}}(t, y_{ij}|\bm b_{i},T_{i}^{*}>s_{ij})=-\frac{1}{\alpha_{ij}}\mbox{log}\left[1+\frac{\left\{\mbox{exp}\left(-\alpha_{ij} U_{t|\bm b_{i},s_{ij}}\right)-1\right\}\left\{\mbox{exp}\left(-\alpha_{ij} U_{y_{ij}|\bm b_{i},s_{ij}}\right)-1\right\}}{\mbox{exp}\left(-\alpha_{ij}\right)-1}\right]
\end{eqnarray}
Therefore its likelihood, depending on censored or not, can be derived as:
\begin{eqnarray}
	\displaystyle
	\nonumber	f_{T_{i}^{*},Y_{ij}}(t, y_{ij}|\bm b_{i},T_{i}^{*}>s_{ij})&=&-\frac{\alpha_{ij}\left\{\mbox{exp}\left(-\alpha_{ij}\right)-1\right\}\mbox{exp}\left\{-\alpha_{ij} \left(U_{t|\bm b_{i},s_{ij}}+U_{y_{ij}|\bm b_{i},s_{ij}}\right)\right\}}{\left[\mbox{exp}\left(-\alpha_{ij}\right)-1+\left\{\mbox{exp}\left(-\alpha_{ij} U_{t|\bm b_{i},s_{ij}}\right)-1\right\}\left\{\mbox{exp}\left(-\alpha_{ij} U_{y_{ij}|\bm b_{i},s_{ij}}\right)-1\right\}\right]^{2}}\\
	&&\times\frac{dU_{t|\bm b_{i},s_{ij}}}{dt}\frac{dU_{y_{ij}|\bm b_{i},s_{ij}}}{dy_{ij}} \label{bivfralikobs}
\end{eqnarray}
or
\begin{eqnarray}
	\displaystyle
	\nonumber&&	f_{T_{i}^{*},Y_{ij}}(T_{i}^{*}>t, y_{ij}|\bm b_{i}, T_{i}^{*}>s_{ij})\\
	\nonumber	&=&\left[1-\frac{\left\{\mbox{exp}\left(-\alpha_{ij} U_{t|\bm b_{i},s_{ij}}\right)-1\right\}\mbox{exp}\left(-\alpha_{ij} U_{y_{ij}|\bm b_{i},s_{ij}}\right)}{\mbox{exp}\left(-\alpha_{ij}\right)-1+\left\{\mbox{exp}\left(-\alpha_{ij} U_{t|\bm b_{i},s_{ij}}\right)-1\right\}\left\{\mbox{exp}\left(-\alpha_{ij} U_{y_{ij}|\bm b_{i},s_{ij}}\right)-1\right\}}\right]\times\frac{dU_{y_{ij}|\bm b_{i},s_{ij}}}{dy_{ij}},\\
	\label{bivfralikcen}
\end{eqnarray}
where $\displaystyle\alpha_{ij}\in(-\infty,\infty)\backslash\{0\}$ controls the strength of dependency and it is a  function of Kendall's correlation via the Debye function. The complete likelihood under the bivariate Frank copula joint model can be obtained by substituting (\ref{bivfralikobs}) and (\ref{bivfralikcen}) back into (\ref{factorlikbi}). Note that the Frank copula can be extended to be comprehensive by allowing $\displaystyle\alpha_{ij}=0,$ which then includes the independence copula. 

\section*{Appendix B}
Firstly, under the assumption in (\ref{factorlikbi}), we have $f_{\bm Y_{i}}(\bm y_{i}|\bm b_{i})=\prod_{j=1}^{n_{i}}f_{ Y_{ij}}( y_{ij}|\bm b_{i})=\prod_{j=1}^{n_{i}}dU_{y_{ij}|\bm b_{i},s_{ij}}/dy_{ij}.$ As $f_{T_{i}^{*},\bm Y_{i}}(t_{i}, \bm y_{i}|\bm b_{i})$ are available in Sections 2.1, 2.2 and Appendix A, we are able to derive $f_{T_{i}^{*}}(t_{i}|\bm y_{i},\bm b_{i})=f_{T_{i}^{*},\bm Y_{i}}(t_{i}, \bm y_{i}|\bm b_{i})/f_{\bm Y_{i}}(\bm y_{i}|\bm b_{i}).$ 

The posterior distribution of $T_{i}^{*}$ under the bivariate Gaussian copula joint model is given by
\begin{eqnarray*}
	\displaystyle
	f_{T_{i}^{*}}(t_{i}|\bm y_{i},\bm b_{i})
	&=&	P_{T_{i}^{*}}(T_{i}^{*}>s_{i1}|\bm b_{i})\left\{\prod_{j=1}^{n_{i}-1}\Phi\left(-\frac{Z_{s_{i(j+1)}|\bm b_{i},s_{ij}}-\alpha_{ij}Z_{y_{ij}|\bm b_{i},s_{ij}}}{\sqrt{1-\alpha_{ij}^{2}}}\right)\right\}^{I(n_{i}\geq2)}\\
	&\times&\left\{\frac{1}{\sqrt{1-\alpha_{in_{i}}^{2}}}\phi\left(\frac{Z_{t_{i}|\bm b_{i},s_{in_{i}}}-\alpha_{in_{i}}Z_{y_{in_{i}}|\bm b_{i},s_{in_{i}}}}{\sqrt{1-\alpha_{in_{i}}^{2}}}\right)\frac{dU_{t_{i}|\bm b_{i},s_{in_{i}}}/dt_{i}}{\phi\left(Z_{t_{i}|\bm b_{i},s_{in_{i}}}\right)}\right\}^{\delta_{i}}\\
	&\times&\Phi\left(-\frac{Z_{t_{i}|\bm b_{i},s_{in_{i}}}-\alpha_{in_{i}}Z_{y_{in_{i}}|\bm b_{i},s_{in_{i}}}}{\sqrt{1-\alpha_{in_{i}}^{2}}}\right)^{(1-\delta_{i})}\\
\end{eqnarray*}

The posterior distribution of $T_{i}^{*}$ under the bivariate $t_{\nu}$ copula joint model is given by
\begin{eqnarray*}
	\displaystyle
	f_{T_{i}^{*}}(t_{i}|\bm y_{i},\bm b_{i})&=&P_{T_{i}^{*}}(T_{i}^{*}>s_{i1}|\bm b_{i})\left[\prod_{j=1}^{n_{i}-1}\Psi\left\{-\frac{W_{s_{i(j+1)}|\bm b_{i},t_{ij}}^{\nu}-\alpha_{ij}W_{y_{ij}|\bm b_{i},s_{ij}}^{\nu}}{\sigma(s_{ij}|\bm b_{i},y_{ij})};\nu+1\right\}\right]^{I(n_{i}\geq2)}\\
	&\times&\left[\frac{1}{\sigma(s_{in_{i}}|\bm b_{i},y_{ij})}\psi\left\{\frac{W_{t_{i}|\bm b_{i},s_{in_{i}}}^{\nu}-\alpha_{in_{i}}W_{y_{in_{i}}|\bm b_{i},s_{in_{i}}}^{\nu}}{\sigma(s_{in_{i}}|\bm b_{i},y_{ij})};\nu+1\right\}\frac{dU_{t_{i}|\bm b_{i},s_{in_{i}}}/dt_{i}}{\psi\left(W_{t_{i}|\bm b_{i},s_{in_{i}}}^{\nu};\nu\right)}\right]^{\delta_{i}}\\
	&\times&\Psi\left\{-\frac{W_{t_{i}|\bm b_{i},s_{in_{i}}}^{\nu}-\alpha_{in_{i}}W_{y_{in_{i}}|\bm b_{i},s_{in_{i}}}^{\nu}}{\sigma(s_{in_{i}}|\bm b_{i},y_{in_{i}})};\nu+1\right\}^{(1-\delta_{i})}\\
\end{eqnarray*}

The posterior distribution of $T_{i}^{*}$ under the bivariate Clayton copula joint model is given by
\begin{eqnarray*}
	\displaystyle
	&&f_{T_{i}^{*}}(t_{i}|\bm y_{i},\bm b_{i})=P_{T_{i}^{*}}(T_{i}^{*}>s_{i1}|\bm b_{i})\\
	&\times&\left(\prod_{j=1}^{n_{i}-1}\left[1-\left(U_{y_{ij}|\bm b_{i},s_{ij}}\right)^{-\alpha_{ij}-1}\left\{\left(U_{s_{i(j+1)}|\bm b_{i},s_{ij}}\right)^{-\alpha_{ij}}+\left(U_{y_{ij}|\bm b_{i},s_{ij}}\right)^{-\alpha_{ij}}-1\right\}^{-\frac{1}{\alpha_{ij}}-1}\right]\right)^{I(n_{i}\geq2)}\\
	&\times&\left[\frac{\left(1+\alpha_{in_{i}}\right)\left(U_{t_{i}|\bm b_{i},s_{in_{i}}}\times U_{y_{in_{i}}|\bm b_{i},s_{in_{i}}}\right)^{-\alpha_{in_{i}}-1}}{\left\{\left(U_{t_{i}|\bm b_{i},s_{in_{i}}}\right)^{-\alpha_{in_{i}}}+\left(U_{y_{in_{i}}|\bm b_{i},s_{in_{i}}}\right)^{-\alpha_{in_{i}}}-1\right\}^{\frac{1}{\alpha_{in_{i}}}+2}}\frac{dU_{t_{i}|\bm b_{i},s_{in_{i}}}}{dt_{i}}\right]^{\delta_{i}}\\
	&\times&\left[1-\left(U_{y_{in_{i}}|\bm b_{i},s_{in_{i}}}\right)^{-\alpha_{in_{i}}-1}\left\{\left(U_{t_{i}|\bm b_{i},s_{in_{i}}}\right)^{-\alpha_{in_{i}}}+\left(U_{y_{in_{i}}|\bm b_{i},s_{in_{i}}}\right)^{-\alpha_{in_{i}}}-1\right\}^{-\frac{1}{\alpha_{in_{i}}}-1}\right]^{(1-\delta_{i})}\\
\end{eqnarray*}

The posterior distribution of $T_{i}^{*}$ under the bivariate Frank copula joint model is given by
\begin{eqnarray*}
	\displaystyle
	&&f_{T_{i}^{*}}(t_{i}|\bm y_{i},\bm b_{i})=P_{T_{i}^{*}}(T_{i}^{*}>s_{i1}|\bm b_{i})\\
	&\times&\left(\prod_{j=1}^{n_{i}-1}\left[1-\frac{\left\{\mbox{exp}\left(-\alpha_{ij} U_{s_{i(j+1)}|\bm b_{i},s_{ij}}\right)-1\right\}\mbox{exp}\left(-\alpha_{ij} U_{y_{ij}|\bm b_{i},s_{ij}}\right)}{\mbox{exp}\left(-\alpha_{ij}\right)-1+\left\{\mbox{exp}\left(-\alpha_{ij} U_{s_{i(j+1)}|\bm b_{i},s_{ij}}\right)-1\right\}\left\{\mbox{exp}\left(-\alpha_{ij} U_{y_{ij}|\bm b_{i},s_{ij}}\right)-1\right\}}\right]\right)^{I(n_{i}\geq2)}\\
	&\times&\left[-\frac{\alpha_{in_{i}}\left\{\mbox{exp}\left(-\alpha_{in_{i}}\right)-1\right\}\mbox{exp}\left\{-\alpha_{in_{i}} \left(U_{t_{i}|\bm b_{i},s_{in_{i}}}+U_{y_{in_{i}}|\bm b_{i},s_{in_{i}}}\right)\right\}}{\left[\mbox{exp}\left(-\alpha_{in_{i}}\right)-1+\left\{\mbox{exp}\left(-\alpha_{in_{i}} U_{t_{i}|\bm b_{i},s_{in_{i}}}\right)-1\right\}\left\{\mbox{exp}\left(-\alpha_{in_{i}} U_{y_{in_{i}}|\bm b_{i},s_{in_{i}}}\right)-1\right\}\right]^{2}}\frac{dU_{t_{i}|\bm b_{i},s_{in_{i}}}}{dt_{i}}\right]^{\delta_{i}}\\
	&\times&\left[1-\frac{\left\{\mbox{exp}\left(-\alpha_{in_{i}} U_{t_{i}|\bm b_{i},s_{in_{i}}}\right)-1\right\}\mbox{exp}\left(-\alpha_{in_{i}} U_{y_{in_{i}}|\bm b_{i},s_{in_{i}}}\right)}{\mbox{exp}\left(-\alpha_{in_{i}}\right)-1+\left\{\mbox{exp}\left(-\alpha_{in_{i}} U_{t_{i}|\bm b_{i},s_{in_{i}}}\right)-1\right\}\left\{\mbox{exp}\left(-\alpha_{in_{i}} U_{y_{in_{i}}|\bm b_{i},s_{in_{i}}}\right)-1\right\}}\right]^{(1-\delta_{i})}\\
\end{eqnarray*}


\begin{thebibliography}{999}
			 	\bibitem{als20}
		Alsefri, M.,  Sudell, M.,
		Garc\'{\i}a-Fi\~{n}ana, M. and Kolamunnage-Dona, R. (2020) Bayesian joint modelling of longitudinal and time to event data: a methodological review. \textit{BMC Med Res Methodol} \textbf{20,} 94.
		
		\bibitem{and03}
		Andersen, P. K., Borgan, O., Gill, R. D. and Kieding, N. (2003) Statistical Models Based on Counting Processes. New York: Springer.
		
		\bibitem{and18}
		Andrinopoulou, E., Eilers, P., Takkenberg, J., and Rizopoulos, D. (2018)  Improved dynamic predictions from joint models of
		longitudinal and survival data with time-varying
		effects using P-Splines. \textit{Biometrics} \textbf{74}, 685-693
		
		\bibitem{bro05}
		Brown, E. R., Ibrahim, J. G. and DeGruttola, V. (2005). A flexible B-spline model for multiple
		longitudinal biomarkers and survival. \textit{Biometrics} \textbf{61}, 64-73.
		
		\bibitem{cho24}
	  Cho, S.,  Psioda,
M. A. and  Ibrahim, J. G. (2024). Bayesian joint modeling of multivariate
longitudinal and survival outcomes using
	  Gaussian copulas. \textit{Biostatistics},  \textbf{25}, 962-977.
	
		
		\bibitem{cox72}
	Cox, D. R. (1972). Regression models and life tables (with discussion). \textit{Journal of the Royal Statistical Society, Series
		B} \textbf{34}, 187–220.	
	
		\bibitem{cox74}
	Cox, D. R. and Hinkley, D. (1974). Theoretical Statistics. Chapman and Hall, London.
	
	
	  	\bibitem{den83}
	Dennis, J. E. and Schnabel, R. B. (1983). Numerical Methods for Unconstrained Optimization and Nonlinear Equations. \textit{Prentice-Hall, Englewood Cliffs, NJ.}
	
	 \bibitem{dut21}
	Dutta, S., Molenberghs, G. and  Chakraborty, A. (2021) Joint modelling of longitudinal response and time-to-event data using conditional distributions: a
	Bayesian perspective. 	\textit{Journal of Applied Statistics,} \textbf{49}, 2228-2245.	
	
	
	 \bibitem{emu17}
	Emura, T., Nakatochi, M., Murotani, K., and Rondeau, V. (2017). A joint frailty-copula model between tumour progression and death for meta-analysis. \textit{Statistical Methods in Medical Research}, \textbf{26(6)}, 2649-2666.
	
	
	 \bibitem{fau96}
	Faucett,  C. L. and Thomas, D. C. (1996). Simultaneously modelling censored survival data and repeatedly
	measured covariates: A Gibbs sampling approach. \textit{Stat Med} \textbf{15,} pp. 1663–1685.
	
	  \bibitem{gan15}
	Ganjali, M. and Baghfalaki, T. (2015). A copula approach to joint modeling of longitudinal measurements
	and survival times using Monte Carlo expectation-maximization with application to AIDS studies. \textit{Journal of Biopharmaceutical Statistics} \textbf{25,} 1077-1099.
	
	\bibitem{Gar08}
	Garre, F. G., Zwinderman, A. H., Geskus, R. B. and Sijpkens, Y. W. J. (2008). A joint latent class changepoint model to improve
the prediction of time to graft failure. \textit{J. R. Statist. Soc. A} 
\textbf{171}, Part 1, pp. 299-308
	
	 \bibitem{guo06}
	 Guo, J., Wall, M., and Amemyia, Y. (2006). Latent class regression on
	latent factors. \textit{Biostatistics} \textbf{7}, 145–163.
	
	    \bibitem{guo04}
	Guo, X. and Carlin, B. P. (2004). Separate and Joint Modelling of Longitudinal
	and Event Time Data Using Standard Computer Packages, \textit{The American Statistician,} \textbf{58:1}, 16-24,
	DOI: 10.1198/0003130042854
	
	 \bibitem{hen00}
	Henderson, R., Diggle, P. and Dobson, A. (2000). Joint modelling of longitudinal measurements and event time data. 	\textit{Biostatistics} \textbf{4}, 465-480.
	
	   \bibitem{hof18}
	Hofert, M., Kojadinovic, I., Machler, M. and Yan, J. (2018). Elements of copula modelling with R. Berlin, Germany: Springer International Publishing AG. 
	
	
	\bibitem{hsi06}
	Hsieh, F., Tseng, Y. K., and Wang, J. L. (2006). Joint modeling of survival
and longitudinal data: likelihood approach revisited. \textit{Biometrics} \textbf{62}, 1037-1043.
	
		\bibitem{ibr10}
	Ibrahim, Joseph, G., Chu, H. and Chen, L. M. (2010). Basic concepts and methods for joint models
	of longitudinal and survival data, \textit{Journal of Clinical Oncology,} Vol. \textbf{28}, pp. 2796-2801.
	
	  \bibitem{jac05}
	J\"{a}ckel, P. (2005). A note on multivariate Gauss-Hermite quadrature.
	
		\bibitem{jac10}
	Jacqmin-Gadda, H., Proust-Lima, C., Taylor, J.M. and Commenges, D. (2010). Score test for conditional independence between longitudinal outcome and time to
	event given the classes in the joint latent class model. \textit{Biometrics} \textbf{66}, 11–19.
	
		
   \bibitem{li21}
    Li, C., Xiao, L. and Luo, S. (2021). Joint model for survival and multivariate sparse functional
   data with application to a study of Alzheimer’s Disease. \textit{Biometrics,} \textbf{78}, 435-447
  
  	  \bibitem{li17}
  Li, K. and Luo, S. (2017). Functional joint model for longitudinal
  and time-to-event data: an application to Alzheimer’s disease. \textit{Stat Med} \textbf{36}, 3560-3572.
  
  \bibitem{li19}
  Li, K. and Luo, S. (2019). Bayesian functional joint models for multivariate longitudinal and time-to-event data. \textit{Computational Statistics and Data Analysis} 129:14–29, https://doi.org/10.1016/j.csda.2018.07.015
  
  \bibitem{lin04}
  Lin, H., McCulloch, C. E., and Rosenheck, R. A. (2004). Latent pattern mixture models for informative intermittent missing data in
longitudinal studies. \textit{Biometrics} \textbf{60}, 295–305.
  
  \bibitem{lin02}
  Lin, H., Turnbull, B. W., McCulloch, C. E., and Slate, E. H. (2002). Latent class models for joint analysis of longitudinal biomarker and
event process data: Application to longitudinal prostate-specific
  antigen readings and prostate cancer. \textit{Journal of the American Statistical Association} \textbf{97}, 53–65.
  
  \bibitem{mal15}
  Malehi, A. S., Hajizadehb, E., Ahmadia, K. A., and Mansouric, P. (2015). Joint modelling of longitudinal biomarker and gap time between recurrent events:
  copula-based dependence. \textit{Journal of Applied Statistics,} 2015
  Vol. \textbf{42,} No. 9, 1931–1945.
  
	\bibitem{mur94}
	Murtaugh, P., Dickson, E., Van Dam, G., Malincho, M., Grambsch, P., Lang worthy, A., and Gips, C. (1994). Primary biliary cirrhosis: prediction of
short-term survival based on repeated patient visits. \textit{Hepatology} \textbf{20}, 126-134.


\bibitem{nel65}
Nelder, J. A. and Mead, R. (1965). A simplex algorithm for function minimization. \textit{Computer Journal,} \textbf{7}, 308--313. 
	
\bibitem{pap19}
	Papageorgiou, G., Mauff, K., Tomer, A., and Rizopoulos, D. (2019). An Overview
	of Joint Modelling of Time-to-Event and Longitudinal Outcomes. \textit{Annual Review of Statistics and Its Application.}
	
	\bibitem{pro09}
	Proust-Lima, C., Joly, P., and Jacqmin-Gadda, H. (2009). Joint modelling of multivariate longitudinal outcomes and a time-to-event:
A nonlinear latent class approach. \textit{Computational Statistics and
Data Analysis} \textbf{53}, 1142–1154.

    \bibitem{riz08a}
Rizopoulos, D., Verbeke, G., and Lesaffre, E. (2008a). A two-part joint model for the analysis of survival and longitudinal binary data with excess zeros. \textit{Biometrics} \textbf{64,} 611-619.

\bibitem{riz08b}
Rizopoulos D., Verbeke, G., and Molenberghs, G. (2008b). Shared parameter models under random effects misspecification. \textit{Biometrika} \textbf{95}, 63-74.


		\bibitem{riz10}
	Rizopoulos, D. (2010). \verb|JM|: An \verb|R| package for the joint modelling of longitudinal and time-to-event data. \textit{Journal of Statistical Software} \textbf{35 (9),} 1-33.
	
	
	
	 \bibitem{riz11}
	Rizopoulos, D. (2011). Dynamic predictions and prospective accuracy in joint models for longitudinal and time-to-event data. \textit{Biometrics} \textbf{67}, 819-829.
	
      
  
      \bibitem{ron15}
      Rondeau, V., Pignon, J. P. and Michiels, S. (2015) A joint model for dependence between clustered times to tumour progression and
      deaths: A meta-analysis of chemotherapy in head and neck cancer. \textit{Stat Meth Med Res} \textbf{24}: 711-729.
      
       \bibitem{roy03}
      	Roy, J. (2003). Modelling longitudinal data with nonignorable dropouts
      using a latent dropout class model. \textit{Biometrics} \textbf{59}, 829-836. 
      
      	\bibitem{sur21a}
      Suresh, K., Taylor, J. M. G., and Tsodikov, A. (2021). A Gaussian copula approach for dynamic prediction of survival with a longitudinal biomarker. \textit{Biostatistics}  Volume \textbf{22}, Issue 3, 504-521.
      

            
      \bibitem{tsi04}
      Tsiatis, A. A. and Davidian, M. (2004). Joint modelling of longitudinal and time-to-event data: An overview. \textit{Statistica Sinica} \textbf{14,} 809-834.
      
       \bibitem{tsi95}
      Tsiatis, A. A., DeGruttola, V., and Wulfsohn, M. (1995). Modelling the relationship of survival to longitudinal data measured with error: Applications to
      survival and CD4 counts in patients with AIDS. \textit{Journal of the American Statistical Association} \textbf{90}, 27-37.
      
      \bibitem{wan01}
      Wang, Y. and Taylor, J. M. G. (2001). Jointly modelling longitudinal and event time data with application to acquired immunodeficiency syndrome. \textit{Journal of the American Statistical Associatio} \textbf{96},
      895-905.
      
      \bibitem{wan24a}
      Wang, C., Shen, J. Charalambous C., and Pan, J. (2024a). Modeling biomarker variability in joint analysis of
      longitudinal and time-to-event data. \textit{Biostatistics} \textbf{25,} 577-596.
      
      \bibitem{wan24b}
      Wang, C., Shen, J. Charalambous C., and Pan, J. (2024b). Weighted biomarker variability in joint analysis of longitudinal and time-to-event data. \textit{The Annals of Applied Statistics} \textbf{18,} 2576-2595.
      
       \bibitem{whi82}
      White, H. (1982). Maximum likelihood estimation of misspecified models. \textit{Econometrica} \textbf{50}, 1-26.
      
      \bibitem{wul97}
      Wulfsohn, M. S. and Tsiatis, A. A. (1997). A joint model for survival and longitudinal data measured with error. \textit{Biometrics} \textbf{53,} 330-339.
      
      \bibitem{yao07}
      Yao, F. (2007). Functional Principal Component Analysis for Longitudinal and Survival Data, \textit{Statistica Sinica} \textbf{17}, 965-983.
      
      \bibitem{yua07}
      Yan, J. (2007). Enjoy the joy of copulas: With a package copula, \textit{Journal of Statistical Software,} \textbf{21}, issue 4.
      
     	\bibitem{zha23a}
     Zhang, Z., Charalambous, C., and Foster, P. (2023a). Joint modelling of longitudinal measurements and survival times via a multivariate copula approach. \textit{Journal of Applied Statistics,}  \textbf{50}, 2739-2759.
      
      \bibitem{zha23b}
      Zhang, Z., Charalambous, C., and Foster, P. (2023b). A Gaussian copula joint model for longitudinal and time-to-event data with random effects. \textit{Computational Statistics and Data Analysis,}  \textbf{181}.
\end{thebibliography}
\end{document}